\documentclass[aps,twocolumn,amsmath,amssymb,floatfix,superscriptaddress,nofootinbib]{revtex4-2}

\usepackage{graphicx}
\usepackage{dcolumn}
\usepackage{bm} 
\usepackage{mathptmx}
\usepackage{mathtools} 
\usepackage{soul}
\usepackage{etoolbox}
\usepackage{dcolumn}
\usepackage{color}
\usepackage{xcolor}
\usepackage[colorlinks=true,linkcolor=blue,citecolor=blue]{hyperref}
\usepackage[utf8]{inputenc}
\usepackage[T1]{fontenc}
\usepackage[english]{babel}
\usepackage{booktabs} 
\usepackage{verbatim}

\usepackage{subcaption}
\usepackage{ragged2e} 
\DeclareCaptionJustification{justified}{\justifying}
\usepackage{pict2e}
\usepackage{lipsum} 
\usepackage{todonotes} 
\usepackage{physics}
\usepackage{adjustbox}
\usepackage{makecell}
\usepackage{float}
\usepackage[none]{hyphenat}

\draft 

\begin{document}


\title{Laser-driven ion and electron acceleration from near-critical density gas targets: towards high-repetition rate operation in the 1~PW, sub-100~fs laser interaction regime} 

\author{V.~Ospina-Boh\'{o}rquez}
\email{vale1504@usal.es}
\altaffiliation[\\ Currently at: ]{Focused Energy GmbH, Darmstadt, Germany}\email{valeria.ospina@focused-energy.world}
\affiliation{Universit\'{e} de Bordeaux, CNRS, CEA, CELIA (Centre Lasers Intenses et Applications), UMR 5107, Talence, France}
\affiliation{CEA, DAM, DIF - 91297 Arpajon, France}
\affiliation{Universit\'e Paris-Saclay, CEA, LMCE, 91680 Bruy\`eres-le-Ch\^atel, France}
\affiliation{Departamento de F\'{i}sica Fundamental, Universidad de Salamanca, 37008 Salamanca, Spain}

\author{C.~Salgado-L\'{o}pez}
\affiliation{Centro de L\'{a}seres Pulsados (CLPU), Parque Cient\'{i}fico, E-37185 Villamayor, Salamanca, Spain}

\author{M.~Ehret}
\affiliation{Centro de L\'{a}seres Pulsados (CLPU), Parque Cient\'{i}fico, E-37185 Villamayor, Salamanca, Spain}

\author{S.~Malko}
\affiliation{Princeton Plasma Physics Laboratory, 100 Stellarator Road, Princeton, NJ 08540}

\author{M.~Salvadori}
\altaffiliation[Currently at: ]{INO-CNR, Pisa, Italy}
\affiliation{ENEA, Fusion and Technologies for Nuclear Safety and Security Department, C.R. Frascati, Via Enrico Fermi 45, Frascati, Italy}

\author{T.~Pisarczyk}
\affiliation{Institute of Plasma Physics and Laser Microfusion, 23 Hery St., 00-908 Warsaw, Poland}

\author{T.~Chodukowski}
\affiliation{Institute of Plasma Physics and Laser Microfusion, 23 Hery St., 00-908 Warsaw, Poland}

\author{Z.~Rusiniak}
\affiliation{Institute of Plasma Physics and Laser Microfusion, 23 Hery St., 00-908 Warsaw, Poland}

\author{M.~Krupka}
\affiliation{Institute of Plasma Physics CAS, Prague, Czech Republic}

\author{P.~Guillon}
\affiliation{Universit\'{e} de Bordeaux, CNRS, CEA, CELIA (Centre Lasers Intenses et Applications), UMR 5107, Talence, France}

\author{M.~Lendrin}
\affiliation{Universit\'{e} de Bordeaux, CNRS, CEA, CELIA (Centre Lasers Intenses et Applications), UMR 5107, Talence, France}

\author{G.~P\'{e}rez-Callejo}
\affiliation{Universit\'{e} de Bordeaux, CNRS, CEA, CELIA (Centre Lasers Intenses et Applications), UMR 5107, Talence, France}
\affiliation{Departamento de F\'{i}sica Te\'{o}rica, At\'{o}mica y \'{O}ptica, Universidad de Valladolid, 47011 Valladolid, Spain.}

\author{C.~Vlachos}
\affiliation{Universit\'{e} de Bordeaux, CNRS, CEA, CELIA (Centre Lasers Intenses et Applications), UMR 5107, Talence, France}

\author{F.~Hannachi}
\affiliation{Universit\'{e} de Bordeaux, CNRS, LP21 Bordeaux, UMR 5797, Gradignan, France}

\author{M.~Tarisien}
\affiliation{Universit\'{e} de Bordeaux, CNRS, LP21 Bordeaux, UMR 5797, Gradignan, France}

\author{F.~Consoli}
\affiliation{ENEA, Fusion and Technologies for Nuclear Safety and Security Department, C.R. Frascati, Via Enrico Fermi 45, Frascati, Italy}

\author{C.~Verona}
\affiliation{Department of Industrial Engineering, University of Rome “Tor Vergata”, Rome, Italy}

\author{G.~Prestopino}
\affiliation{Department of Industrial Engineering, University of Rome “Tor Vergata”, Rome, Italy}

\author{J.~Dostal}
\affiliation{Institute of Plasma Physics CAS, Prague, Czech Republic}

\author{R.~Dudzak}
\affiliation{Institute of Plasma Physics CAS, Prague, Czech Republic}

\author{J.~L.~Henares}
\affiliation{Centro de L\'{a}seres Pulsados (CLPU), Parque Cient\'{i}fico, E-37185 Villamayor, Salamanca, Spain}

\author{J.~I.~Api\~{n}aniz}
\affiliation{Centro de L\'{a}seres Pulsados (CLPU), Parque Cient\'{i}fico, E-37185 Villamayor, Salamanca, Spain}

\author{D.~De~Luis}
\affiliation{Centro de L\'{a}seres Pulsados (CLPU), Parque Cient\'{i}fico, E-37185 Villamayor, Salamanca, Spain}

\author{A.~Debayle}
\affiliation{CEA, DAM, DIF - 91297 Arpajon, France}
\affiliation{Universit\'e Paris-Saclay, CEA, LMCE, 91680 Bruy\`eres-le-Ch\^atel, France}

\author{J.~Caron}
\affiliation{Universit\'{e} de Bordeaux, CNRS, CEA, CELIA (Centre Lasers Intenses et Applications), UMR 5107, Talence, France}

\author{T.~Ceccotti}
\affiliation{Université Paris-Saclay, CEA, LIDYL, 91191 Gif-sur-Yvette, France}

\author{R.~Hern\'{a}ndez-Mart\'{i}n}
\affiliation{Centro de L\'{a}seres Pulsados (CLPU), Parque Cient\'{i}fico, E-37185 Villamayor, Salamanca, Spain}

\author{J.~Hern\'{a}ndez-Toro}
\affiliation{Centro de L\'{a}seres Pulsados (CLPU), Parque Cient\'{i}fico, E-37185 Villamayor, Salamanca, Spain}

\author{M.~Huault}
\affiliation{Departamento de F\'{i}sica Fundamental, Universidad de Salamanca, 37008 Salamanca, Spain}
\affiliation{Centro de L\'{a}seres Pulsados (CLPU), Parque Cient\'{i}fico, E-37185 Villamayor, Salamanca, Spain}

\author{A. Mart\'{i}n-L\'{o}pez}
\affiliation{Centro de L\'{a}seres Pulsados (CLPU), Parque Cient\'{i}fico, E-37185 Villamayor, Salamanca, Spain}

\author{C.~Méndez}
\affiliation{Centro de L\'{a}seres Pulsados (CLPU), Parque Cient\'{i}fico, E-37185 Villamayor, Salamanca, Spain}

\author{T.-H.~Nguyen-Bui}
\affiliation{Universit\'{e} de Bordeaux, CNRS, CEA, CELIA (Centre Lasers Intenses et Applications), UMR 5107, Talence, France}

\author{J.~A.~Perez-Hern\'{a}ndez}
\affiliation{Centro de L\'{a}seres Pulsados (CLPU), Parque Cient\'{i}fico, E-37185 Villamayor, Salamanca, Spain}

\author{X.~Vaisseau}
\affiliation{Centro de L\'{a}seres Pulsados (CLPU), Parque Cient\'{i}fico, E-37185 Villamayor, Salamanca, Spain}

\author{O.~Varela}
\affiliation{Centro de L\'{a}seres Pulsados (CLPU), Parque Cient\'{i}fico, E-37185 Villamayor, Salamanca, Spain}

\author{L.~Volpe}
\affiliation{Centro de L\'{a}seres Pulsados (CLPU), Parque Cient\'{i}fico, E-37185 Villamayor, Salamanca, Spain}
\affiliation{ETSIAE Universidad Politecnica de Madrid, 28006 Madrid, Spain}

\author{L.~Gremillet}
\email{laurent.gremillet@cea.fr}
\affiliation{CEA, DAM, DIF - 91297 Arpajon, France}
\affiliation{Universit\'e Paris-Saclay, CEA, LMCE, 91680 Bruy\`eres-le-Ch\^atel, France}

\author{J.~J.~Santos}
\email{joao.santos@u-bordeaux.fr}
\affiliation{Universit\'{e} de Bordeaux, CNRS, CEA, CELIA (Centre Lasers Intenses et Applications), UMR 5107, Talence, France}

\begin{abstract}
Ion acceleration from gaseous targets driven by relativistic-intensity lasers was demonstrated as early as the late 90s, yet most of the experiments conducted to date have involved
picosecond-duration, Nd:glass lasers operating at low repetition rate.
Here, we present measurements on the interaction of ultraintense ($\sim~10^{20}\,\rm W\,cm^{-2}$, 1~PW), ultrashort ($\sim 70\,\rm fs$) Ti:Sa laser pulses with near-critical ($\sim 10^{20}\,\rm cm^{-3}$) helium gas jets, a debris-free targetry compatible with high ($\sim 1\,\rm Hz$) repetition rate operation. We provide evidence of $\alpha$ particles being forward accelerated up to $\sim 2.7\,\rm MeV$ energy with a total flux of $\sim 10^{11}\,\rm sr^{-1}$ as integrated over $>0.1 \,\rm MeV$ energies and detected within a $0.5\,\rm mrad$ solid angle. We also report on on-axis emission of relativistic electrons with an exponentially decaying spectrum characterized by a $\sim 10\,\rm MeV$ slope, i.e., five times larger than the standard ponderomotive scaling. The total charge of these electrons with energy above 2~MeV is estimated to be of $\sim 1 \,\rm nC$, corresponding to $\sim 0.1\,\%$ of the laser drive energy. 
In addition, we observe the formation of a plasma channel, extending longitudinally across the gas density maximum and expanding radially with time. These results are well captured by large-scale particle-in-cell simulations, which reveal that the detected fast ions most likely originate from reflection off the rapidly expanding channel walls. The latter process is predicted to yield ion energies in the MeV range, which compare well with the measurements. Finally, direct laser acceleration is shown to be the dominant mechanism behind the observed electron energization.
\end{abstract}

\maketitle

\section{Introduction}
\label{sec:Introduction}

Laser-driven ion beams \cite{Daido_2012, Macchi_2013, Schreiber_2016} are spurring increasing interest because of their many established or potential uses in science and industry. Not only can they serve as ultrafast diagnostic tools of dynamic plasma systems  \cite{Borghesi_2003,Borghesi_2004,Santos_2015,Ehret_2023}, they can also generate high-energy-density states of matter \cite{Patel_2003, Roth_2009, Mancic_2010} or secondary particle sources such as neutrons \cite{Roth_2013, Kleinschmidt_2018, Horny_2022} or radioiosotopes \cite{Spencer_2001, Fritzler_2003} with possible medical spinoffs \cite{Ledingham_2014}. Most of these applications exploit the unique properties of laser-driven ion beams, notably their short ($\sim \rm ps$) duration, high number density, low emittance, high laminarity and compactness \cite{Cowan_2004}. Yet some of those uses
require substantial advances in repetition rate, which should at least approach that of the laser system. This is the case, for example, for achieving the high time-averaged particle fluxes needed for nuclear astrophysical studies \cite{Chen_2019, Horny_2023} or for producing the pulsed ion sources used for neutron-related applications and isotope creation \cite{Kleinschmidt_2018, Gunther_2022, Iwamoto_2020}.

A major current effort in the experimental laser-plasma community is thus geared towards developing ion accelerator setups that make the most of the high-repetition-rate (HRR) capability of modern 1-PW-class, few-femtosecond Ti:Sa laser systems, such as LULI Apollon (France) \cite{Burdonov_2021}, CLPU VEGA-3 (Spain) \cite{Salvadori_2022}, CoReLS (Korea) \cite{Sung_2017}, ELI-Beamlines L4 Aton (Czech Republic) \cite{Rus_2017}, ELI-NP HPLS (Romania) \cite{Gales_2018}, BELLA PW iP2 (USA) \cite{Hakimi_2022} or University of Michigan ZEUS (USA) \cite{Maksimchuk_2021}. These are indeed designed to run at a rate (at least one shot per second) much higher than that (about one shot per hour) accessible to older picosecond, Nd:glass laser systems such as LLNL Titan (USA), RAL Vulcan (UK), GSI PHELIX (Germany) or LULI PICO2000 (France). Not only would such a progress allow for a considerable enhancement of the time-averaged particle flux over the state of the art \cite{Horny_2022}, it would also enable much more statistically robust studies of diverse physical phenomena \cite{Prencipe_2017}.

High-density gas jets are able to provide electron densities close to the critical density of $0.8\,\rm \mu m$ wavelength Ti:Sa lasers ($n_c = 1.74 \times 10^{21} \ \mathrm{cm^{-3}}$) after ionization. Compared to commonly employed solid foils, these gas systems offer a debris-free, HRR-compatible targetry option while also ensuring efficient energy coupling with the laser pulse. Their practical use, however, is made difficult by the need to achieve well controlled, reproducible density profiles that do not lead to premature absorption of the laser pulse before reaching the peak density region. Moreover, the laser should be shot far enough from the nozzle not to damage it by the plasma plume
\cite{Ehret_2020} and electromanetic pulse perturbations on the gas valves, nozzles and general electronics need to be understood and controlled \cite{Bradford_2023}.

Most previous experiments on laser-driven ion acceleration from gases have made use of low-repetition-rate ($\sim~1$~shot/hour), $\sim 1\,\rm ps$ duration, $1\,\rm \mu m$ wavelength Nd:glass lasers. As early as in 1999, Krushelnick \textit{et al.} measured He ions transversely accelerated up to 4~MeV energies from a He gas jet (of peak electron density $n_{e,\rm max} \simeq 0.045\,n_c$) acted upon by the Vulcan laser ($I_L \simeq 6\times 10^{19}\,\rm W\,cm^{-2}$) \cite{Krushelnick_1999}. Coulomb explosion \cite{Sentoku_2000} in the laser-drilled channel was then invoked  as the main acceleration mechanism. More recently, coupling the Titan laser ($I_L \simeq 2\times 10^{19}\,\rm W\,cm^{-2}$) with a high-density ($n_{e,\rm max}~\simeq~2.5\,n_c$) hydrogen gas jet, Chen \textit{et al.} reported on forward acceleration of protons up to $0.6\,\rm MeV$ and with a low ($\sim 2\%$) energy spread \cite{Chen_2017}, which they ascribed to ion reflection from a collisionless electrostatic shock \cite{Fiuza_2012, *Fiuza_2013, Dieckmann_2013a, *Dieckmann_2013b, Lecz_2015, Liu_2016} as previously observed using a CO$_2$ laser \cite{Haberberger_2012}. Next, Puyuelo-Vald\'{e}s \textit{et al.} \cite{Puyuelo-Valdes_2019} leveraged the PICO2000 laser ($I_L \simeq 5\times 10^{19}\, \rm W\,cm^{-2}$) and a H$_2$ gas jet ($n_{e,\rm max}~\simeq~2.3\,n_c$) to generate slightly peaked proton spectra extending up to $\sim 6\,\rm MeV$. Laser hole boring \cite{Wilks_1992} was held responsible for these results. In a follow-up experiment making use of a $\rm H_2/He$ gas mixture, they detected transversely accelerated $\alpha$ particles up to 15~MeV energies \cite{Puyuelo-Valdes_2019b}.

Much fewer experiments have been performed on HHR-compatible, sub-$100\,\rm fs$, $0.8\,\rm \mu m$ Ti:Sa laser systems. In 2013 Sylla \textit{et al.} observed He$^+$ ions transversely accelerated up to $\sim 0.25\,\rm MeV$ in a $n_{e,\rm max} \simeq 0.2\,n_c$ He gas jet delivered by a supersonic conical nozzle and irradiated by LOA's Salle Jaune laser ($I_L \simeq 10^{19}\,\rm W\,cm^{-2}$) \cite{Sylla_2013}. Lately, Singh \textit{et al.} were the first to investigate ion acceleration from gas jets in the PW regime at the CoReLS facility ($I_L \simeq 10^{20}\, \rm W\,cm^{-2}$), using a gas jet (delivered by a Laval nozzle) made of He and a small fraction of H$_2$ with $n_{e,\rm max} \simeq 0.2\,n_c$ \cite{Singh_2020}. Radial collisionless shock acceleration (CSA) was invoked to explain the detection in the transverse direction of energetic protons and $\alpha$ ($\mathrm{He^{2+}}$) ions up to $\sim 0.75\,\rm MeV$, and characterized by exponentially decaying spectra. These proof-of-concept experiments, however, were not intended to assess the HHR capability of the implemented gas-based setups.

In this article, we present and analyze the results of one of the first experiments on particle acceleration from a near-critical ($n_{e,\rm max} \simeq 0.1\,n_c$) supersonic Helium gas jet irradiated by a PW-class, $\sim 10^{20}\,\rm W\,cm^{-2}$, sub-100~fs, Ti:Sa laser pulse. By means of extensive diagnostics and large-scale particle-in-cell (PIC) modeling, we examine the ion and electron acceleration processes at play as well as the bulk plasma dynamics resulting from the strong laser-gas coupling. In particular, we provide evidence for forward oblique acceleration of $\alpha$ particles up to \mbox{$\sim 2.7\,\rm MeV$}, which we interpret as mainly originating from the observed fast-expanding laser-created channel, formed all across the dense gas region. This mechanism is similar to that accounting for the transverse ion acceleration seen in Ref.~\cite{Singh_2020}. Moreover, it better agrees with our measurements than the target normal sheath acceleration (TNSA) that our PIC simulations also predict to arise at the edge of the gas down-ramp. In addition, electron energization well above the ponderomotive scaling is detected, and attributed to direct laser acceleration (DLA) \cite{Arefiev_2016, Hussein_2021, Shaw_2017, Shaw_2018} inside the laser-created channel. Our simulations further indicate that the ions may undergo significant deflections in the magnetic fields induced by the electron currents at the plasma boundary. Finally, we discuss the current repetition-rate capability of our laser-gas setup and identify laser-induced nozzle damage \cite{Ehret_2020} as its main limitation factor.

\section{Experimental setup}
\label{sec:Experimental_setup}

\begin{figure*}
    \centering
    \includegraphics[width=1\linewidth]{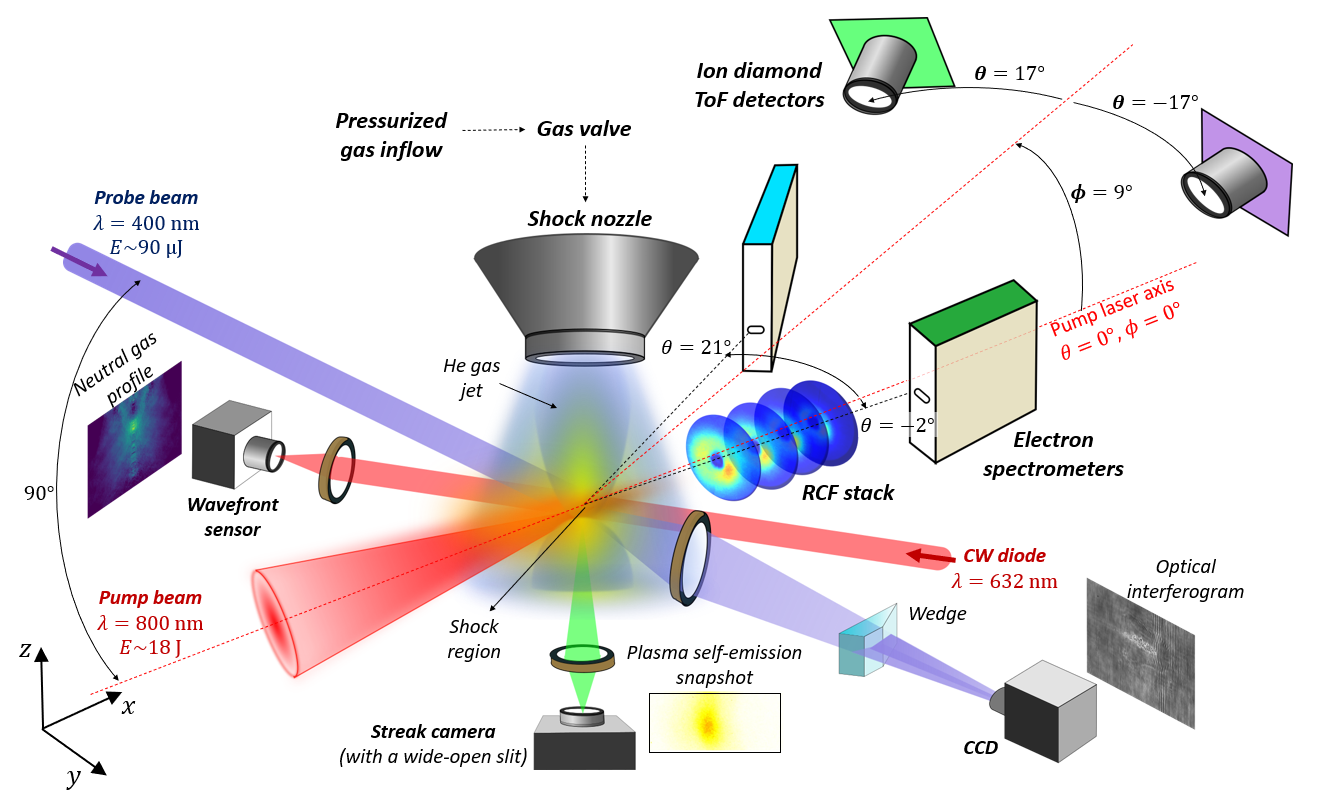}
	\caption{Experimental setup. The gas jet system consists of pressurized gas bottles connected to a gas valve, itself attached to a shock nozzle, located near the target chamber center (TCC). The main VEGA-3 laser beam, propagating in the equator plane of the chamber, is focused in the shocked gas region, formed by the converging gas flows at $\simeq 500\,\rm \mu m$ from the nozzle exit surface. A $\simeq 70\,\rm fs$ probe beam, crossing the TCC at $90^{\circ}$ from the main beam, allows one to acquire on-shot optical interferograms of the ionized gas at different temporal delays using a folding-wave interferometer. Interferograms of the neutral gas prior to each UHI shot are also obtained using a continuous He-Ne laser sent through TCC and imaged onto a wavefront sensor. A bottom-view optical line is used for nozzle alignment. It is also connected to a streak camera operated in gate mode to record snapshots of the plasma self-emission. Particle diagnostics comprise two ion time-of-flight diamond detectors [located in an inclined plane with respect to the equatorial plane at the angles $(\theta,\phi) = (+17^{\circ},9^{\circ})$ and $(-17^{\circ},9^{\circ})$, as defined in the schematic], two permanent magnet electron spectrometers [placed in the equatorial place at $(\theta,\phi) = (-2^{\circ},0^{\circ})$ and $(21^{\circ},0^{\circ})$, and a stack of radiochromic films (located on axis at $6\,\rm cm$ from TCC). In some shots, the latter diagnostic was replaced by a Thomson parabola placed $10\,\rm cm$ from TCC.
 }
	\label{fig:setup} 
\end{figure*}

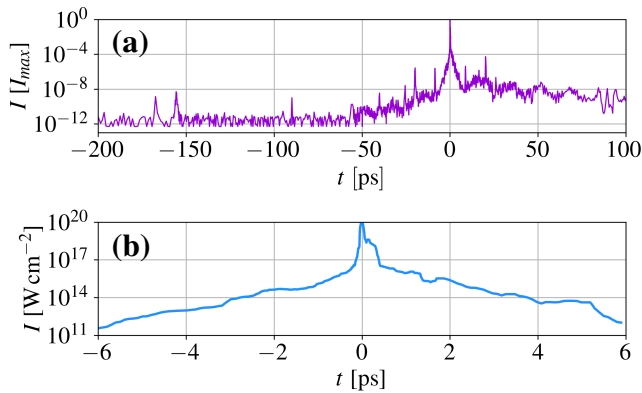
\begin{figure}
    \centering
	\includegraphics[width=\linewidth]{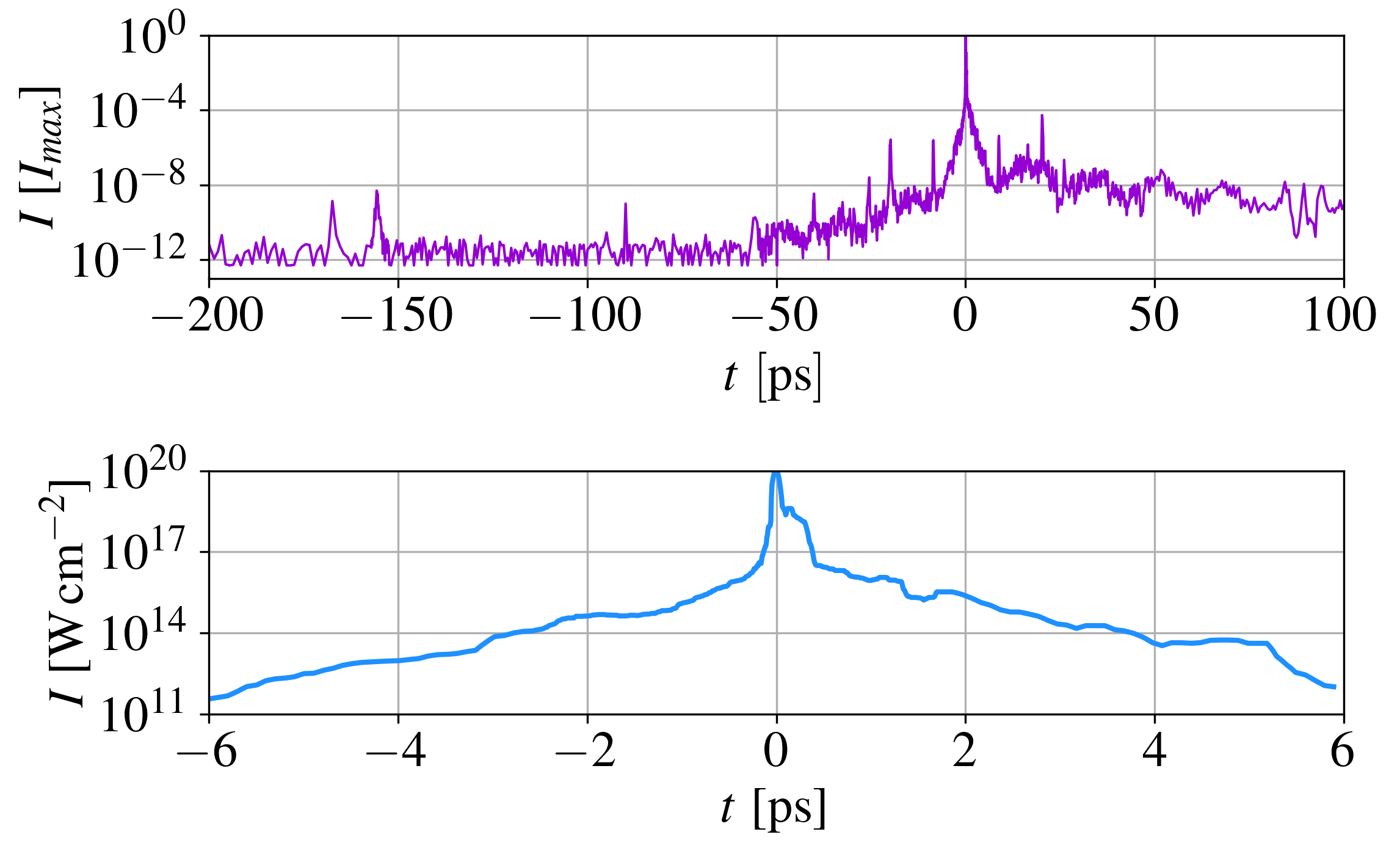}
    \begin{picture}(20,20)
    \put(-70,152){\large\textbf{(a)}}
    \put(-70,75){\large\textbf{(b)}}
    \end{picture}
    \vspace{-10mm}
	\caption{(a) Laser temporal contrast measurement. (b) Laser intensity profile used in the PIC simulation. The laser intensity peaks at the time origin ($t=0$).}
	\label{fig:laser}
\end{figure}

The experiment was performed at the Centro de L\'{a}seres Pulsados (CLPU) facility using the PW-class, ultrahigh-intensity (UHI) Ti:Sa VEGA-3 laser system. The experimental setup is depicted in Fig.~\ref{fig:setup}. 

The $\lambda_L=0.8\,\rm \mu m$ wavelength laser pulse delivered an energy of $18.4\pm 2.3\,\rm J$ on target. It had a full-width-at-half-maximum (FWHM) duration of $\tau_L=72.4\pm~23.6\,\rm fs$ (monitored on each shot) and was focused using an $f/10$ off-axis parabola to a $D_L=14.1\pm1.2 \,\rm \mu m$ FWHM spot size (monitored on a daily basis). These characteristics represent the mean and standard deviation values from more than 40 measurements taken during the campaign. The pulse energy is extrapolated from calibrations at low energy. The same energy level is also used to image the focal spot, which is assume invariant at high energy. A peak pulse intensity of $I_L=(8.7 \pm 2.1)\times 10^{19}\,\rm W\,cm^{-2}$ was then inferred, corresponding to a normalized laser amplitude of $a_L \simeq 6.3$. The laser contrast was measured to be $\sim 10^{12}$ up to $\sim -100\,\rm ps$ prior to the intensity peak, and $\sim 10^{8}$ up to $-5\,\rm ps$ [see Fig.~\ref{fig:laser}(a)].

The gas jet was produced by a so-called shock nozzle, consisting of a cylindrical Laval nozzle with an extra straight conduct added at its exit \cite{Rovige_2021}. Upon bouncing off the latter, the converging hydrodynamic flow produces a peaked density profile along the laser propagation direction. Both the nozzle and gas valve were located near the target chamber center (TCC), their position being adjusted vertically to make the laser pulse interact with the shocked gas region. The high-pressure gas system comprised a SL-GT-10 gas compressor (with 400~bar backing pressure) and a gas valve, both commercialized by SourceLAB \cite{SourceLab}. The valve was designed to avoid leakages in harsh UHI laser environments by means of a normally-closed opening system adapted to the millimeter-sized throat of the nozzle. Moreover, a security system \cite{MartinLopez_2021} automatically closed the valves between the vacuum chambers to protect the laser transport pipes and vacuum components from strong pressure rises during UHI shots.

Two nozzle types were fielded, one provided by SourceLAB, the other developed within our collaboration group \cite{Puyuelo-Valdes_2020, Henares_2019}. Both were designed to produce shocked gas regions with sharp density gradients and located as far as possible from the nozzle to mitigate laser damage to the latter. 
However, albeit not severe enough to disrupt the laser-plasma interaction, such damage turned out to be significant enough to gradually smooth the density profiles (see orange shaded area in Fig.~\ref{fig:profile}), which hence ended up looking similar to those achieved with Laval nozzles. 

A wavefront sensor and a folding-wave interferometer \cite{Pisarczyk_2019,Zaras_2020} allowed us to obtain two-dimensional profiles of, respectively, the pre-shot neutral gas atomic density
and on-shot plasma electron density.
The first diagnostic used a continuous He-Ne laser while the second used a $\sim 90\,\rm \mu J$ probe beam (a pickoff of the pump beam) sent through TCC perpendicularly to the main laser's path and at different temporal delays. 

\begin{figure}
    \centering
    \includegraphics[width=0.8\linewidth]{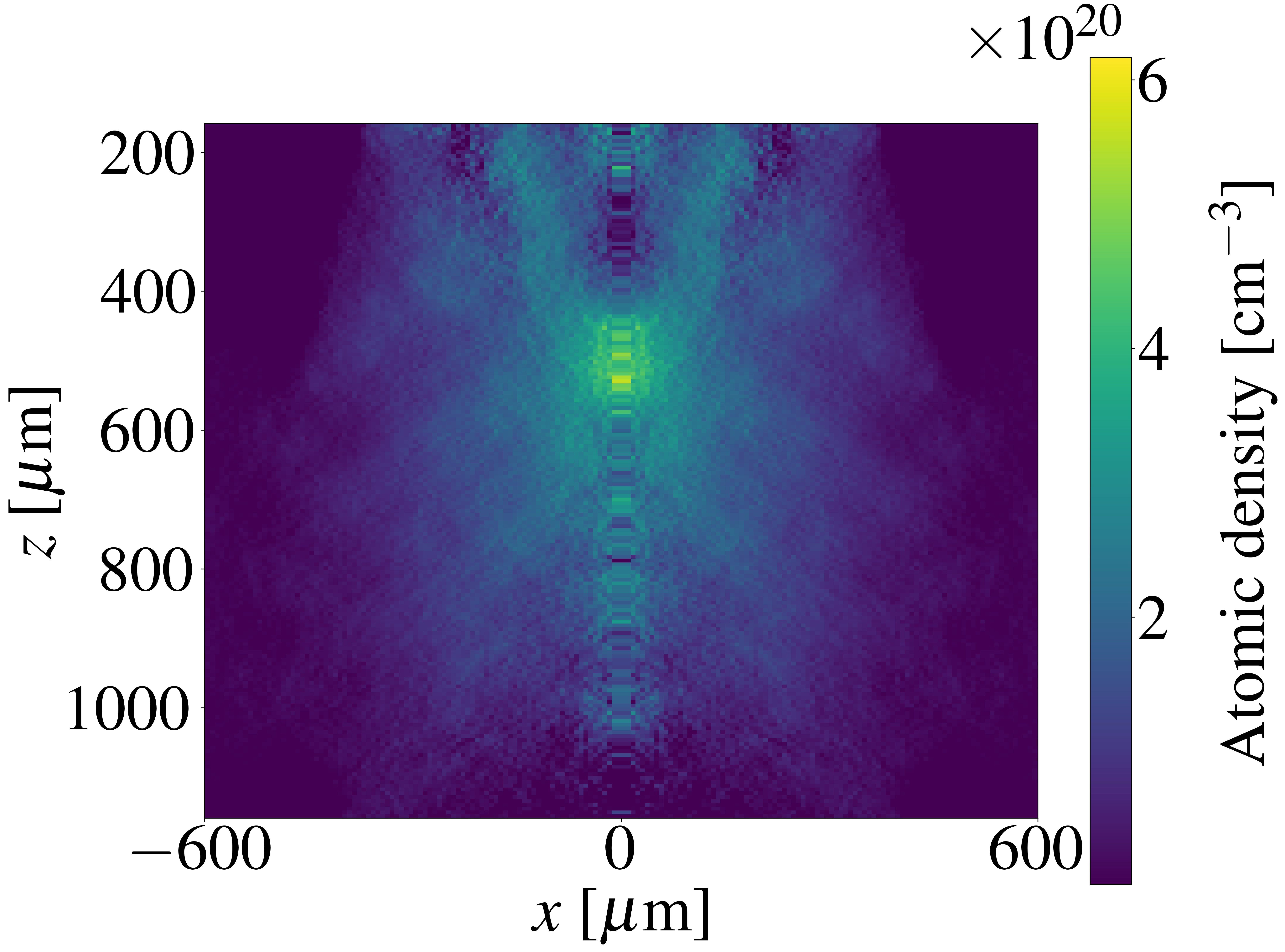}
	\caption{Atomic density map of the neutral He gas jet as obtained before a UHI laser shot.
 In this case, the laser pulse was injected along $x$ at $z \simeq 650\,\mu \rm m$, targeting a maximum atomic density of $\sim 10^{20}\,\rm cm^{-3}$.
 }
	\label{fig:neutral_gas} 
\end{figure}

Optical self-emission images of the plasma, integrated over a $180\,\rm ps$ time window, were also recorded using a Hamamatsu S20 streak camera operating in gate mode (with a wide-open slit), combined with a $532\,\rm nm$ band-pass filter. The same imaging system was connected to a charged coupled device (CCD) camera to monitor the nozzle position between shots and adjust it if needed. The laser axis was controlled separately by imaging the defocused beam $15\,\rm cm$ after TCC along the expected axis.

The particle diagnostic suite included two diamond time-of-flight (ToF) detectors \cite{Verona_2020, Salvadori_2021, Salvadori_2022} placed on the chamber flanges, at angles $(\theta, \phi) = (+17^{\circ},9^{\circ})$ and $(\theta, \phi) = (-17^{\circ}, 9^{\circ})$ relative to the laser axis, as defined in Fig.~\ref{fig:setup} (negative $\theta$ values correspond to clockwise rotation as seen from top view). In addition, two magnet electron spectrometers (MES) \cite{Krupka_2021} were positioned at $(\theta, \phi)=(-2^{\circ},0^{\circ})$ and $(\theta,\phi) = (21^{\circ},0^{\circ})$. Both ToF \cite{Salvadori_2021} and MES \cite{Krupka_2021} systems were fully calibrated. Moreover, to characterize the transverse profile of the accelerated particles on some shots, a stack of radiochromic films (RCF) was placed $6\,\rm cm$ from TCC, normally to the laser axis. It was composed of a $10\,\rm \mu m$ thick aluminum foil followed by five unlaminated EBT-3 films and ten standard EBT-3 Gafchromic dosimetric films. Each layer had a 4~cm diameter exposed surface. A motorized holder allowed several shots to be performed in a row without opening the vacuum chamber. Finally, on some shots, an angle-resolved Thomson parabola (TP) \cite{Salgado_2022}, coupled with a Fuji BAS-TR imaging plate, was fielded on axis at $100\,\rm cm$ from TCC, with three horizontally aligned, $200\,\rm \mu m$ diameter pinholes at its entrance. These were spaced $3\,\rm mm$ apart to capture ion spectra at $\theta = \pm 1.7^{\circ}$ and $0^{\circ}$. A manual extraction system allowed the IPs to be replaced between shots without breaking vacuum. All those diagnostics were successfully tested on laser-solid shots (using $6\,\rm \mu m$ thick aluminium foil targets) \cite{Ospina_2022}, before moving on to laser-gas shots.

In the following, we will examine the experimental results from five shots on a He gas jet. Figure~\ref{fig:neutral_gas} displays a typical atomic density map of the pre-shot neutral gas jet while  Fig.~\ref{fig:profile} shows (as an orange shaded area) the range of the atomic density profiles recovered along the laser path. Maximum atomic densities of $\simeq (1-2) \times 10^{20}\,\rm cm^{-3}$, corresponding to maximum electron densities of $n_{e,\rm max} \simeq 0.1-0.2\,n_c$ at full ionization, were achieved with a $\sim 500-600\,\rm \mu m$ FWHM. The two shock nozzle types that were used delivered similar gas profiles. Shot-to-shot variations due to laser-induced nozzle damage were mitigated by adjusting the vertical ($z$) position of the shocked region before each UHI shot \cite{Ospina_2022}. Nevertheless, after a couple of shots, one observed the appearance of low-density ($\sim (1-2)\times 10^{19}\,\rm cm^{-3}$) spatial ``elbows'' (shaded areas beyond $\vert x \vert \gtrsim 700\,\rm \mu m$ in Fig.~\ref{fig:profile}). The non-damaged nozzle profiles were similar to the one (red dashed curve) used as input to the PIC simulations. 
The laser pulse was focused either to the TCC, where the gas density peak (GDP) was located, or \mbox{$\sim 250\,\rm \mu m$} in front of it so as to weaken laser self-focusing and filamentation \cite{Sprangle_1987}.

\section{PIC simulation setup}
\label{sec:PIC_simulation}

\begin{figure}
    \centering
	\includegraphics[width=0.85\linewidth]{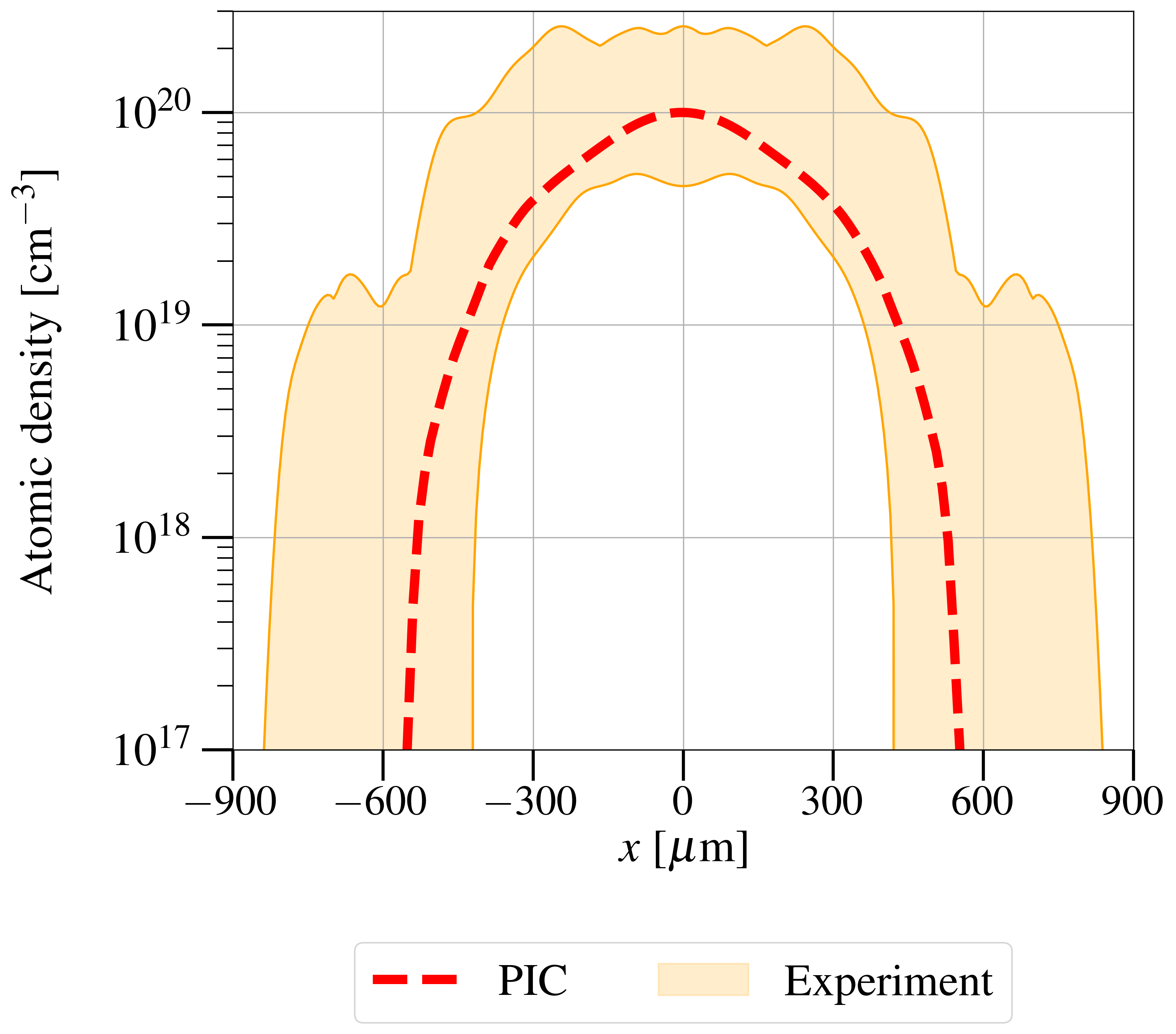}
	\caption{Atomic density profile of the neutral gas along the laser propagation ($x$) axis. Light orange shaded area: scatter of five density profiles acquired before different laser shots (from which the experimental data shown below has been extracted). Red dashed curve: gas density profile used in the PIC simulations.}
    \label{fig:profile}
\end{figure}

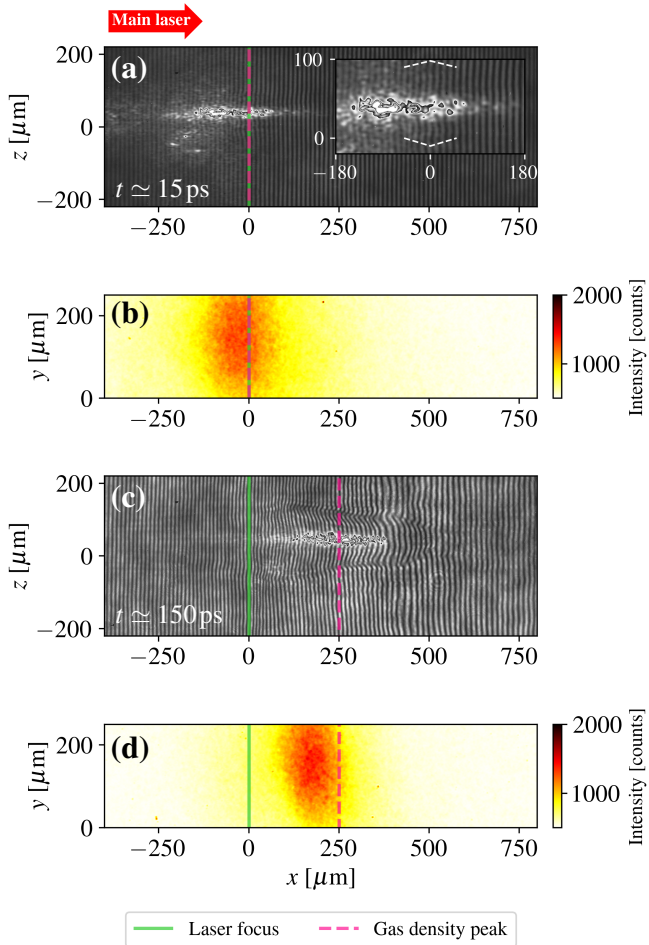
\begin{figure}
	\centering
	\begin{subfigure}[b]{0.49\textwidth}
		\centering
		\includegraphics[width=\textwidth]{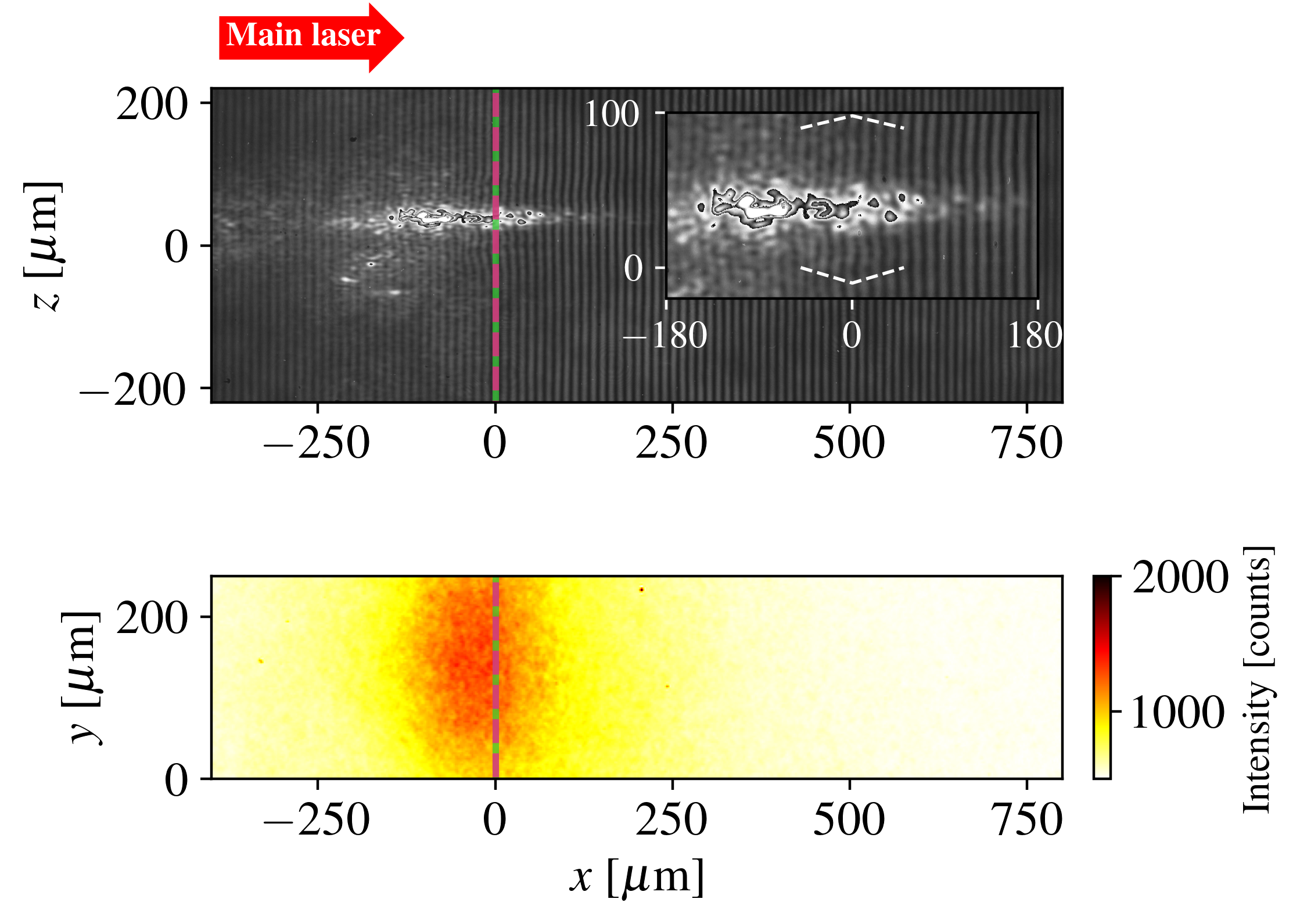}
		\begin{picture}(20,20)
		\put(-72,170){\color{white}\large\textbf{(a)}}%
            \put(-70,125){\color{white}\normalsize{$t \simeq 15 \, \rm{ps}$}}%
            \put(-72,77){\large\textbf{(b)}}
		\end{picture}	
	\end{subfigure}
	\begin{subfigure}[b]{0.49\textwidth}
		\centering
        \vspace{-13mm}
		\includegraphics[width=\textwidth]{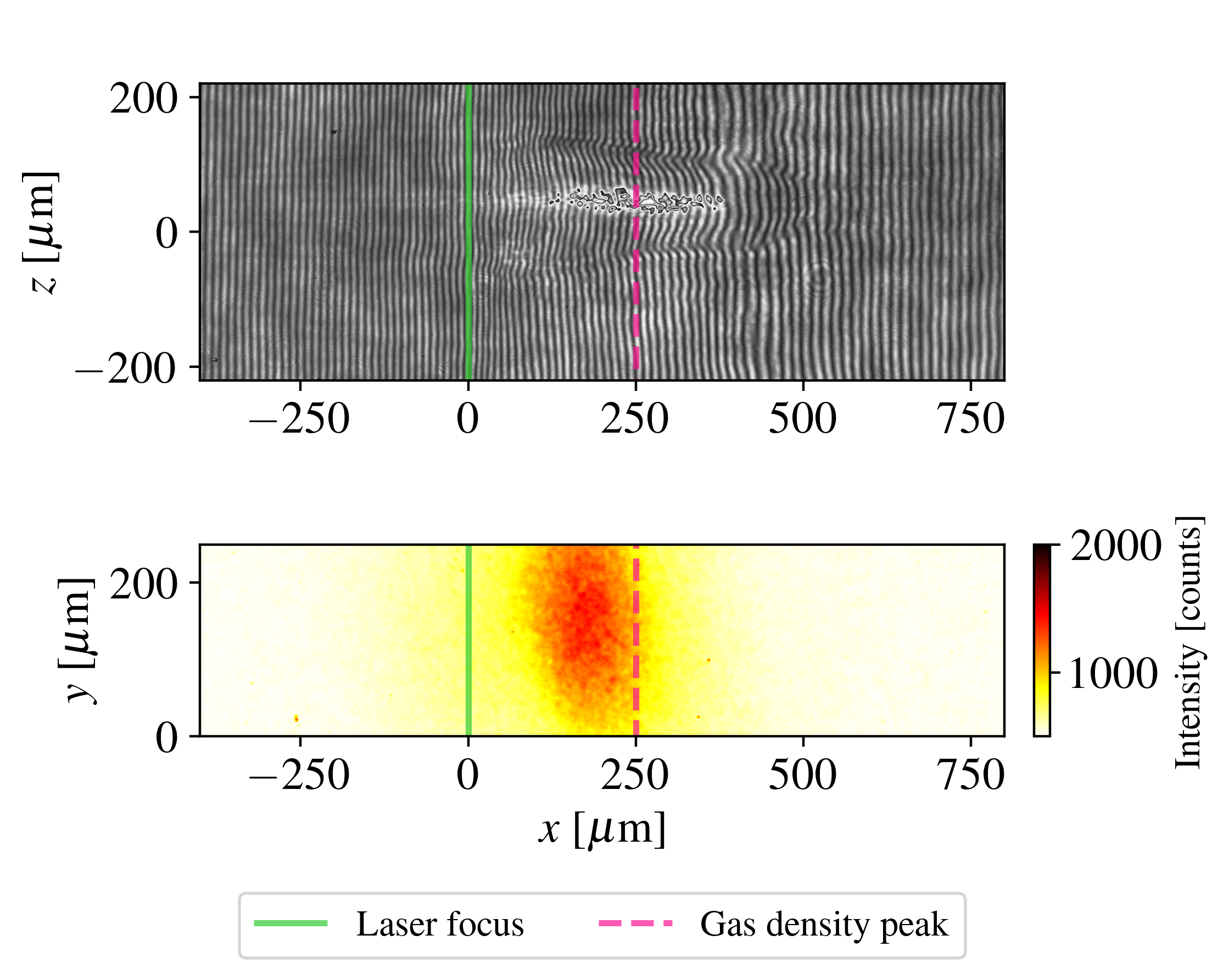}
		\begin{picture}(20,20)
		\put(-72,192){\color{white}\large\textbf{(c)}}%
            \put(-70,148){\color{white}\normalsize{$t \simeq 150 \, \rm{ps}$}}%
            \put(-72,97){\large\textbf{(d)}}
		\end{picture}
	\end{subfigure}
    \vspace{-12mm}
    \caption{(a) Raw interferogram acquired $ \sim 15\,\rm ps$ after the laser pulse maximum at the focal plane. Here, as in all panels, the green solid and pink dashed lines mark the longitudinal ($x$) positions of the laser focus and gas density peak, respectively. The inset zooms in on the region of brightest plasma self-emission. The white dashed curves indicate the off-axis fringe displacement.   
    (b) Corresponding self-emission image of the plasma synchronized with the main laser, integrated over a 180-ps time window. The emission is brightest around the gas density peak which, in this case, coincides with the laser focus. (c), (d) Same as (a) and (b) but from a different shot and recorded $\sim 150\,\rm ps$ after the laser pulse maximum. In (c) one clearly observes a laser-driven channel extending longitudinally across the gas density peak, which is here displaced by $\sim 250\,\rm \mu m$ from the laser focal plane. The plasma self-emission peaks in a $\sim 100\,\rm \mu m$ long region preceding the density maximum.}
    \label{fig:optical_interf}
\end{figure}

Before discussing the experimental data, we detail the two PIC simulations performed to interpret them. These simulations, conducted in 2D3V geometry (two-dimensional in configuration space and three-dimensional in momentum space) using the fully relativistic and electromagnetic PIC \textsc{calder} code \cite{Lefebvre_2003, Debayle_2017}, describe the interaction of a $0.8\,\rm \mu m$ wavelength laser pulse with $10^{20}\,\rm W\,cm^{-2}$ peak intensity and $15\,\rm \mu m$ FWHM focal spot. This pulse is linearly polarized along the $y$-axis and injected along $x$ from the left-hand side ($x=0$) of the simulation box. In the reference simulation,  the vacuum focal plane is located $250\,\rm \mu m$ in front of the GDP, as in the shots corresponding to the particle spectra reported below. The temporal laser intensity profile, extracted from an experimental measurement, is shown in Fig.~\ref{fig:laser}(b). It comprises a $\sim 6\,\rm ps$-long, low-intensity ($I_L \lesssim 10^{17}\,\rm W\,cm^{-2}$) up-ramp, a $<~100\,\rm fs$ main pulse and a $\sim 6\,\rm ps$-long, low-intensity down-ramp. In order to assess possible laser filamentation effects, a second simulation was performed with the same setup, but with the laser focal plane located at the GDP.

The simulation domain, of dimensions $L_x \times L_y = 1920 \times 979\,\rm \mu m^2$, is discretized into $40\,000 \times 20\,400$ cells with a mesh size of $\Delta x = \Delta y = 0.048\,\rm \mu m$. The time step is set to $\Delta t = 0.15\,\rm fs$. The density profile of the initially neutral He gas is taken to be uniform along the transverse ($y$) direction. Its longitudinal (along $x$) profile is extracted from experimental data (see red dashed curve in Fig.~\ref{fig:profile}) and extrapolated below $10^{19}\,\rm cm^{-3}$ (the minimum measurable density value) down to $10^{17}\,\rm cm^{-3}$. The GDP is located at $x \simeq 955 \,\rm \mu m$ in the simulation box.
Owing to their computational cost, our simulations are restricted to specific density conditions. As the shot-to-shot variations in the gas profile displayed in Fig.~\ref{fig:profile} did not profoundly alter the experimental characteristics of the particle spectra or the bulk gas behavior, we are confident that the simulation setup captures well the relevant interaction physics.

The gas is initially represented by two macro-atoms per cell and its temperature is set to a low ($1\,\rm eV$) value. Electron impact \cite{Perez_2012} and field-induced \cite{Nuter_2011} ionization processes are taken into account together with Coulomb collisions between all charged particles. Absorbing boundary conditions are applied for both fields and particles in all directions. 

Each simulation was run on $40\,000$ cores during 72 hours, for a total number of $98\,000$ time steps, corresponding to $\sim~15\,\rm ps$ physical time. 

\section{Experimental results and discussion}
\label{sec:experimental_results}

Let us first examine the bulk response of the ionized gas jet to the laser drive. Figure~\ref{fig:optical_interf}(a) shows a raw interferogram recorded $\sim 15\,\rm ps$ after the arrival of the UHI pulse to the GDP. The green solid and pink dashed lines mark the locations of the focal plane and GDP, respectively. One cannot discern the fringes around the laser path ($y \simeq 0$) at a distance less than $\sim 100\,\rm \mu m$ from the GDP (here located at $x=0$) due to intense plasma self-emission (integrated over the millisecond exposure time of the CCD imaging the fringe pattern). This observation suggests strong laser-gas coupling, as confirmed by the time-integrated plasma image of Fig.~\ref{fig:optical_interf}(b) which reveals bright optical emission within a $\sim 100\,\rm \mu m$ long region encompassing the GDP. Fringe displacement, however, is visible in Fig.~\ref{fig:optical_interf}(a) a few $10\,\rm \mu m$ off axis (as indicated by the dashed white curves in the inset), evidencing formation of a plasma (the borders of which are marked by dashed lines) around the bright laser channel.

Figure~\ref{fig:optical_interf}(c) shows an interferogram taken at a later time ($\sim 150\,\rm ps$ after the laser maximum), in the case where the laser is focused $\sim 250\,\rm \mu m$ before the GDP. Fringes are clearly displaced in a $\sim 500\,\rm \mu m$ length, $\sim 100\,\rm \mu m$ radius region extending across the GDP, as a result of plasma expulsion from the laser path (see below). The plasma self-emission is then most intense in a $\sim 100\,\rm \mu m$-long region of the gas up-ramp nearing the GDP.

It should be realized that the interferometric patterns in Figs.~\ref{fig:optical_interf}(a) and (c) correspond to snapshots of the plasma with a temporal resolution given by the $\sim 100\,\rm fs$ probe beam duration. Note further that the 180~ps integration time (starting at the laser pulse's arrival time) of the plasma images in Figs.~\ref{fig:optical_interf}(b) and (d) is likely much larger than the timescale of intense plasma self-emission, ascribed to nonlinear spectral broadening of the scattered laser light \cite{Giuletti_2013}.

It did not prove possible to reconstruct the free electron density distribution from the early-time interferogram of Fig.~\ref{fig:optical_interf}(a) because of too intense plasma emission in the vicinity of the laser path. In the later-time interferogram of Fig.~\ref{fig:optical_interf}(c), by contrast, the plasma channel has radially expanded outside the brightest emission zone, allowing the fringe pattern to be deconvolved in the region of interest.

\begin{figure}
	\begin{subfigure}[b]{0.47\textwidth}
		\centering
		\includegraphics[width=\textwidth]{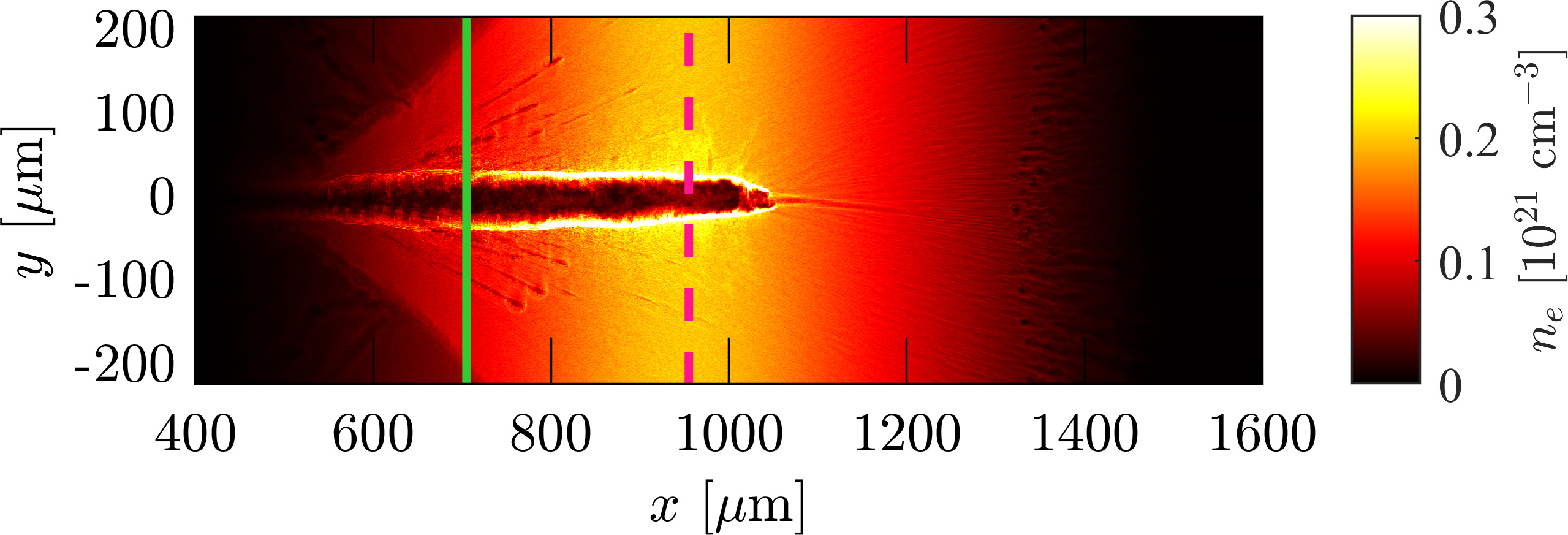}
		\begin{picture}(20,20)
        \put(-77,90){\color{white}\large\textbf{(a)}}
        \put(-77,50){\color{white}\normalsize{$t \simeq 5.6\,\rm{ps}$}}
		\end{picture}	
	\end{subfigure}
	\begin{subfigure}[b]{0.47\textwidth}
		\centering
        \vspace{-6mm}
		\includegraphics[width=\textwidth]{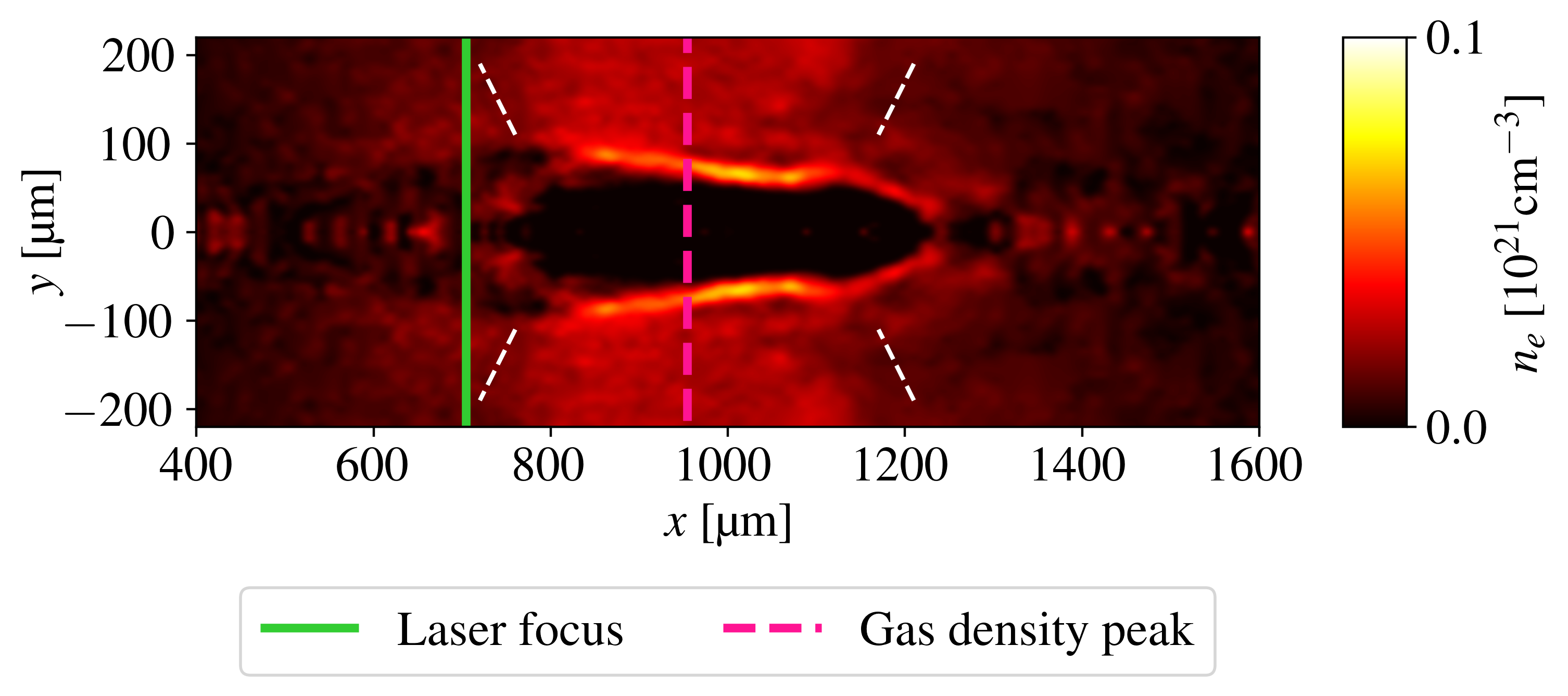}
		\begin{picture}(20,20)
        \put(-77,110){\color{white}\large\textbf{(b)}}
        \put(-77,65){\color{white}\normalsize{$t \simeq 150\,\rm{ps}$}}
		\end{picture}
	\end{subfigure}
    \vspace{-9mm}
    \caption{(a) Free electron density distribution as predicted from the reference 2D PIC simulation, $\sim 5.6\,\rm ps$ after the laser pulse has reached the GDP.
    (b) Free electron density distribution as inferred from the interferogram recorded at $t\simeq 150\,\rm ps$ [Fig.~\ref{fig:setup}(c)]. In both panels, the green solid and pink dashed lines mark the positions of the vacuum focal plane ($x \simeq 705\,\rm \mu m$) and gas density peak ($x \simeq 955\,\rm \mu m$), respectively. The white dashed lines in panel (b) visualize the numerical artefacts discussed in the main text.}
    \label{fig:interf}
\end{figure}
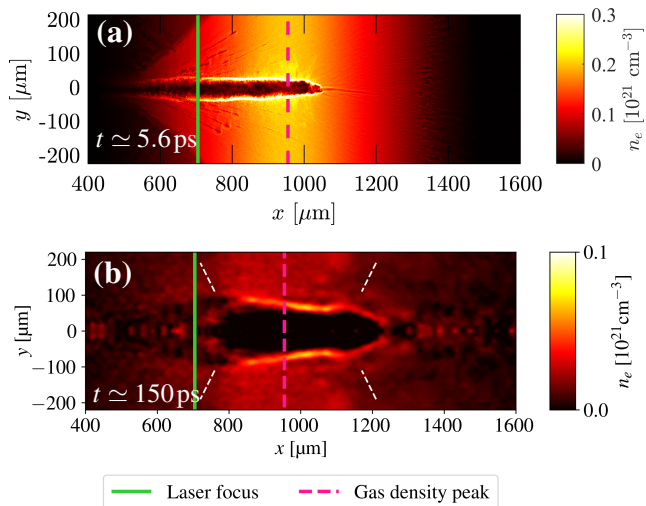

The retrieved electron density distribution is shown in Fig.~\ref{fig:interf}(b). As was guessed from the raw interferogram, one clearly sees
an electron-depleted channel, extending longitudinally across the GDP (pink dashed curve) over a $\sim 500\,\rm \mu m$ total length and radially $\sim 100\,\rm \mu m$ from the laser axis. The white dashed lines serve to guide the eye along outward ``wings'', namely, artifacts resulting from the cylindrical symmetry around the $y = 0$ axis assumed when Abel inverting the phase map deduced from the interferogram. Yet this assumption is only approximate as it conflicts with the normally directed gas flow. Note that in Fig.~\ref{fig:interf}(b) the $z$-ordinate of the original interferogram has been changed to $y$ (the reconstructed density profiles along $y$ and $z$ being supposed identical) for consistency with the $x-y$ coordinates used in the PIC simulation.

For qualitative comparison, we display in Fig.~\ref{fig:interf}(a) the electron density map predicted by the reference PIC simulation $\sim~5.6\,\rm ps$ after the laser pulse maximum reached the GDP (pink dashed line). Fair agreement is found between the simulated and measured plasma channels: they share about the same ($\sim 500\,\rm \mu m$) length and both extend across the GDP. The simulated channel, though, forms earlier in the gas up-ramp and terminates at a shorter depth in the gas down-ramp, where the laser pulse ends up being fully absorbed. Moreover, as expected given the earlier time of probing considered, the channel has expanded over a shorter transverse distance ($\sim 33\,\rm \mu m$ vs $\sim 100\,\rm \mu m$) than observed in the experiment at $t\simeq 150\,\rm ps$. Note that the reduced 2D geometry of the simulation may also affect the late-time transverse dynamics of the channel \cite{Nicholls_1974, Borghesi_1998}. Furthermore, the He gas has been fully ionized over most of the simulation domain as a result of field and impact ionization by the laser-accelerated electrons spraying from the channel. The diverging envelope of the electron density distribution in the gas up-ramp visualizes the emission cone of the ionizing fast electrons from the plasma channel.

\begin{figure}
	\begin{subfigure}[b]{0.47\textwidth}
		\centering
		\includegraphics[width=\textwidth]{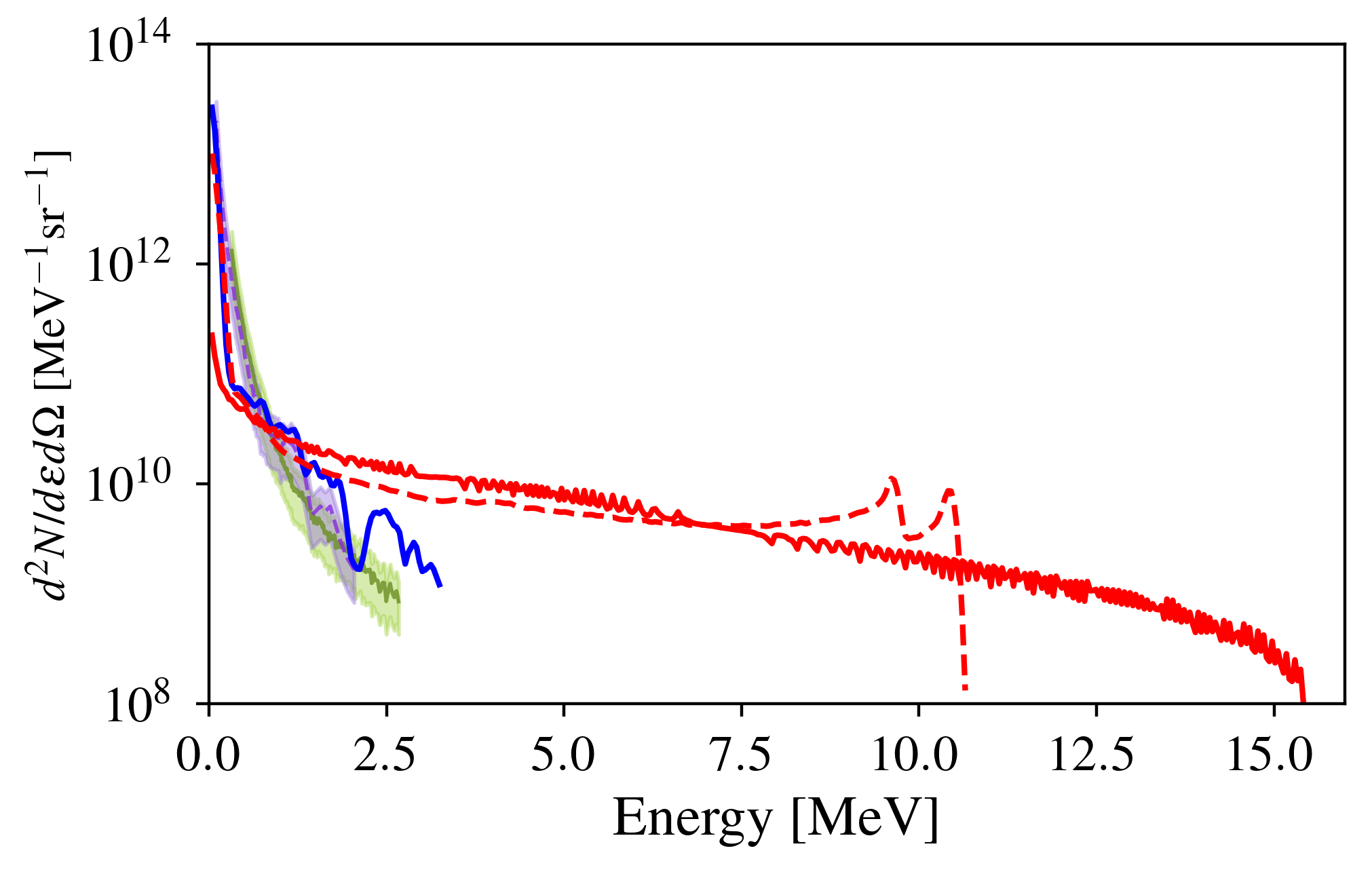}
		\begin{picture}(20,20)
        \put(-70,155){\color{black}\large\textbf{(a)}}
		\end{picture}	
	\end{subfigure}
	\begin{subfigure}[b]{0.49\textwidth}
		\centering
        \vspace{-6mm}
		\includegraphics[width=\textwidth]{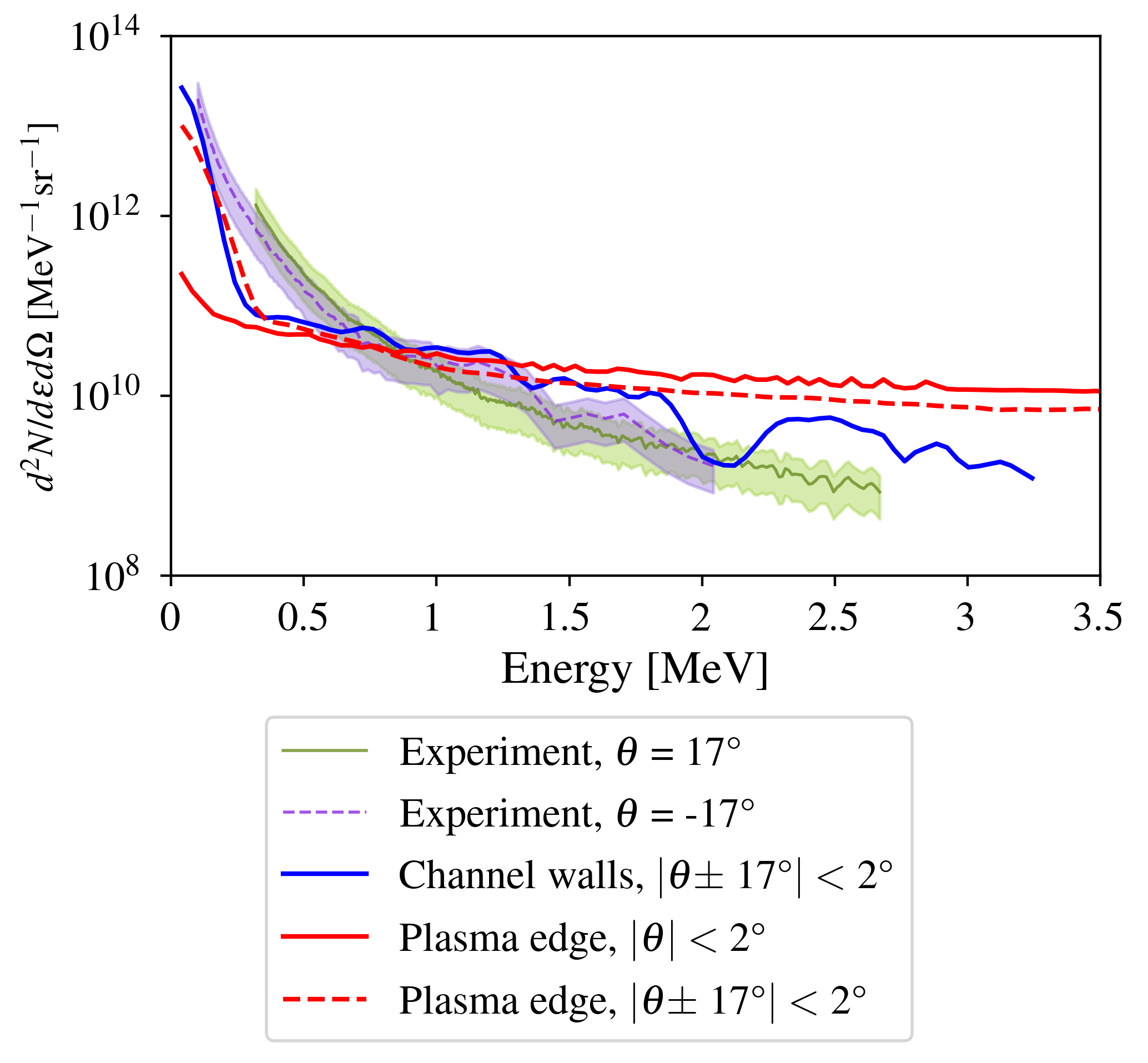}
		\begin{picture}(20,20)
        \put(-70,230){\color{black}\large\textbf{(b)}}
		\end{picture}
	\end{subfigure}
    \vspace{-12mm}
    \caption{(a) Experimental energy spectra of $\alpha$ particles as inferred from the ToF signals detected at (green solid line) $(\theta,\phi)=(17^{\circ},9^{\circ})$ and (purple dashed line) $(\theta,\phi)=(-17^{\circ},9^{\circ})$ relative to the laser axis (see setup in Fig.~\ref{fig:setup}). Overlaid are $\alpha$ particle spectra extracted from the PIC simulation, either around (blue curve) the right end of the plasma channel (integrated over the $\vert \theta \pm 17^{\circ} \vert < 2^{\circ}$ angular range),
    around (red dashed curve) the right edge of the gas profile (integrated over $\vert \theta \pm 17^{\circ} \vert < 2^{\circ}$)
    or around (red curve) the right edge of the gas profile (integrated over $\vert \theta \vert < 2^{\circ}$).
    They have been rescaled to be plotted along with the experimental data.
    (b) Closeup of (a) in the $<3.5\,\rm MeV$ energy range. 
    }
	\label{fig:ions_spectra}
\end{figure}
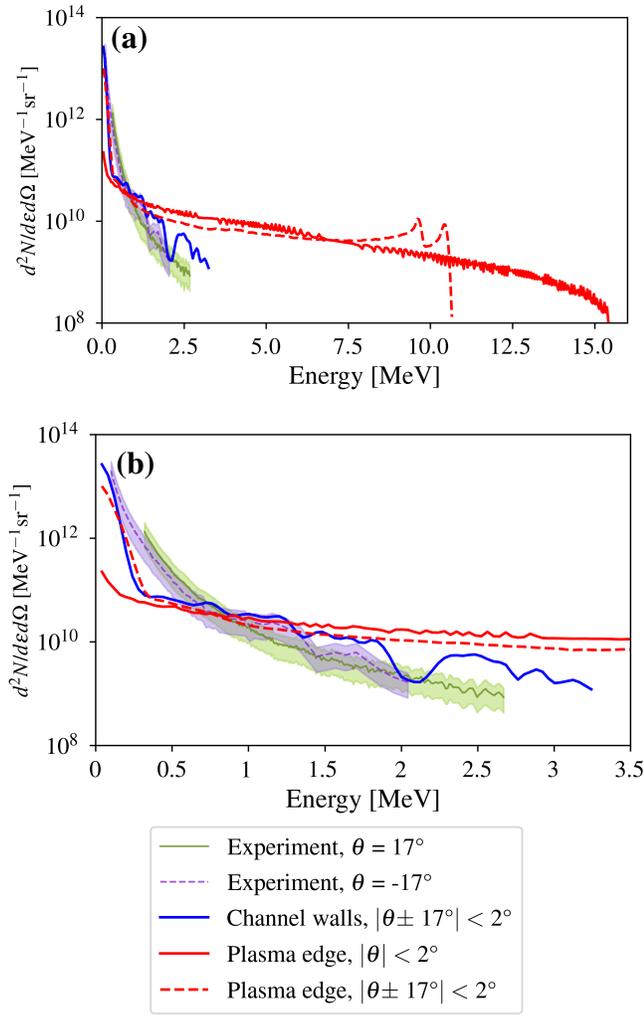

Significant ion acceleration was detected by the two ToF diamond detectors fielded at $(\theta,\phi) = (\pm 17^{\circ}, 9^{\circ})$ relative to the laser axis, with a collection solid angle of $\sim 0.5\,\rm mrad$. Figure~\ref{fig:ions_spectra} plots (as green solid and purple dashed curves) the $\alpha$ (He$^{2+}$) particle spectra inferred via the method detailed in Refs.~\cite{Salvadori_2021, Salvadori_2022}. Similar spectra were detected on the two channels. They extend up to $\sim 2.7\,\rm MeV$ with a total flux of $\sim 10^{11} \,\rm sr^{-1}$, as integrated over $>0.1\,\rm MeV$ energies and seen within the solid angle of the detectors. In the $0.5-2.5\,\rm MeV$ range, the energy-differential flux varies between $\sim 10^{11} \,\rm MeV^{-1} sr^{-1}$ and $\sim 10^{9} \,\rm MeV^{-1} sr^{-1}$ , while it reaches $\sim 10^{13}\,\rm MeV^{-1} sr^{-1}$ around the lower detection limit ($\sim 0.1\,\rm MeV$).
Although ToF detectors cannot differentiate between charged species, we are confident that the measured spectra are mainly associated with He$^{2+}$ (even though a minor contribution of He$^+$ ions due to recombination cannot be strictly ruled out). Our PIC simulations indeed indicate that the gas is fully ionized where ion acceleration mainly takes place (see below) and that there is no energetic $\mathrm{He}^{+}$ ions in the experimentally detected $\sim 17^{\circ}$ emission cone.
It should also be stressed that the two spectra reported in Fig.~\ref{fig:ions_spectra} were acquired on different shots. No simultaneous ion signals on the two ToF detectors could indeed be recorded during the campaign, suggesting strongly anisotropic ion emissions. Moreover, on the TP-dedicated shots, no ion signal was retrieved on the exposed IP; only the three zero-deflection points imprinted by the x~rays coming from the interaction zone were visible.

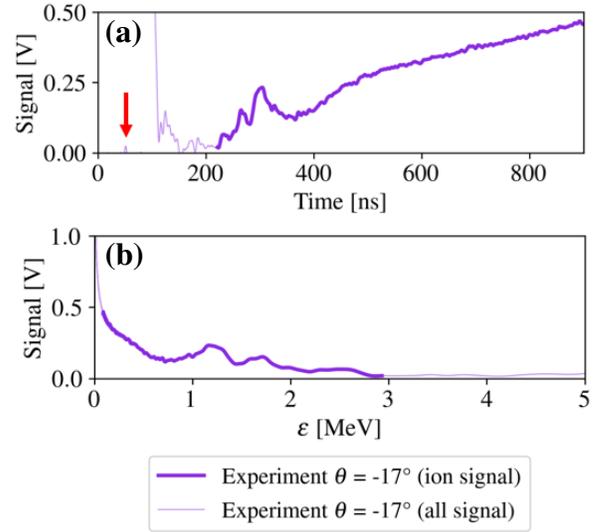
\begin{figure}
	\begin{subfigure}[b]{0.45\textwidth}
		\centering
		\includegraphics[width=\textwidth]{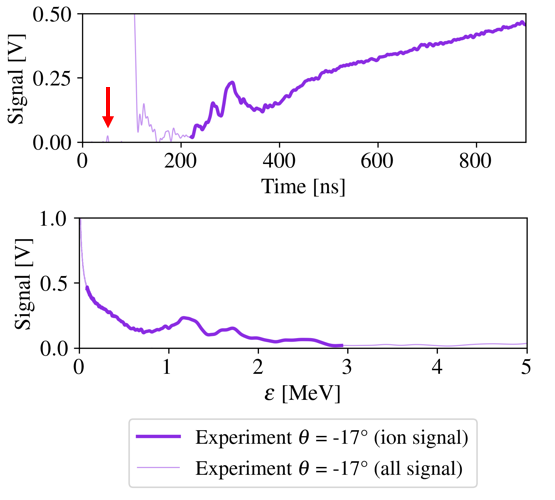}
		\begin{picture}(20,20)
            \put(-68,211){\color{black}\large\textbf{(a)}}
            \put(-68,127){\color{black}\large\textbf{(b)}}
		\end{picture}	
	\end{subfigure}
    \vspace{-10mm}
    \caption{Raw time-of-flight (ToF) data acquired at $\theta~=~-17^{\circ}$. The ordinate represents the signal level (in V) as recorded by a digital oscilloscope. The abscissa in (a) represents the time of arrival of the particles to the detector, which is converted into energy in (b), see details in the main text. The thick curves correspond to the ion part of the signal. The photopeak is marked with a red arrow in (a). The signal in (b) corresponds to the spectrum plotted in purple in Figs.~\ref{fig:ions_spectra}(a) and \ref{fig:ions_spectra}(b).
    }
	\label{fig:tof_raw}
\end{figure}

Figure~\ref{fig:tof_raw}(a) shows the raw ToF data associated with the $\theta = -17^{\circ}$ spectrum plotted (in purple) in Figs.~\ref{fig:ions_spectra}(a) and \ref{fig:ions_spectra}(b). 
The abscissa corresponds to the time of arrival of the particles to the detector while the ordinate represents the signal voltage, proportional to the collected particle charge. The red arrow points to the signal due to x rays that first reach the detector. This so-called photopeak marks the origin of time and allows the subsequent ion-induced signal to be  deconvolved \cite{Salvadori_2021, Salvadori_2022}. In Fig.~\ref{fig:tof_raw}(b) the abscissa has been converted into energy knowing the particle mass and distance traveled. Notice how the signal in Fig.~\ref{fig:tof_raw}(b) already reproduces the spectral shape seen in Fig.~\ref{fig:ions_spectra}(b), where the detector's calibration was used to obtain absolute particle numbers.

\begin{figure}
	\begin{subfigure}[b]{0.47\textwidth}
		\centering
		\includegraphics[width=\textwidth]{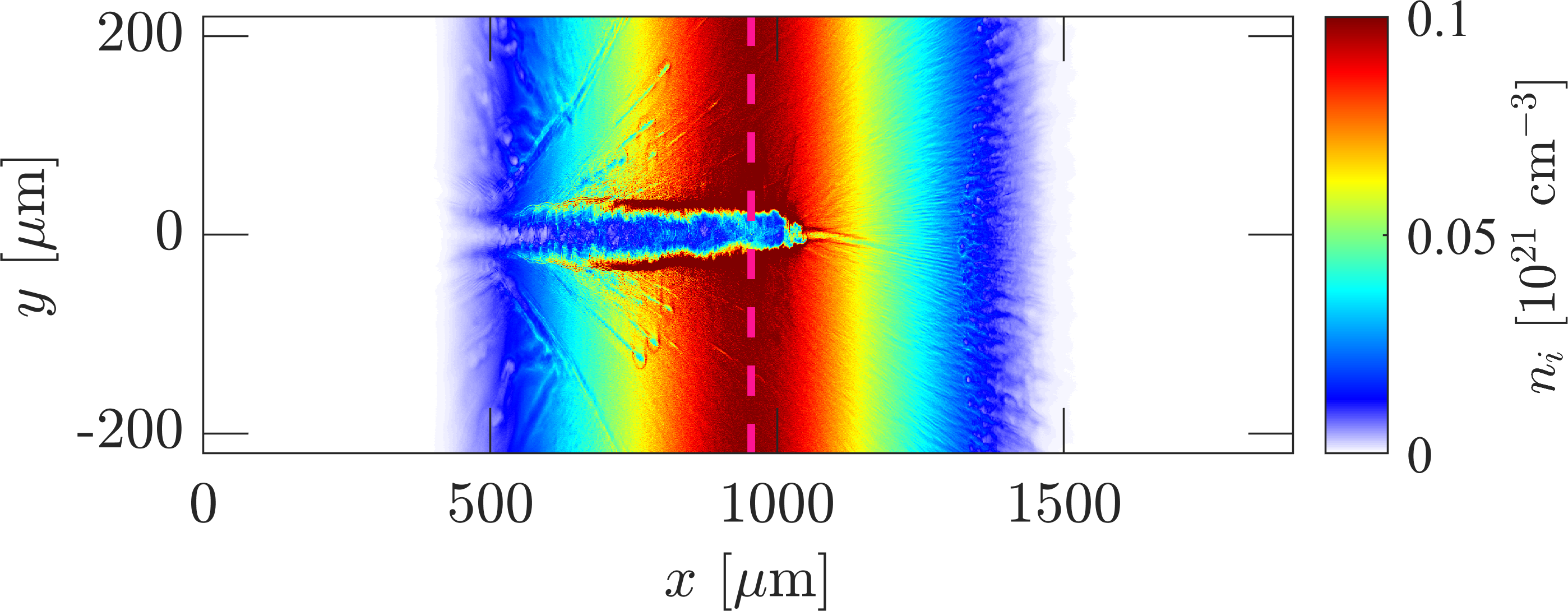}
		\begin{picture}(20,20)
        \put(-76,99){\color{black}\large\textbf{(a)}}
		\end{picture}	
	\end{subfigure}
	\begin{subfigure}[b]{0.47\textwidth}
		\centering
        \vspace{-6mm}
		\includegraphics[width=\textwidth]{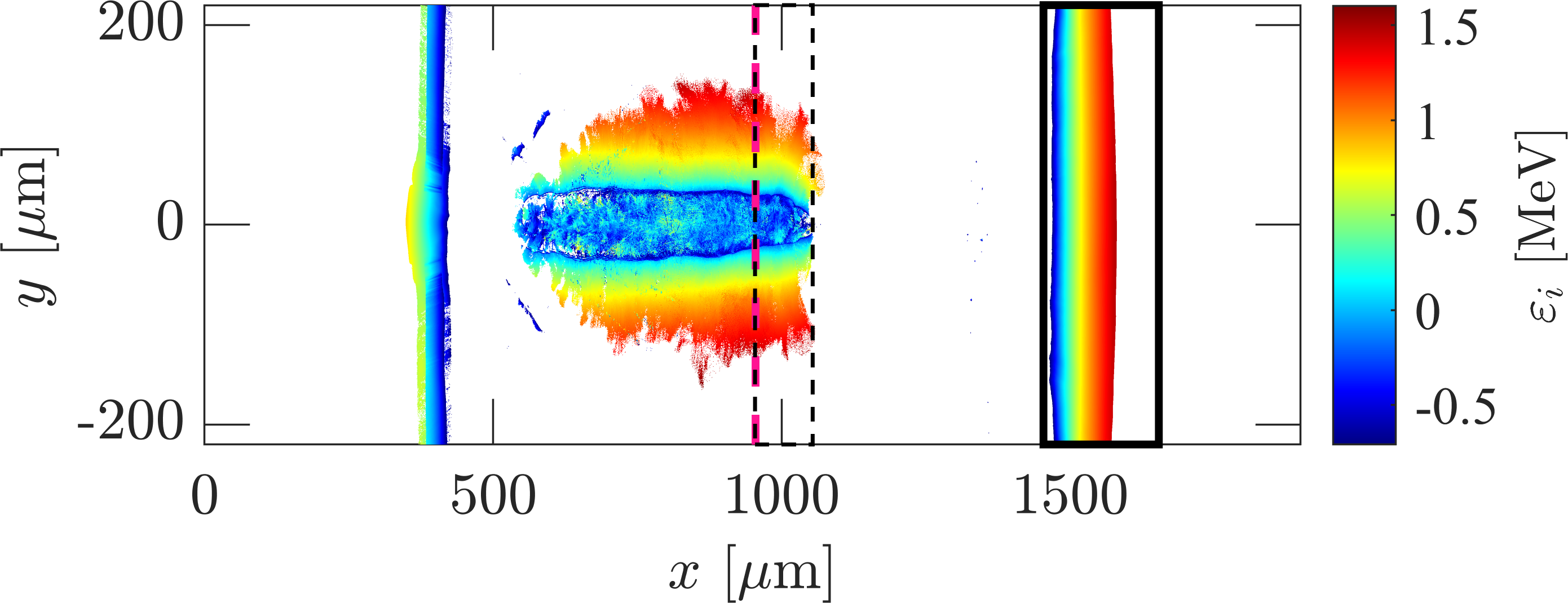}
		\begin{picture}(20,20)
        \put(-76,100){\color{black}\large\textbf{(b)}}
		\end{picture}
	\end{subfigure}
	\begin{subfigure}[b]{0.47\textwidth}
		\centering
        \vspace{-6mm}
		\includegraphics[width=\textwidth]{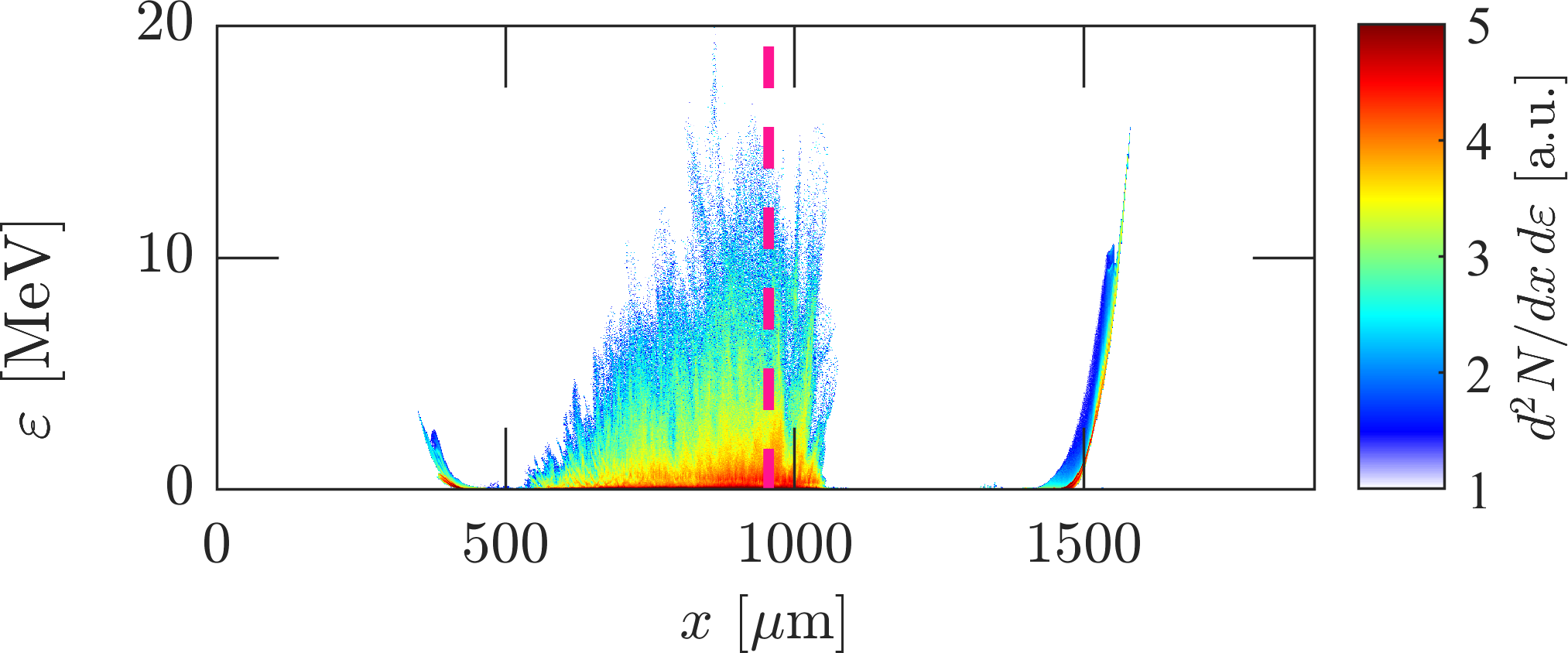}
		\begin{picture}(20,20)
        \put(-74,104){\color{black}\large\textbf{(c)}}
		\end{picture}
	\end{subfigure}
	\begin{subfigure}[b]{0.28\textwidth}
		\centering
        \vspace{-6mm}
		\includegraphics[width=\textwidth]{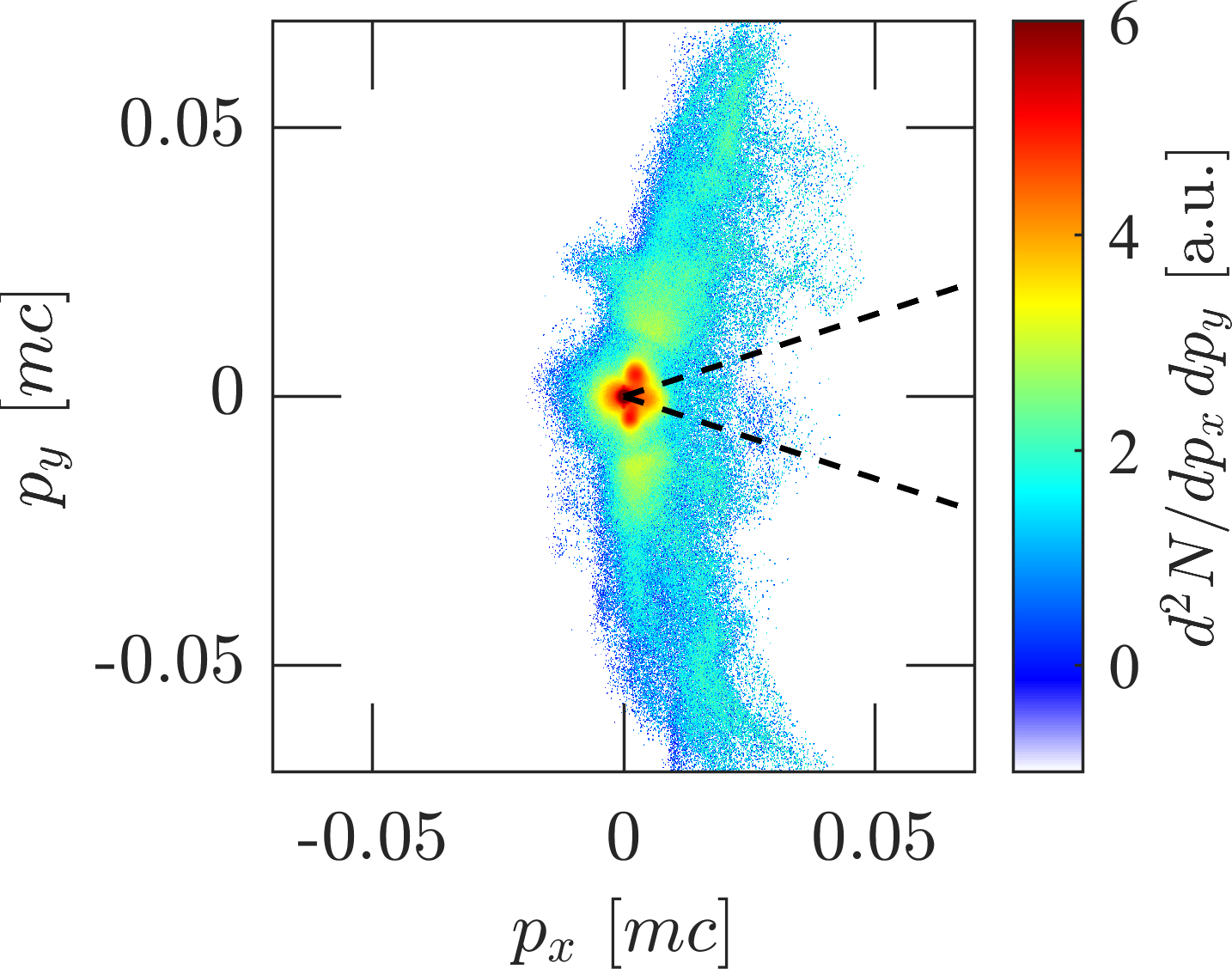}
		\begin{picture}(20,20)
        \put(-74,120){\color{black}\large\textbf{(d)}}
		\end{picture}
	\end{subfigure}
	\begin{subfigure}[b]{0.28\textwidth}
		\centering
        \vspace{-6mm}
		\includegraphics[width=\textwidth]{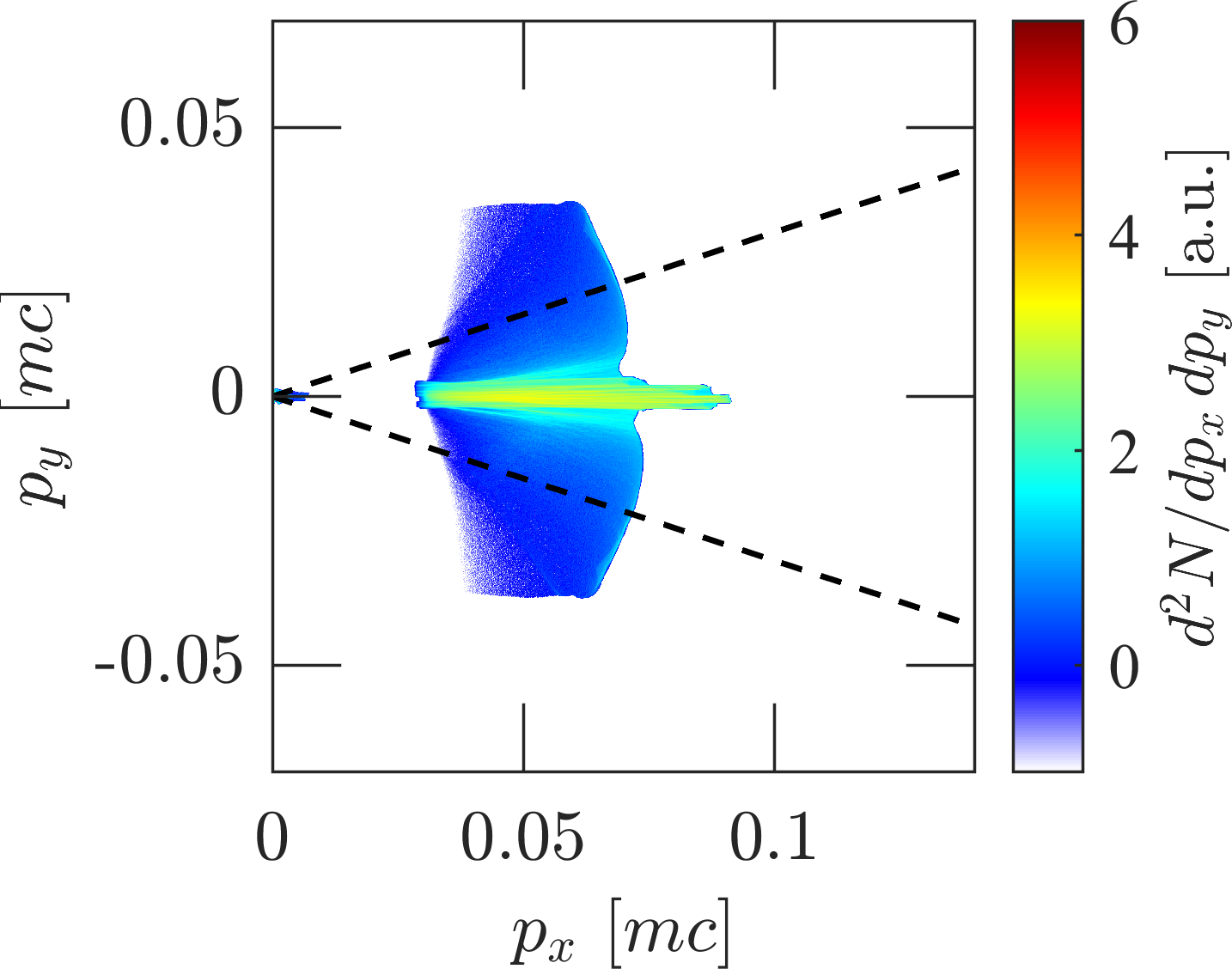}
		\begin{picture}(20,20)
        \put(-74,120){\color{black}\large\textbf{(e)}}
		\end{picture}
	\end{subfigure}
    \vspace{0mm}
    \caption{Phase-space projections of the simulated He ion distribution $\sim 5.6\,\rm ps$ after the passage of the laser pulse through the gas density peak plane. (a) Spatial density distribution (linear scale) (b) Spatial distribution of the mean kinetic energy (log$_{10}$ scale). (c) Kinetic energy distribution (log$_{10}$ scale) as a function of longitudinal position. (d,e) $p_x-p_y$ momentum distribution (log$_{10}$ scale) at the (d) right end of the channel [dashed black box in panel (b)] and (e) the plasma-vacuum interface [solid black box in panel (b)]. The black dashed lines in (d) and (e) indicate the $\pm 17^{\circ}$ lines of sight of the ToF detectors in the experiment.}
    \label{fig:PIC_ions}
\end{figure}
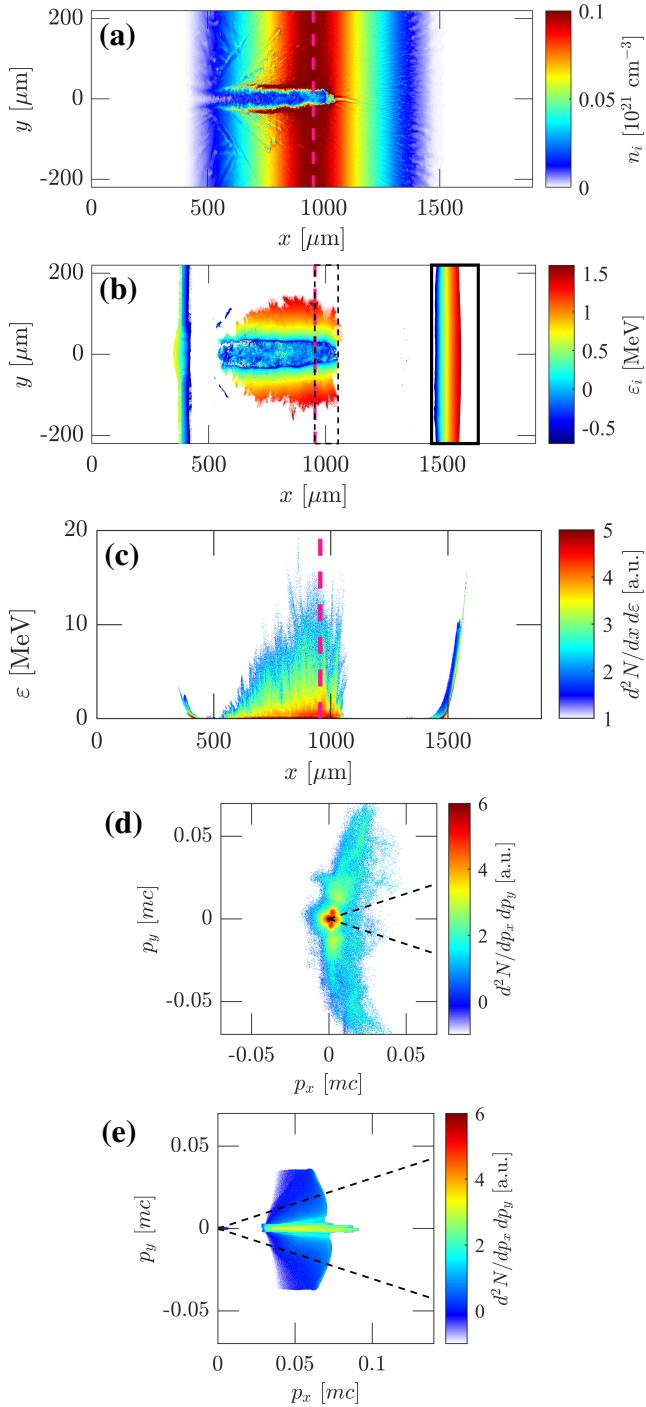

To explain the origin of the detected MeV-range $\alpha$ particles, we present in Figs.~\ref{fig:PIC_ions}(a)-(e) various phase-space projections of the simulated He ion distribution. Figure~\ref{fig:PIC_ions}(a) shows the spatial ion density distribution at the final simulation time ($\sim~5.6\,\rm ps$ after the laser pulse has crossed the GDP) and, notably, the ion depletion within the plasma channel already depicted in Fig.~\ref{fig:interf}(a) through the electron density distribution.

Figures~\ref{fig:PIC_ions}(b) and \ref{fig:PIC_ions}(c), which depict, respectively, the spatial distribution of the mean ion kinetic energy and the $x$-resolved ion energy distribution, reveal three main sites of ion acceleration. The first is located at the (mainly) transversely expanding walls of the laser-drilled channel, where electrostatic ion reflection can occur as reported in Ref.~\cite{Singh_2020}. Maximum ion energies of $\sim 20 \,\rm MeV$ are found in this region.
While most of the ions swept up by the channel are reflected at near-normal angles, those located in the vicinity of the channel's head [see dashed box in Fig.~\ref{fig:PIC_ions}(b)] are accelerated over a broad forward-directed cone. Their $p_x-p_y$ momentum distribution, displayed in Fig.~\ref{fig:PIC_ions}(d), appears to be highly anisotropic, with the fastest ($\sim 0.07c$) ions propagating at $\sim 75^{\circ}$ from the laser propagation direction 

\begin{figure}
	\begin{subfigure}[b]{0.47\textwidth}
		\centering
		\includegraphics[width=\textwidth]{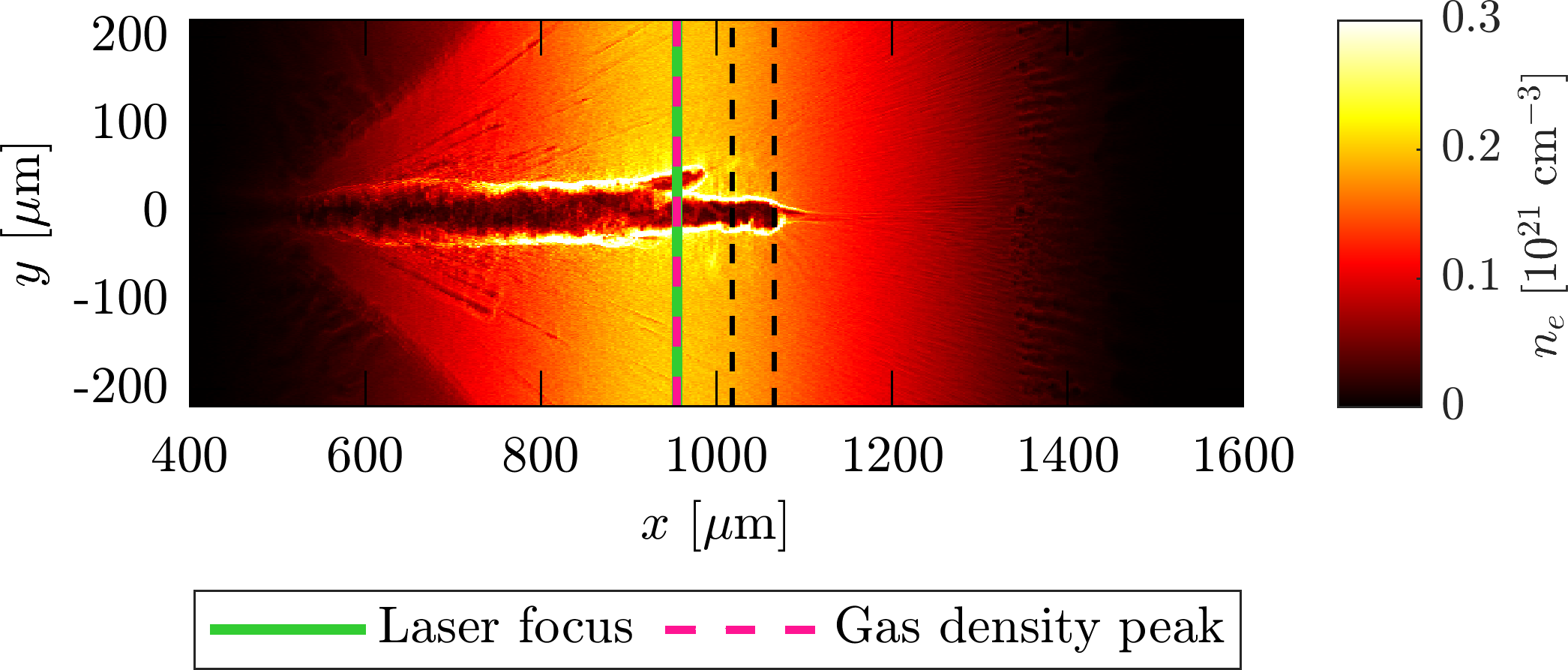}
		\begin{picture}(20,20)
        \put(-77,110){\color{white}\large\textbf{(a)}}
		\end{picture}	
	\end{subfigure}
	\begin{subfigure}[b]{0.47\textwidth}
		\centering
        \vspace{-6mm}
		\includegraphics[width=\textwidth]{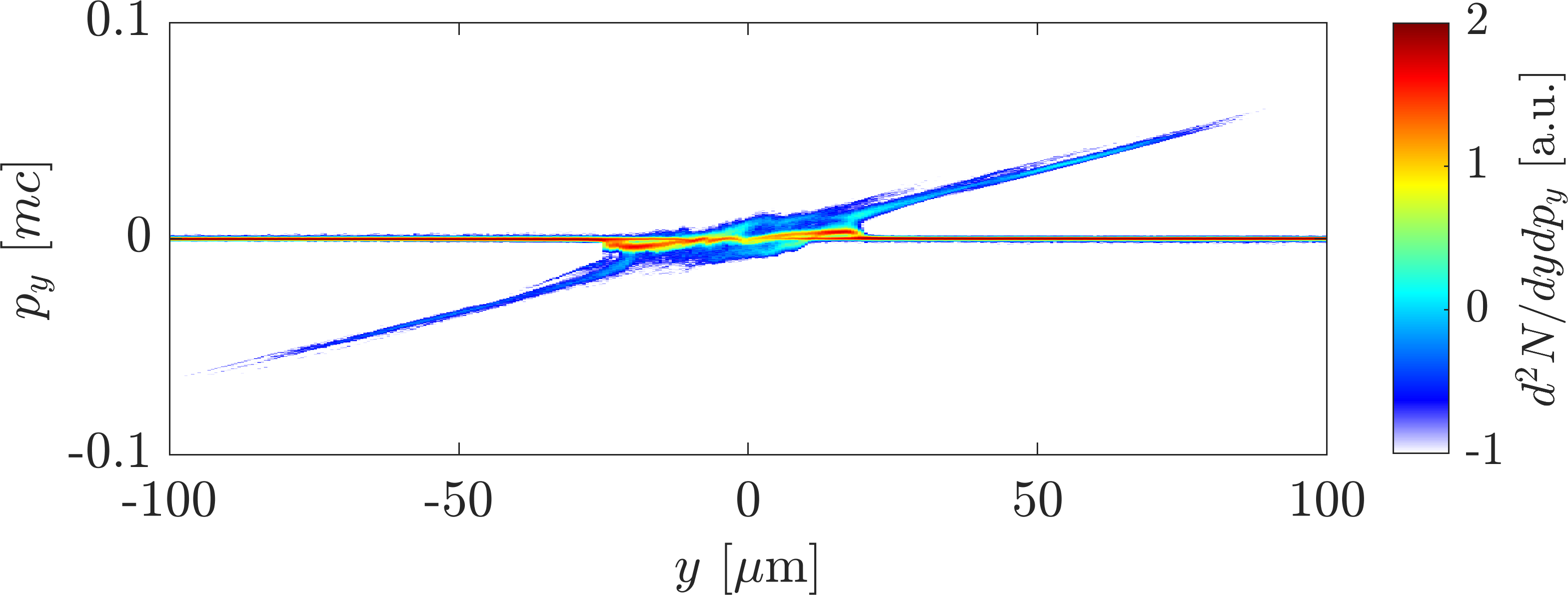}
		\begin{picture}(20,20)
        \put(-75,95){\color{black}\large\textbf{(b)}}
		\end{picture}
	\end{subfigure}
    \vspace{-8mm}
    \caption{(a) Free electron density distribution, extracted $\sim~5.6\,\rm ps$ after the laser pulse has reached the GDP, as predicted from a PIC simulation using a laser pulse focused at the GDP. (b) $y-p_y$ phase space of the He ions (log$_{10}$ scale) near the right end of the laser-created channel [dashed black box in panel (a)].
    }
    \label{fig:channel_refl}
\end{figure}
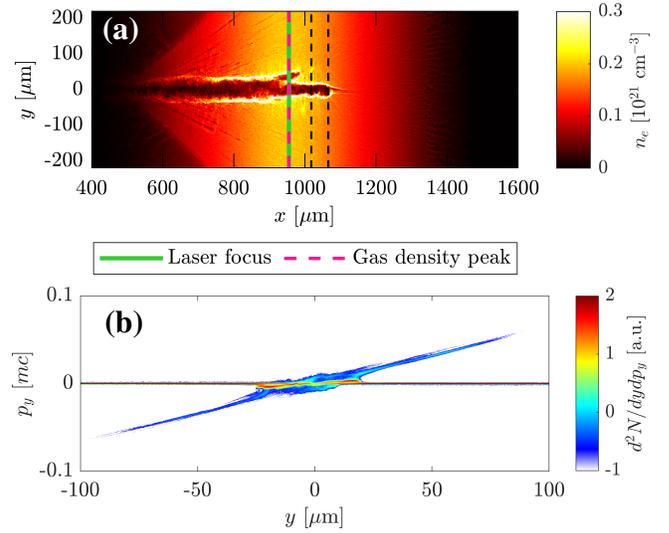

To illustrate the channel reflection mechanism, Fig.~\ref{fig:channel_refl}(a) shows the free electron density distribution, extracted $\sim~5.6\,\rm ps$ after the laser pulse has reached the GDP, from the PIC simulation using a laser pulse focused at the GDP. Two distinct laser-driven channels are now visible as a consequence of laser filamentation. This contrasts with the single channel observed in the previous simulation, where the pulse was focused $250\,\rm \mu m$ before the GDP [Fig.~\ref{fig:interf}(a)]. 
The ion acceleration associated with the transverse expansion of the main channel is evidenced in Fig.~\ref{fig:channel_refl}(b), which displays the ion $y-p_y$ phase space around the right end of the channel [dashed black box in Fig.~\ref{fig:channel_refl}(a)]. A close examination shows that the ions reflected (up to {$v_y \simeq \pm 0.03c$}) off the expanding channel boundaries ($y \simeq \pm 20\,\rm \mu m$) subsequently experience TNSA-type acceleration in the charge-separation field set up by the hot electrons outside of the channel. This leads to maximum transverse velocities of {$v_y \simeq \pm 0.07c$ (corresponding to $\sim 9\,\rm MeV$ energies)} at the time considered.

\begin{figure}
    \centering
    \includegraphics[width=0.47\textwidth]{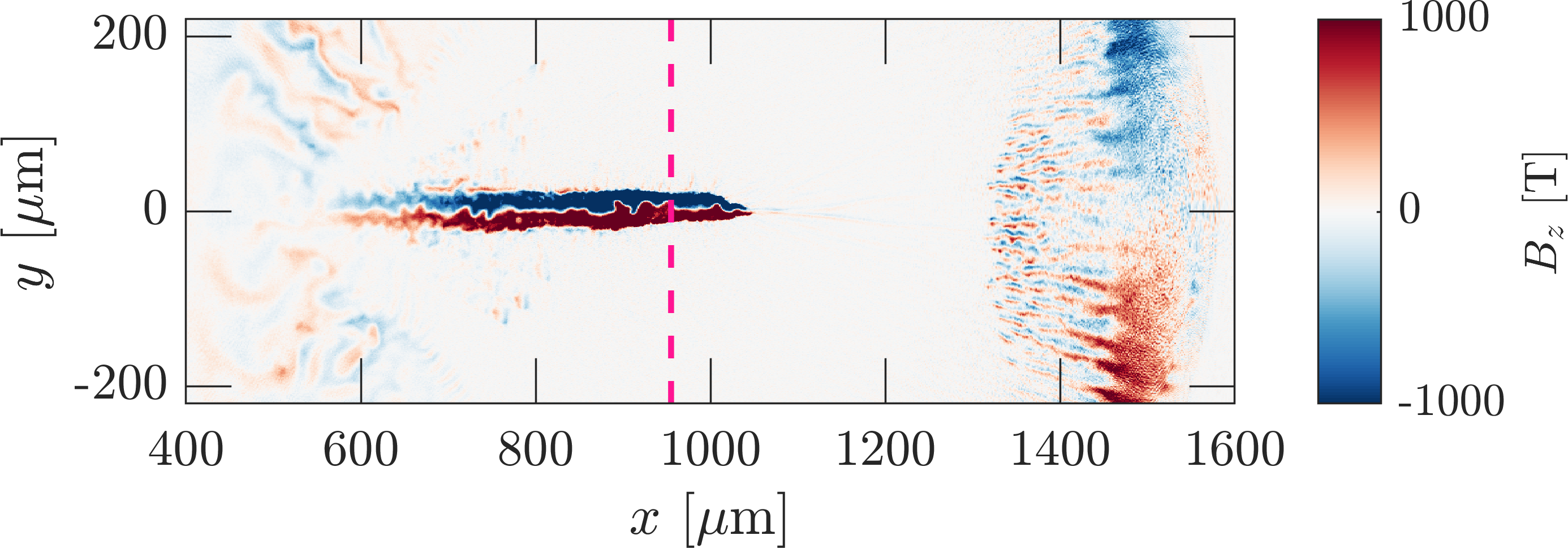}
    \caption{Spatial distribution of $B_z$ magnetic field recorded in the simulation $2.1\,\rm ps$ after the laser pulse has crossed the GDP.
    }
    \label{fig:b_field}
\end{figure}

Figures~\ref{fig:PIC_ions}(b) and (c) disclose two other potential sources of energetic ions, namely, at the left- and right-hand boundaries of the gas profile ($x\simeq 400\,\rm \mu m$ and $x\simeq 1500\,\rm \mu m$), where TNSA sets in following the arrival of the laser-generated relativistic electrons. The ions accelerated via this process (most efficient at the right border of the gas profile) are predicted to reach about the same maximum energies as the ions reflected from the channel walls, but with a more collimated angular distribution [see Fig.~\ref{fig:PIC_ions}(e) corresponding to the solid black box in Fig.~\ref{fig:PIC_ions}(b)]. The fastest ($\sim 15\,\rm MeV$) ``TNSA ions'' are contained in a very narrow cone ($\theta \lesssim 0.7^{\circ}$), yet a dilute halo of quite energetic ($\sim 2-10\,\rm MeV$) ions propagating at much larger angles ($\theta \lesssim 30^{\circ}$) can also be seen. These divergent ions, which could, in principle, be collected by the ToF detectors, originate from deflections in the strong (as high as $\sim 1000\,\rm T$) transverse magnetic fields induced at the rear edge of the gas, as shown in Fig.~\ref{fig:b_field}. This figure displays the spatial distribution of the out-of-plane ($B_z$) magnetic field $2.1\,\rm ps$ after the laser pulse has reached the GDP. The laser pulse has then been fully absorbed in the gas down-ramp, as evidenced by the relatively short longitudinal extent of the magnetized plasma channel past the GDP. The strong $B$ fields that have developed at the plasma backside are ascribed to both the fountain-type motion of the fast electrons exiting the gas \cite{Sarri_2012} -- which induces coherent fields of opposite polarity across the symmetry $x$-axis -- and the unstable interpenetration of the exiting and space-charge-reflected electron streams \cite{Tatarakis_2003, Gode_2017, Zhou_2018, Ruyer_2020} -- which induces transverse field modulations with a $\sim 10-30\,\rm \mu m$ wavelength increasing at lower densities.

An ion of mass $m_i$, charge $Z$, longitudinal velocity $v_x$ and energy $\varepsilon_i = m_i v_x^2/2$ travelling across a magnetic field of amplitude $B_z$ and longitudinal extent $l_B$ will undergo a transverse deflection $\delta \theta \simeq \arctan (Z e B_z l_B/\sqrt{2 m_i \varepsilon_i})$ in the weak-deflection limit. Taking $B_z \simeq 500\,\rm T$ and $l_B \sim 100\,\rm \mu m$ as typical values (see Fig.~\ref{fig:b_field}), one thus expects deflection angles of $\sim 6-14^{\circ}$ for $2-10\,\rm MeV$ He$^{2+}$ ions, roughly consistent with the angular spread seen in Fig.~\ref{fig:PIC_ions}(e).

The $\sim 10\,\rm MeV$ maximum energies predicted to be reached by the TNSA ions along the lines of sight of the ToF detectors are, however, inconsistent with the experimentally inferred \mbox{$\sim 2.7\,\rm MeV$} cutoff energy (Fig.~\ref{fig:ions_spectra}). Worse, the dominant highly collimated component of those ions should have been detected by the TP fielded on axis on some shots. These discrepancies, therefore, cast serious doubt on the effective operation of the TNSA mechanism in the experiment. It is indeed well known \cite{Mora_2003, Grismayer_2006} that the efficiency of TNSA is sensitive to the shape of the plasma profile where the sheath field develops, i.e., the outer region where the background electron density becomes lower than the local hot-electron density or where the local Debye length becomes larger than the density scale length. 


Thus, the absolute value of the hot-electron density not only controls the accelerating field strength but also the position of the acceleration site. Here, owing to a gas profile truncated below $10^{17}\,\rm cm^{-3}$ (the lower bound of the detectable gas density) and a likely overestimated hot-electron density (due to the 2D geometry considered) at the remote gas edge, $\sim~500\,\rm \mu m$ away from the plasma channel where most of the hot-electron generation takes place, one may expect TNSA to be greatly enhanced in the simulation compared to what occurs in the experiment. Likewise, the $B$ fields induced by the fast-electron currents in the TNSA region are very likely overestimated. By contrast, the numerical modeling of the channel-expansion-induced ion acceleration should be more reliable because this process operates in the vicinity of the laser path, because our simulation captures fairly well the shape of this channel and, finally, because the foreseen $\sim 2\,\rm MeV$ cutoff energies associated with this mechanism compare rather well with the measurements.

\begin{figure}
    \centering
	\includegraphics[width=0.9\linewidth]{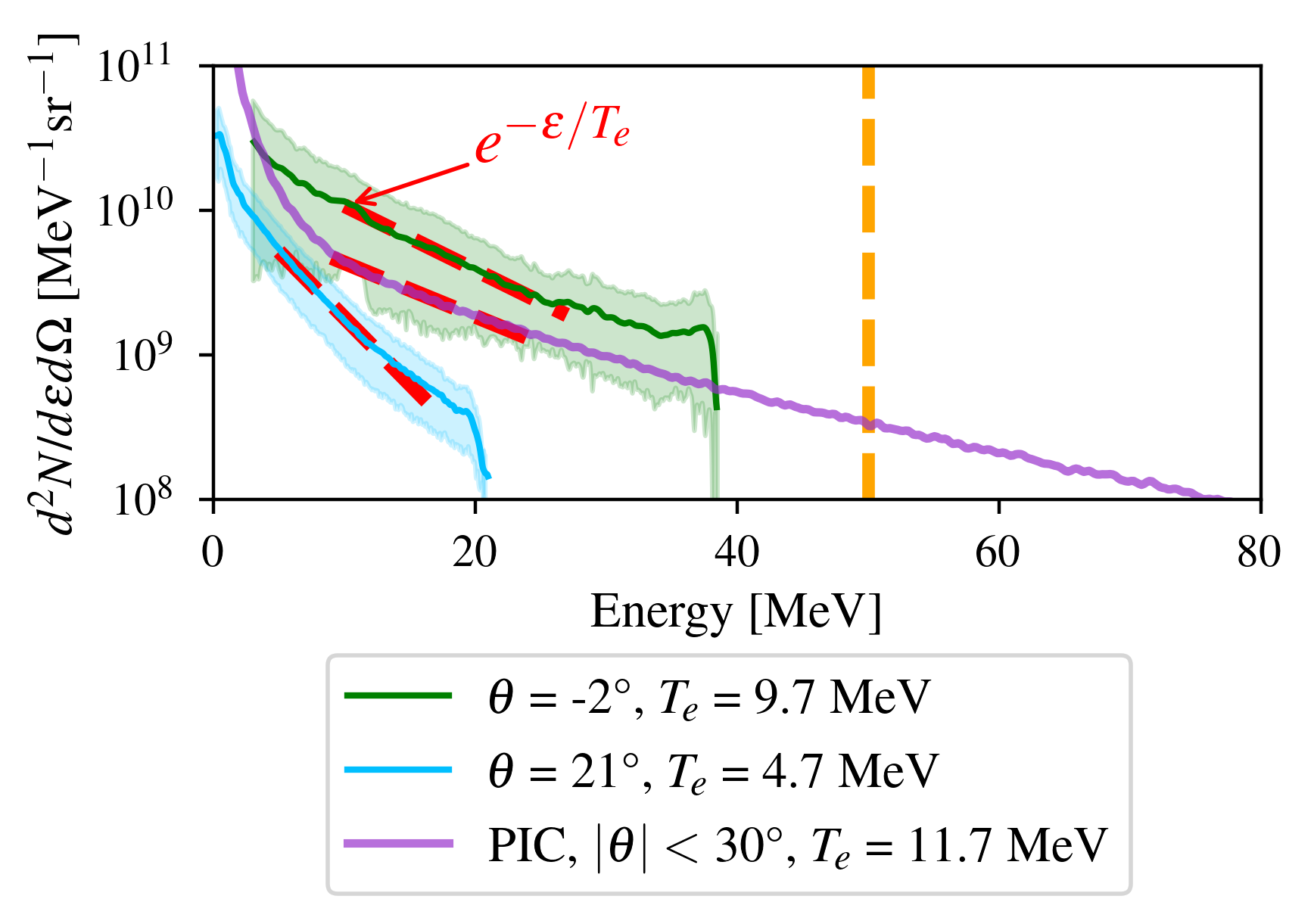}
	\caption{Experimental electron energy spectra measured at (green) $-2^{\circ}$ and (light blue) $21^{\circ}$ from the laser axis, compared with the spectrum extracted from the PIC simulation (violet) and rescaled to fit the experimental data range. The experimental error bars are evaluated over three different shots. The vertical dashed line indicates the upper energy limit of the spectrometers. The legend indicates, for each spectrum, the temperature associated with the best-fitting exponential function (red dashed lines).
    }
	\label{fig:electron_results_spectra}
\end{figure}

We now discuss the electron acceleration measurements.
Figure~\ref{fig:electron_results_spectra} plots representative electron spectra collected by the MES fielded on axis ($\theta = -2^{\circ}$) and off axis ($\theta = 21^{\circ}$). The error bars are computed over three different shots. The vertical dashed line indicates the $\sim 50\,\rm MeV$ upper detection limit. The typical spectra recorded at $-2^{\circ}$ (green curve) exhibit a quasi-exponential shape, decreasing from $\sim 5\times 10^{11}\,\rm MeV^{-1}\,sr^{-1}$ at $\sim 5\,\rm MeV$ down to $\sim 2\times 10^9\,\rm MeV^{-1}\,sr^{-1}$ at the $\sim 38\,\rm MeV$ cutoff energy. Assuming those spectra scale as $\propto\, e^{-\varepsilon/T_e}$, this corresponds to a best-fitting ``temperature'' of $T_e \simeq 9.7\,\rm MeV$, i.e., about $5\times$ larger than the standard ponderomotive scaling \cite{Wilks_1992} $T_e \simeq \left(\sqrt{1+a_L^2/2}-1 \right)m_ec^2 \simeq 2\,\rm MeV$.
This trend qualitatively agrees with previous related numerical studies \cite{Robinson_2011, Debayle_2017, Rosmej_2019}.

The typical electron spectra recorded at $21^{\circ}$ from the laser axis are plotted in light blue in Fig.~\ref{fig:electron_results_spectra}. Their significantly lower temperature ($\sim 4.7\,\rm MeV$) and energy cutoff ($\sim 20\,\rm MeV$) give a measure of the directionality of the fast electrons exiting the plasma. The purple curve represents the simulated electron spectrum, integrated over a $\vert \theta \vert < 30^{\circ}$ angular range (corresponding to the full forward-emission ``cone'' of the hot electrons in the simulation). Its quasi exponential shape with $\sim 11.7\,\rm MeV$ temperature is consistent with the measurements.

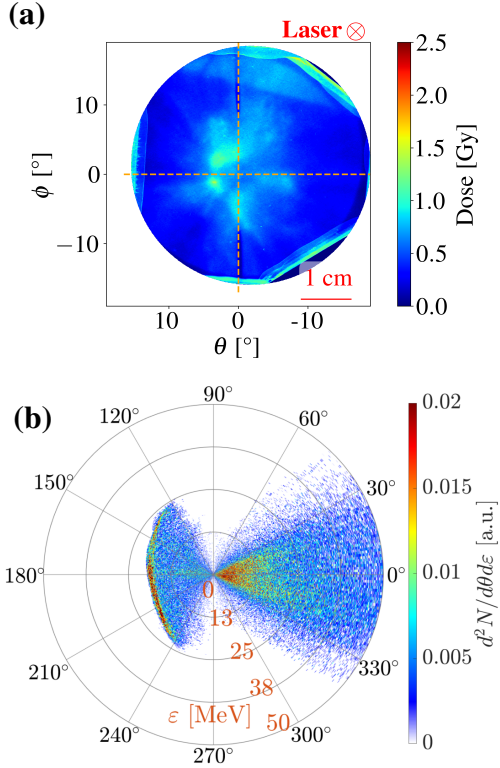
\begin{figure}
	\begin{subfigure}[b]{0.34\textwidth}
		\centering
		\includegraphics[width=\textwidth]{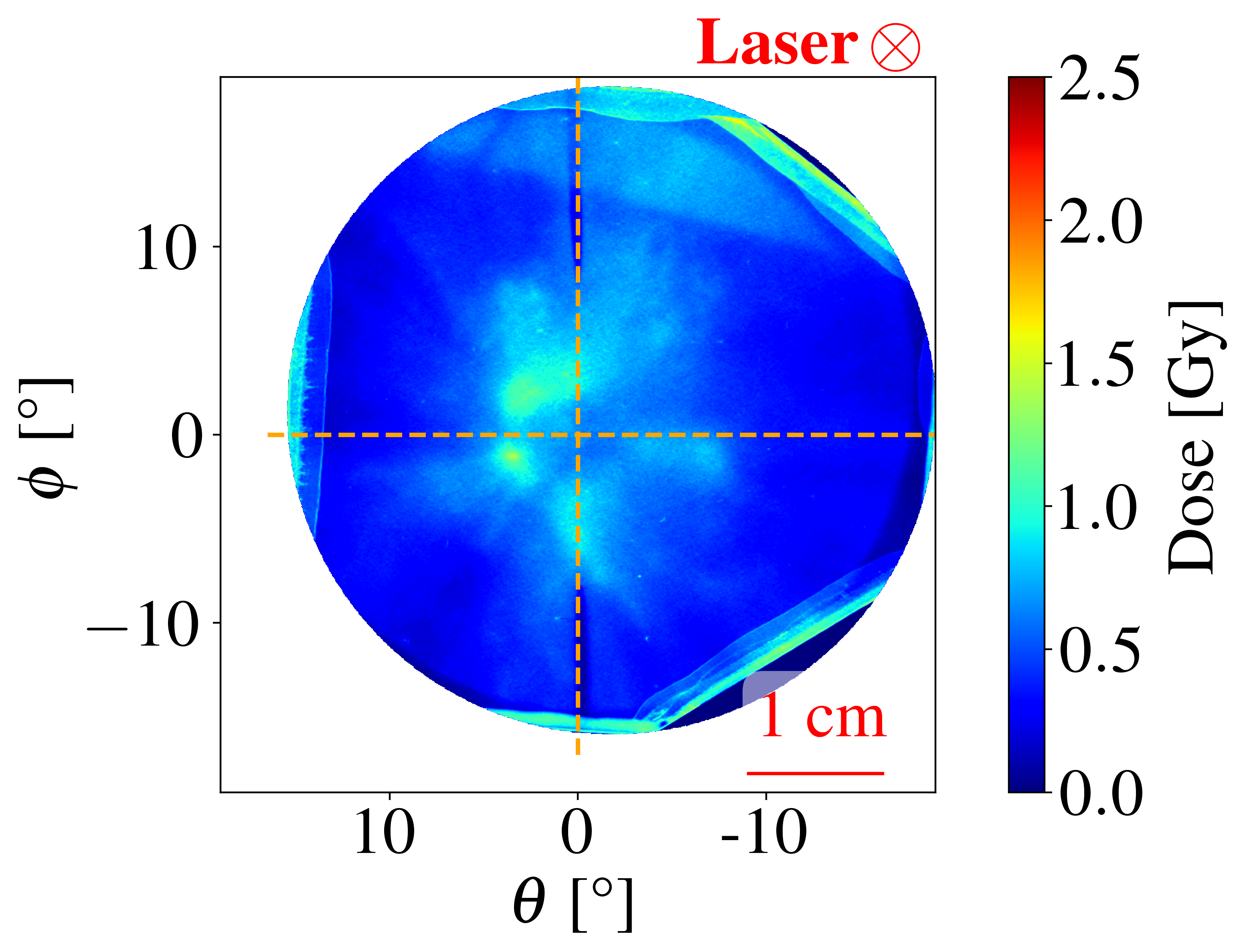}
		\begin{picture}(20,20)
        \put(-83,150){\large\textbf{(a)}}
		\end{picture}	
	\end{subfigure}
	\begin{subfigure}[b]{0.37\textwidth}
		\centering
        \vspace{-6mm}
		\includegraphics[width=\textwidth]{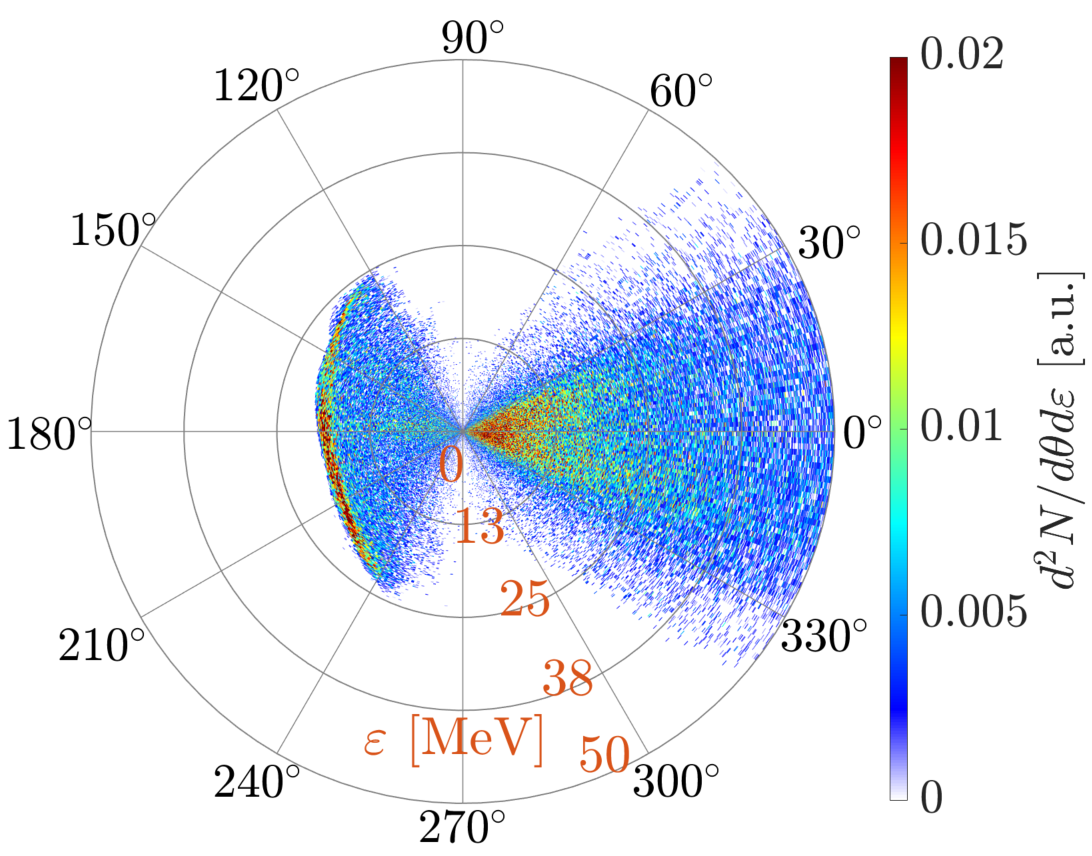}
		\begin{picture}(20,20)
        \put(-80,150){\large\textbf{(b)}}
		\end{picture}
	\end{subfigure}
    \vspace{-7mm}
    \caption{(a) Angular dose distribution in the (deepest) 15th layer of the RCF stack located $6\,\rm cm$ beyond TCC along the laser axis (center of the dashed cross). This layer is mainly sensitive to $>2\,\rm MeV$ electrons. (b) Simulated energy-angle distribution (linear scale) of the electrons having reached the $x = 1600\,\mu \rm m$ longitudinal PIC coordinate. Here, the laser propagates along the $\theta=0^{\circ}$ axis.
    }

    \label{fig:RCF_results}
\end{figure}

The angular distribution of the outgoing fast electrons can be further assessed from the RCF data. Indeed, given the $\sim 2.7\,\rm MeV$ cutoff ion energies inferred from the ToF data and the absence of ion signal on the on-axis TP, one can predict through Monte Carlo simulations performed with PySTarT \cite{Ehret_2021} (a python wrapper for the \textsc{srim} package \cite{Ziegler_2010}) that the emitted ions will be fully stopped by the second RCF stack layer, and hence that the dose deposited on subsequent layers is essentially due to fast electrons.
The spatial dose distributions measured deep (beyond the 10th layer) inside the RCF stack typically exhibit several hot spots,
as illustrated in Fig.~\ref{fig:RCF_results}(a).  The signal shown was recorded on the deepest (15th) layer (mainly sensitive to $>2\,\rm MeV$ electrons according to \textsc{geant4} \cite{Agostinelli_2003} Monte Carlo modeling) during a single shot (not associated with the electron spectra shown in Fig.~\ref{fig:electron_results_spectra}, as the MES and RCF stack could not be fielded simultaneously). The dose is mainly deposited over a $\sim 8^{\circ}$ FWHM cone, with
several hot spots surrounding the laser axis (center of dashed cross), corresponding to electron beamlets emitted at a few ($\sim 3^{\circ}-5^{\circ}$) degrees. This suggests that the MES located at $(\theta,\phi)=(-2^{\circ},0^{\circ})$ and $(21^{\circ},0^{\circ})$ may miss the dominant components of the outgoing energetic electron population. An estimate of the outgoing electron charge can be obtained from the electron spectrum in green in Fig.~\ref{fig:electron_results_spectra} considering the $\sim 5^{\circ}$ dose-deposition spot centered at $(\theta,\phi)=(2^{\circ},2^{\circ})$ in the RCF layer depicted in Fig.~\ref{fig:RCF_results}(a). The typical outgoing charge above $2\,\rm MeV$ is $\sim 1 \,\rm nC$ corresponding to $\sim0.02 \, \rm J$, i.e., $\sim 0.1\,\%$ of the laser drive energy.



We display in Fig.~\ref{fig:RCF_results}(b) the energy-angle distribution of the fast electrons having reached the $x = 1600\,\mu \rm m$ ``plane'' in the PIC simulation. 
The backward-directed part of this distribution corresponds to the electrons reflected by the TNSA field. The purpose of this numerical diagnostic is to compare the angular distribution of the outgoing energetic electrons with that inferred from the RCF signals. The angular distribution of the fast electrons is quite inhomogeneous and mainly contained in the $\vert \theta \vert \lesssim 30^{\circ}$ range. 





\begin{figure}
    \centering
	\begin{subfigure}[b]{0.32\textwidth}
		\centering
		\includegraphics[width=\textwidth]{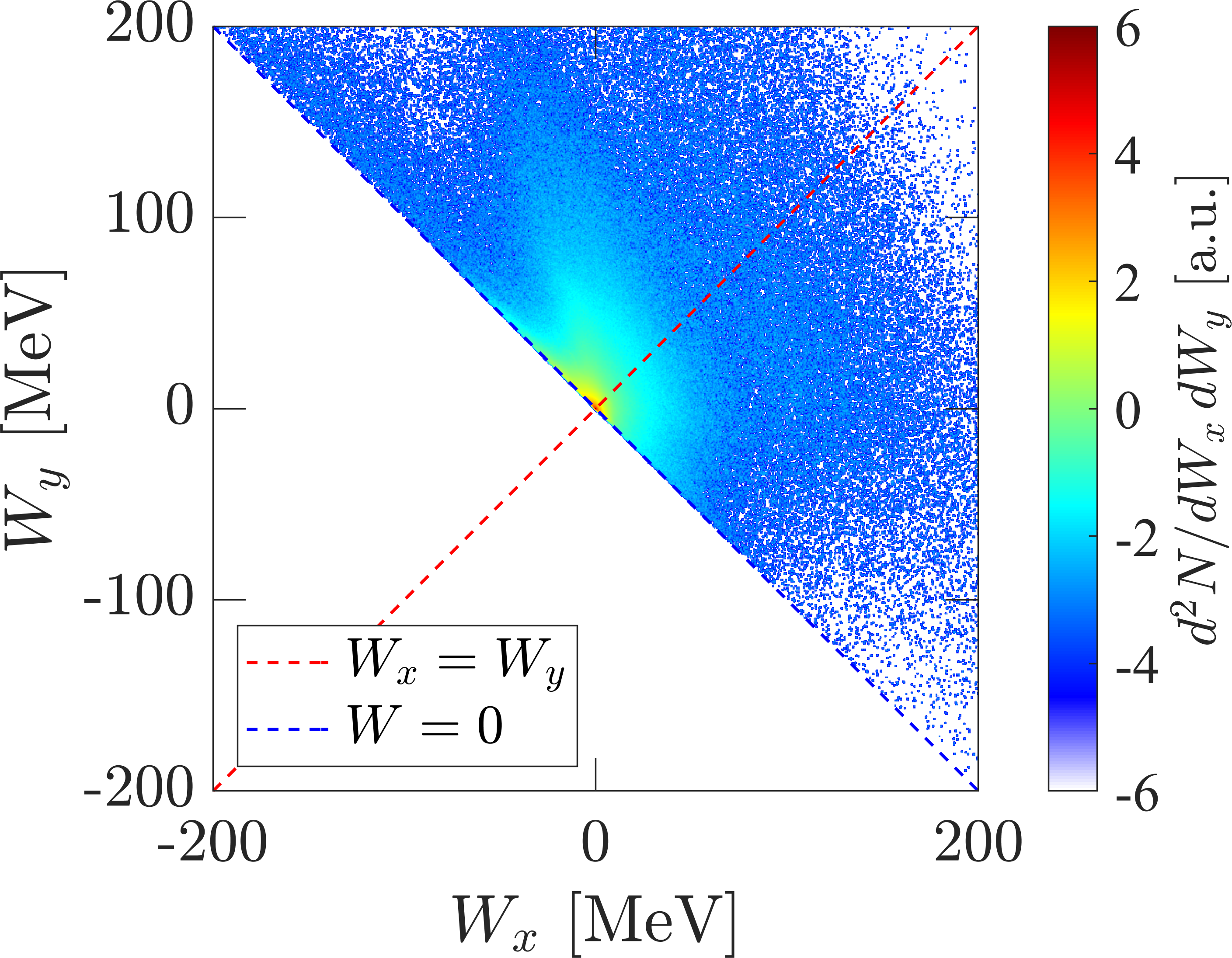}
		\begin{picture}(20,20)
        \put(-90,140){\large\textbf{(a)}}
		\end{picture}	
	\end{subfigure}
	\begin{subfigure}[b]{0.35\textwidth}
		\centering
        \vspace{-3mm}
		\includegraphics[width=\textwidth]{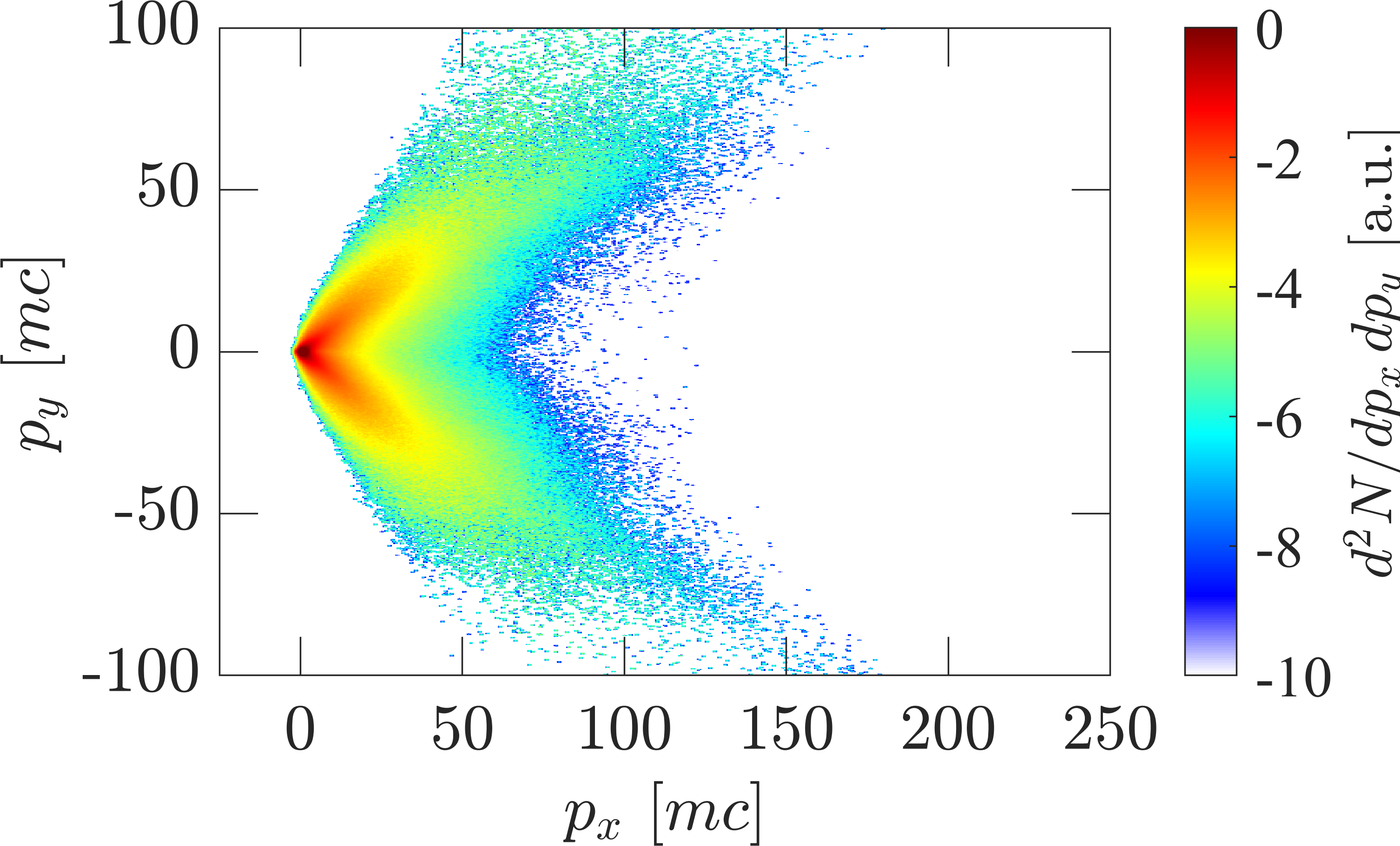}
		\begin{picture}(20,20)
        \put(-90,110){\large\textbf{(b)}}
		\end{picture}
	\end{subfigure}
    \vspace{-8mm}
    \caption{(a) Electron distribution as a function of the works performed by the longitudinal ($W_x$) and transverse ($W_y$) electric fields, extracted $\sim 0.5\,\rm ps$ before the laser pulse crosses the GDP.
    (b) $p_x-p_y$ momentum distribution of the electrons having exited the plasma channel but not yet crossed the magnetized right-hand gas region ($1400 \, \mu \rm m \leq x \leq 1500 \, \mu \rm m$), as recorded at $t \simeq 1.8\, \rm ps$. 
    All figures are in log$_{10}$ scale.
    }
    \label{fig:DLA}
    \vspace{-5mm}
\end{figure}
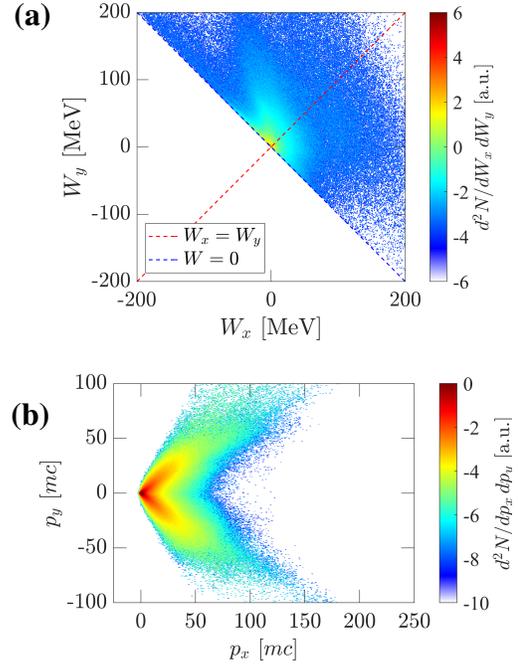

To identify the dominant electron acceleration mechanism, we plot in Fig.~\ref{fig:DLA}(a) the electron distribution resolved as a function of the works performed by the longitudinal ($W_x$) and transverse ($W_y$) electric fields defined as $W_{x,y} = -e \int_0^t dt'\,E_{x,y}(t',\textbf{r}(t'))v_{x,y}(t')$. This diagnostic is extracted \mbox{$\sim 0.5\,\rm ps$} before the laser pulse has reached the GDP. Since all plasma electrons have gained energy from the laser, we have $W_x+W_y \ge 0$, which explains the sharp linear lower boundary of the distribution. The laser-accelerated electrons are then mainly concentrated around the head of the laser-created plasma channel, located around $x \simeq 850\,\rm \mu m$. Importantly, a significant majority of them satisfy $W_y > W_x$; this means that they have been mainly energized by transverse $E_y$ fields -- largely dominated by the laser field -- rather than by longitudinal, laser-wakefield-type $E_x$ fields, a process known as direct laser acceleration (DLA) \cite{Arefiev_2016, Shaw_2017, *Shaw_2018, Hussein_2021}.

Figure~\ref{fig:DLA}(b) shows the $p_x-p_y$ momentum distribution of the energetic electrons outgoing from the plasma channel and then lying in the $1400 \leq x \leq 1500 \,\mu\rm m$ spatial range (in the gas down-ramp), $\sim 1.8\,\rm ps$ after the laser pulse has crossed the GDP. The forked shape of this distribution is typical of DLA.

\section{Conclusions}
\label{sec:conclusions}

Here, we have presented the results of an experiment in which we coupled a $\sim~70\,\rm fs$, $\sim~10^{20}\,\rm W\,cm^{-2}$ laser pulse to a near-critical He gas jet, and probed the interaction region with an extensive diagnostic suite. Our main findings are (i) the acceleration at forward oblique angles ($\sim 17^{\circ}$) of $\alpha$ particles up to $2.7\,\rm MeV$ energies ($\sim~0.67\,\rm MeV/amu$) with a total flux of $\sim 10^{11}\,\rm sr^{-1}$ above $\sim 0.1\,\rm MeV$ energies; (ii) the emission of a quasi-exponential distribution of forward-directed hot electrons with a $\sim 1\,\rm nC$ total charge and a $\sim 10\,\rm MeV$ temperature well above the ponderomotive scaling. To our knowledge, the only other experiment on particle acceleration in near-critical gas jets ($n_e \gtrsim 0.1n_c$) driven by PW-class, ultrashort ($<100\,\rm fs$) laser pulses was reported by Singh \emph{et al.} \cite{Singh_2020}. 

According to large-scale PIC modeling, the observed forward ion acceleration most likely arises from electrostatic reflection of the ions swept up by the radially expanding laser-induced plasma channel -- the formation of which across the gas density peak was also evidenced via interferometry. The same collisionless-shock-based mechanism was held responsible for the transverse emission of $\sim 0.8\,\rm MeV$ H$^+$ and He$^{2+}$ ions in Ref.~\cite{Singh_2020}. While this mechanism preferentially drives ions perpendicularly to the laser path \cite{Singh_2020}, our simulations reveal that the ions initially located near the end of the channel, in the gas down-ramp, can be accelerated over a broad forward directed cone, including the $\pm 17^{\circ}$ lines of the sight of the ToF detectors, and to energies consistent with the measurements. The simulations also predict TNSA-type ion acceleration at the rear edge of the gas, as well as deflection of the lower-energy ions by the local magnetic fields. However, the high energies ($\gtrsim 10\,\rm MeV$) reached by the fastest TNSA ions are not corroborated by the measurements. We ascribe this disagreement to an improper description of the fast-electron dilution far from the plasma channel and/or to a possibly too sharp truncation of the gas down-ramp. Furthermore, an experimental characterization of the gas-edge $B$ fields is needed to further assess the ion deflection scenario. Nevertheless, our modeling satisfactorily reproduces the quasi-exponential spectrum of the fast electrons and shows that they are primarily produced by the direct action of the laser's electric field.

Finally, it is worth noting that, although our experimental suite was not fully adapted to HRR operation, we were able to achieve a frequency of one ``UHI physics shot'' every 20 minutes. This time was required to (i) obtain an interferogram of the pre-shot neutral gas, (ii) translate the nozzle vertically to have the laser interact with the shocked gas region, and (iii) perform two different acquisitions with the on-shot interferometry CCD (one with the probe only and another one with both gas and probe), which were needed to deconvolve the subsequent on-shot interferogram.
Such a meticulous procedure was necessary to achieve well-controlled interaction conditions. We also leveraged this time to replace the passive particle detectors (i.e., the imaging plates and radiochromic films). In addition, we succeeded in performing up to four UHI laser shots in a row without the laser-induced nozzle damage severely altering the gas shape. Although the latter exhibited fluctuations in both peak density ($\simeq~(1~-~2)~\times 10^{20}\,\rm cm^{-3}$) and FWHM ($\sim~500-600\,\rm \mu m$), along with the formation of low-density ($\lesssim 10^{19}\,\rm cm^{-3}$) ``elbows'' after a certain degree of damage, such shot-to-shot variations remained moderate enough not tocompromise the typical properties of the accelerated particles or the bulk gas response to the laser drive.

Although very far from the ultimate goal of approaching the $>1-10\,\rm Hz$ shot rate of modern Ti:Sa laser systems, these results represent, to our knowledge, the best performance ever reported using a PW-class Ti:Sa laser coupled with a dense gas jet. As such, they open up encouraging prospects for future applications requiring large statistics.
For instance, once the transverse size and emittance of the ion source have been characterized, high-precision stopping-power measurements \cite{Malko_2022} could be made. Compared to laser-based accelerators using solid targets, a prominent advantage of laser-gas setups would be the ability to deliver fast ions of any chemical element simply by changing the gas composition. The case of $\alpha$ particles, as considered in this work, would be of particular interest for accurate predictions of the performance of inertial confinement fusion designs \cite{Zylstra_2019, *Zylstra_2022}, a topic with far-reaching implications given the fusion breakthroughs reported at the National Ignition Facility in December 2022 and July 2023. \\   


\section{Acknowledgments}

Thanks are due to R. Nuter for useful discussions.
We received financial support from the French State, managed by the French National Research Agency (ANR) in the frame of the Investments for the future Programme IdEx Bordeaux - LAPHIA (ANR-10-IDEX-03-02).
This work has received funding from the European Union's Horizon 2020 research and innovation programs: Laserlab V (grant agreement No. 871124 INFRAIA and Joint Research Activity 2.4), IMPULSE (grant agreement No. 871161 INFRADEV).
V.~O-B and C.~V. acknowledge the support from the LIGHT S\&T Graduate Program (PIA3 Investment for the Future Program, ANR-17-EURE-0027).
We acknowledge GENCI for providing us access to the Joliot-Curie supercomputer (grants 2021-A0130512993 and 2022-A0130512993). 
This scientific paper is published as part of the international project called ``PMW'', co-financed by the Polish Ministry of Science and Higher Education within the framework of the scientific financial resources for 2021-2022 under the contract no 5205/CELIA/2021/0 (project CNRS No. 239915).
This work has also been supported by the Research Grant No. PID2019-108764RB-I00 from the Spanish Ministry of Science and Innovation and from the Unidad de Investigaci{\'o}n Consolidada de Castilla y Le{\'o}n No. CLP087U16.

\section*{Data Availability Statement}

The data that support the findings of this study are available from the corresponding authors upon reasonable request.



\bibliography{Laser-driven_ion_and_electron}

\providecommand{\noopsort}[1]{}\providecommand{\singleletter}[1]{#1}%
\begin{thebibliography}{85}%
\makeatletter
\providecommand \@ifxundefined [1]{%
 \@ifx{#1\undefined}
}%
\providecommand \@ifnum [1]{%
 \ifnum #1\expandafter \@firstoftwo
 \else \expandafter \@secondoftwo
 \fi
}%
\providecommand \@ifx [1]{%
 \ifx #1\expandafter \@firstoftwo
 \else \expandafter \@secondoftwo
 \fi
}%
\providecommand \natexlab [1]{#1}%
\providecommand \enquote  [1]{``#1''}%
\providecommand \bibnamefont  [1]{#1}%
\providecommand \bibfnamefont [1]{#1}%
\providecommand \citenamefont [1]{#1}%
\providecommand \href@noop [0]{\@secondoftwo}%
\providecommand \href [0]{\begingroup \@sanitize@url \@href}%
\providecommand \@href[1]{\@@startlink{#1}\@@href}%
\providecommand \@@href[1]{\endgroup#1\@@endlink}%
\providecommand \@sanitize@url [0]{\catcode `\\12\catcode `\$12\catcode
  `\&12\catcode `\#12\catcode `\^12\catcode `\_12\catcode `\%12\relax}%
\providecommand \@@startlink[1]{}%
\providecommand \@@endlink[0]{}%
\providecommand \url  [0]{\begingroup\@sanitize@url \@url }%
\providecommand \@url [1]{\endgroup\@href {#1}{\urlprefix }}%
\providecommand \urlprefix  [0]{URL }%
\providecommand \Eprint [0]{\href }%
\providecommand \doibase [0]{https://doi.org/}%
\providecommand \selectlanguage [0]{\@gobble}%
\providecommand \bibinfo  [0]{\@secondoftwo}%
\providecommand \bibfield  [0]{\@secondoftwo}%
\providecommand \translation [1]{[#1]}%
\providecommand \BibitemOpen [0]{}%
\providecommand \bibitemStop [0]{}%
\providecommand \bibitemNoStop [0]{.\EOS\space}%
\providecommand \EOS [0]{\spacefactor3000\relax}%
\providecommand \BibitemShut  [1]{\csname bibitem#1\endcsname}%
\let\auto@bib@innerbib\@empty
\bibitem [{\citenamefont {{Daido}}\ \emph {et~al.}(2012)\citenamefont
  {{Daido}}, \citenamefont {{Nishiuchi}},\ and\ \citenamefont
  {{Pirozhkov}}}]{Daido_2012}%
  \BibitemOpen
  \bibfield  {author} {\bibinfo {author} {\bibfnamefont {H.}~\bibnamefont
  {{Daido}}}, \bibinfo {author} {\bibfnamefont {M.}~\bibnamefont
  {{Nishiuchi}}},\ and\ \bibinfo {author} {\bibfnamefont {A.~S.}\ \bibnamefont
  {{Pirozhkov}}},\ }\bibfield  {title} {\bibinfo {title} {{Review of
  laser-driven ion sources and their applications}},\ }\href
  {https://doi.org/10.1088/0034-4885/75/5/056401} {\bibfield  {journal}
  {\bibinfo  {journal} {Rep. Prog. Phys.}\ }\textbf {\bibinfo {volume} {75}},\
  \bibinfo {eid} {056401} (\bibinfo {year} {2012})}\BibitemShut {NoStop}%
\bibitem [{\citenamefont {Macchi}\ \emph {et~al.}(2013)\citenamefont {Macchi},
  \citenamefont {Borghesi},\ and\ \citenamefont {Passoni}}]{Macchi_2013}%
  \BibitemOpen
  \bibfield  {author} {\bibinfo {author} {\bibfnamefont {A.}~\bibnamefont
  {Macchi}}, \bibinfo {author} {\bibfnamefont {M.}~\bibnamefont {Borghesi}},\
  and\ \bibinfo {author} {\bibfnamefont {M.}~\bibnamefont {Passoni}},\
  }\bibfield  {title} {\bibinfo {title} {Ion acceleration by superintense
  laser-plasma interaction},\ }\href
  {https://doi.org/10.1103/RevModPhys.85.751} {\bibfield  {journal} {\bibinfo
  {journal} {Rev. Mod. Phys.}\ }\textbf {\bibinfo {volume} {85}},\ \bibinfo
  {pages} {751} (\bibinfo {year} {2013})}\BibitemShut {NoStop}%
\bibitem [{\citenamefont {{Schreiber}}\ \emph {et~al.}(2016)\citenamefont
  {{Schreiber}}, \citenamefont {{Bolton}},\ and\ \citenamefont
  {{Parodi}}}]{Schreiber_2016}%
  \BibitemOpen
  \bibfield  {author} {\bibinfo {author} {\bibfnamefont {J.}~\bibnamefont
  {{Schreiber}}}, \bibinfo {author} {\bibfnamefont {P.~R.}\ \bibnamefont
  {{Bolton}}},\ and\ \bibinfo {author} {\bibfnamefont {K.}~\bibnamefont
  {{Parodi}}},\ }\bibfield  {title} {\bibinfo {title} {{Invited Review Article:
  ``Hands-on'' laser-driven ion acceleration: A primer for laser-driven source
  development and potential applications}},\ }\href
  {https://doi.org/10.1063/1.4959198} {\bibfield  {journal} {\bibinfo
  {journal} {Rev.Sci. Instrum.}\ }\textbf {\bibinfo {volume} {87}},\ \bibinfo
  {eid} {071101} (\bibinfo {year} {2016})}\BibitemShut {NoStop}%
\bibitem [{\citenamefont {{Borghesi}}\ \emph {et~al.}(2003)\citenamefont
  {{Borghesi}}, \citenamefont {{Schiavi}}, \citenamefont {{Campbell}},
  \citenamefont {{Haines}}, \citenamefont {{Willi}}, \citenamefont
  {{Mackinnon}}, \citenamefont {{Patel}}, \citenamefont {{Galimberti}},\ and\
  \citenamefont {{Gizzi}}}]{Borghesi_2003}%
  \BibitemOpen
  \bibfield  {author} {\bibinfo {author} {\bibfnamefont {M.}~\bibnamefont
  {{Borghesi}}}, \bibinfo {author} {\bibfnamefont {A.}~\bibnamefont
  {{Schiavi}}}, \bibinfo {author} {\bibfnamefont {D.~H.}\ \bibnamefont
  {{Campbell}}}, \bibinfo {author} {\bibfnamefont {M.~G.}\ \bibnamefont
  {{Haines}}}, \bibinfo {author} {\bibfnamefont {O.}~\bibnamefont {{Willi}}},
  \bibinfo {author} {\bibfnamefont {A.~J.}\ \bibnamefont {{Mackinnon}}},
  \bibinfo {author} {\bibfnamefont {P.}~\bibnamefont {{Patel}}}, \bibinfo
  {author} {\bibfnamefont {M.}~\bibnamefont {{Galimberti}}},\ and\ \bibinfo
  {author} {\bibfnamefont {L.~A.}\ \bibnamefont {{Gizzi}}},\ }\bibfield
  {title} {\bibinfo {title} {{Proton imaging detection of transient
  electromagnetic fields in laser-plasma interactions (invited)}},\ }\href
  {https://doi.org/10.1063/1.1534390} {\bibfield  {journal} {\bibinfo
  {journal} {Rev. Sci. Instrum.}\ }\textbf {\bibinfo {volume} {74}},\ \bibinfo
  {pages} {1688} (\bibinfo {year} {2003})}\BibitemShut {NoStop}%
\bibitem [{\citenamefont {{Borghesi}}\ \emph {et~al.}(2004)\citenamefont
  {{Borghesi}}, \citenamefont {{MacKinnon}}, \citenamefont {{Campbell}},
  \citenamefont {{Hicks}}, \citenamefont {{Kar}}, \citenamefont {{Patel}},
  \citenamefont {{Price}}, \citenamefont {{Romagnani}}, \citenamefont
  {{Schiavi}},\ and\ \citenamefont {{Willi}}}]{Borghesi_2004}%
  \BibitemOpen
  \bibfield  {author} {\bibinfo {author} {\bibfnamefont {M.}~\bibnamefont
  {{Borghesi}}}, \bibinfo {author} {\bibfnamefont {A.~J.}\ \bibnamefont
  {{MacKinnon}}}, \bibinfo {author} {\bibfnamefont {D.~H.}\ \bibnamefont
  {{Campbell}}}, \bibinfo {author} {\bibfnamefont {D.~G.}\ \bibnamefont
  {{Hicks}}}, \bibinfo {author} {\bibfnamefont {S.}~\bibnamefont {{Kar}}},
  \bibinfo {author} {\bibfnamefont {P.~K.}\ \bibnamefont {{Patel}}}, \bibinfo
  {author} {\bibfnamefont {D.}~\bibnamefont {{Price}}}, \bibinfo {author}
  {\bibfnamefont {L.}~\bibnamefont {{Romagnani}}}, \bibinfo {author}
  {\bibfnamefont {A.}~\bibnamefont {{Schiavi}}},\ and\ \bibinfo {author}
  {\bibfnamefont {O.}~\bibnamefont {{Willi}}},\ }\bibfield  {title} {\bibinfo
  {title} {{Multi-MeV Proton Source Investigations in Ultraintense Laser-Foil
  Interactions}},\ }\href {https://doi.org/10.1103/PhysRevLett.92.055003}
  {\bibfield  {journal} {\bibinfo  {journal} {Phys. Rev. Lett.}\ }\textbf
  {\bibinfo {volume} {92}},\ \bibinfo {eid} {055003} (\bibinfo {year}
  {2004})}\BibitemShut {NoStop}%
\bibitem [{\citenamefont {Santos}\ \emph {et~al.}(2015)\citenamefont {Santos},
  \citenamefont {Bailly-Grandvaux}, \citenamefont {Giuffrida}, \citenamefont
  {Forestier-Colleoni}, \citenamefont {Fujioka}, \citenamefont {Zhang},
  \citenamefont {Korneev}, \citenamefont {Bouillaud}, \citenamefont {Dorard},
  \citenamefont {Batani}, \citenamefont {Chevrot}, \citenamefont {Cross},
  \citenamefont {Crowston}, \citenamefont {Dubois}, \citenamefont {Gazave},
  \citenamefont {Gregori}, \citenamefont {d'Humi{\`{e}}res}, \citenamefont
  {Hulin}, \citenamefont {Ishihara}, \citenamefont {Kojima}, \citenamefont
  {Loyez}, \citenamefont {Marqu{\`{e}}s}, \citenamefont {Morace}, \citenamefont
  {Nicolaï}, \citenamefont {Peyrusse}, \citenamefont {Poy{\'{e}}},
  \citenamefont {Raffestin}, \citenamefont {Ribolzi}, \citenamefont {Roth},
  \citenamefont {Schaumann}, \citenamefont {Serres}, \citenamefont
  {Tikhonchuk}, \citenamefont {Vacar},\ and\ \citenamefont
  {Woolsey}}]{Santos_2015}%
  \BibitemOpen
  \bibfield  {author} {\bibinfo {author} {\bibfnamefont {J.~J.}\ \bibnamefont
  {Santos}}, \bibinfo {author} {\bibfnamefont {M.}~\bibnamefont
  {Bailly-Grandvaux}}, \bibinfo {author} {\bibfnamefont {L.}~\bibnamefont
  {Giuffrida}}, \bibinfo {author} {\bibfnamefont {P.}~\bibnamefont
  {Forestier-Colleoni}}, \bibinfo {author} {\bibfnamefont {S.}~\bibnamefont
  {Fujioka}}, \bibinfo {author} {\bibfnamefont {Z.}~\bibnamefont {Zhang}},
  \bibinfo {author} {\bibfnamefont {P.}~\bibnamefont {Korneev}}, \bibinfo
  {author} {\bibfnamefont {R.}~\bibnamefont {Bouillaud}}, \bibinfo {author}
  {\bibfnamefont {S.}~\bibnamefont {Dorard}}, \bibinfo {author} {\bibfnamefont
  {D.}~\bibnamefont {Batani}}, \bibinfo {author} {\bibfnamefont
  {M.}~\bibnamefont {Chevrot}}, \bibinfo {author} {\bibfnamefont {J.~E.}\
  \bibnamefont {Cross}}, \bibinfo {author} {\bibfnamefont {R.}~\bibnamefont
  {Crowston}}, \bibinfo {author} {\bibfnamefont {J.-L.}\ \bibnamefont
  {Dubois}}, \bibinfo {author} {\bibfnamefont {J.}~\bibnamefont {Gazave}},
  \bibinfo {author} {\bibfnamefont {G.}~\bibnamefont {Gregori}}, \bibinfo
  {author} {\bibfnamefont {E.}~\bibnamefont {d'Humi{\`{e}}res}}, \bibinfo
  {author} {\bibfnamefont {S.}~\bibnamefont {Hulin}}, \bibinfo {author}
  {\bibfnamefont {K.}~\bibnamefont {Ishihara}}, \bibinfo {author}
  {\bibfnamefont {S.}~\bibnamefont {Kojima}}, \bibinfo {author} {\bibfnamefont
  {E.}~\bibnamefont {Loyez}}, \bibinfo {author} {\bibfnamefont {J.-R.}\
  \bibnamefont {Marqu{\`{e}}s}}, \bibinfo {author} {\bibfnamefont
  {A.}~\bibnamefont {Morace}}, \bibinfo {author} {\bibfnamefont
  {P.}~\bibnamefont {Nicolaï}}, \bibinfo {author} {\bibfnamefont
  {O.}~\bibnamefont {Peyrusse}}, \bibinfo {author} {\bibfnamefont
  {A.}~\bibnamefont {Poy{\'{e}}}}, \bibinfo {author} {\bibfnamefont
  {D.}~\bibnamefont {Raffestin}}, \bibinfo {author} {\bibfnamefont
  {J.}~\bibnamefont {Ribolzi}}, \bibinfo {author} {\bibfnamefont
  {M.}~\bibnamefont {Roth}}, \bibinfo {author} {\bibfnamefont {G.}~\bibnamefont
  {Schaumann}}, \bibinfo {author} {\bibfnamefont {F.}~\bibnamefont {Serres}},
  \bibinfo {author} {\bibfnamefont {V.~T.}\ \bibnamefont {Tikhonchuk}},
  \bibinfo {author} {\bibfnamefont {P.}~\bibnamefont {Vacar}},\ and\ \bibinfo
  {author} {\bibfnamefont {N.}~\bibnamefont {Woolsey}},\ }\bibfield  {title}
  {\bibinfo {title} {Laser-driven platform for generation and characterization
  of strong quasi-static magnetic fields},\ }\href
  {https://doi.org/10.1088/1367-2630/17/8/083051} {\bibfield  {journal}
  {\bibinfo  {journal} {New J. Phys.}\ }\textbf {\bibinfo {volume} {17}},\
  \bibinfo {pages} {083051} (\bibinfo {year} {2015})}\BibitemShut {NoStop}%
\bibitem [{\citenamefont {{Ehret}}\ \emph {et~al.}(2023)\citenamefont
  {{Ehret}}, \citenamefont {{Bailly-Grandvaux}}, \citenamefont {{Korneev}},
  \citenamefont {{Api{\~n}aniz}}, \citenamefont {{Brabetz}}, \citenamefont
  {{Morace}}, \citenamefont {{Bradford}}, \citenamefont {{d'Humi{\`e}res}},
  \citenamefont {{Schaumann}}, \citenamefont {{Bagnoud}}, \citenamefont
  {{Malko}}, \citenamefont {{Matveevskii}}, \citenamefont {{Roth}},
  \citenamefont {{Volpe}}, \citenamefont {{Woolsey}},\ and\ \citenamefont
  {{Santos}}}]{Ehret_2023}%
  \BibitemOpen
  \bibfield  {author} {\bibinfo {author} {\bibfnamefont {M.}~\bibnamefont
  {{Ehret}}}, \bibinfo {author} {\bibfnamefont {M.}~\bibnamefont
  {{Bailly-Grandvaux}}}, \bibinfo {author} {\bibfnamefont {P.}~\bibnamefont
  {{Korneev}}}, \bibinfo {author} {\bibfnamefont {J.~I.}\ \bibnamefont
  {{Api{\~n}aniz}}}, \bibinfo {author} {\bibfnamefont {C.}~\bibnamefont
  {{Brabetz}}}, \bibinfo {author} {\bibfnamefont {A.}~\bibnamefont {{Morace}}},
  \bibinfo {author} {\bibfnamefont {P.}~\bibnamefont {{Bradford}}}, \bibinfo
  {author} {\bibfnamefont {E.}~\bibnamefont {{d'Humi{\`e}res}}}, \bibinfo
  {author} {\bibfnamefont {G.}~\bibnamefont {{Schaumann}}}, \bibinfo {author}
  {\bibfnamefont {V.}~\bibnamefont {{Bagnoud}}}, \bibinfo {author}
  {\bibfnamefont {S.}~\bibnamefont {{Malko}}}, \bibinfo {author} {\bibfnamefont
  {K.}~\bibnamefont {{Matveevskii}}}, \bibinfo {author} {\bibfnamefont
  {M.}~\bibnamefont {{Roth}}}, \bibinfo {author} {\bibfnamefont
  {L.}~\bibnamefont {{Volpe}}}, \bibinfo {author} {\bibfnamefont {N.~C.}\
  \bibnamefont {{Woolsey}}},\ and\ \bibinfo {author} {\bibfnamefont {J.~J.}\
  \bibnamefont {{Santos}}},\ }\bibfield  {title} {\bibinfo {title} {{Guided
  electromagnetic discharge pulses driven by short intense laser pulses:
  Characterization and modeling}},\ }\href {https://doi.org/10.1063/5.0124011}
  {\bibfield  {journal} {\bibinfo  {journal} {Phys. Plasmas}\ }\textbf
  {\bibinfo {volume} {30}},\ \bibinfo {eid} {013105} (\bibinfo {year}
  {2023})}\BibitemShut {NoStop}%
\bibitem [{\citenamefont {{Patel}}\ \emph {et~al.}(2003)\citenamefont
  {{Patel}}, \citenamefont {{MacKinnon}}, \citenamefont {{Key}}, \citenamefont
  {{Cowan}}, \citenamefont {{Foord}}, \citenamefont {{Allen}}, \citenamefont
  {{Price}}, \citenamefont {{Ruhl}}, \citenamefont {{Springer}},\ and\
  \citenamefont {{Stephens}}}]{Patel_2003}%
  \BibitemOpen
  \bibfield  {author} {\bibinfo {author} {\bibfnamefont {P.~K.}\ \bibnamefont
  {{Patel}}}, \bibinfo {author} {\bibfnamefont {A.~J.}\ \bibnamefont
  {{MacKinnon}}}, \bibinfo {author} {\bibfnamefont {M.~H.}\ \bibnamefont
  {{Key}}}, \bibinfo {author} {\bibfnamefont {T.~E.}\ \bibnamefont {{Cowan}}},
  \bibinfo {author} {\bibfnamefont {M.~E.}\ \bibnamefont {{Foord}}}, \bibinfo
  {author} {\bibfnamefont {M.}~\bibnamefont {{Allen}}}, \bibinfo {author}
  {\bibfnamefont {D.~F.}\ \bibnamefont {{Price}}}, \bibinfo {author}
  {\bibfnamefont {H.}~\bibnamefont {{Ruhl}}}, \bibinfo {author} {\bibfnamefont
  {P.~T.}\ \bibnamefont {{Springer}}},\ and\ \bibinfo {author} {\bibfnamefont
  {R.}~\bibnamefont {{Stephens}}},\ }\bibfield  {title} {\bibinfo {title}
  {{Isochoric Heating of Solid-Density Matter with an Ultrafast Proton Beam}},\
  }\href {https://doi.org/10.1103/PhysRevLett.91.125004} {\bibfield  {journal}
  {\bibinfo  {journal} {\prl}\ }\textbf {\bibinfo {volume} {91}},\ \bibinfo
  {eid} {125004} (\bibinfo {year} {2003})}\BibitemShut {NoStop}%
\bibitem [{\citenamefont {{Roth}}\ \emph {et~al.}(2009)\citenamefont {{Roth}},
  \citenamefont {{Alber}}, \citenamefont {{Bagnoud}}, \citenamefont {{Brown}},
  \citenamefont {{Clarke}}, \citenamefont {{Daido}}, \citenamefont
  {{Fernandez}}, \citenamefont {{Flippo}}, \citenamefont {{Gaillard}},
  \citenamefont {{Gauthier}}, \citenamefont {{Geissel}}, \citenamefont
  {{Glenzer}}, \citenamefont {{Gregori}}, \citenamefont {{G{\"u}nther}},
  \citenamefont {{Harres}}, \citenamefont {{Heathcote}}, \citenamefont
  {{Kritcher}}, \citenamefont {{Kugland}}, \citenamefont {{Le Pape}},
  \citenamefont {{Li}}, \citenamefont {{Makita}}, \citenamefont {{Mithen}},
  \citenamefont {{Niemann}}, \citenamefont {{N{\"u}rnberg}}, \citenamefont
  {{Offermann}}, \citenamefont {{Otten}}, \citenamefont {{Pelka}},
  \citenamefont {{Riley}}, \citenamefont {{Schaumann}}, \citenamefont
  {{Schollmeier}}, \citenamefont {{Sch{\"u}trumpf}}, \citenamefont {{Tampo}},
  \citenamefont {{Tauschwitz}},\ and\ \citenamefont
  {{Tauschwitz}}}]{Roth_2009}%
  \BibitemOpen
  \bibfield  {author} {\bibinfo {author} {\bibfnamefont {M.}~\bibnamefont
  {{Roth}}}, \bibinfo {author} {\bibfnamefont {I.}~\bibnamefont {{Alber}}},
  \bibinfo {author} {\bibfnamefont {V.}~\bibnamefont {{Bagnoud}}}, \bibinfo
  {author} {\bibfnamefont {C.~R.~D.}\ \bibnamefont {{Brown}}}, \bibinfo
  {author} {\bibfnamefont {R.}~\bibnamefont {{Clarke}}}, \bibinfo {author}
  {\bibfnamefont {H.}~\bibnamefont {{Daido}}}, \bibinfo {author} {\bibfnamefont
  {J.}~\bibnamefont {{Fernandez}}}, \bibinfo {author} {\bibfnamefont
  {K.}~\bibnamefont {{Flippo}}}, \bibinfo {author} {\bibfnamefont
  {S.}~\bibnamefont {{Gaillard}}}, \bibinfo {author} {\bibfnamefont
  {C.}~\bibnamefont {{Gauthier}}}, \bibinfo {author} {\bibfnamefont
  {M.}~\bibnamefont {{Geissel}}}, \bibinfo {author} {\bibfnamefont
  {S.}~\bibnamefont {{Glenzer}}}, \bibinfo {author} {\bibfnamefont
  {G.}~\bibnamefont {{Gregori}}}, \bibinfo {author} {\bibfnamefont
  {M.}~\bibnamefont {{G{\"u}nther}}}, \bibinfo {author} {\bibfnamefont
  {K.}~\bibnamefont {{Harres}}}, \bibinfo {author} {\bibfnamefont
  {R.}~\bibnamefont {{Heathcote}}}, \bibinfo {author} {\bibfnamefont
  {A.}~\bibnamefont {{Kritcher}}}, \bibinfo {author} {\bibfnamefont
  {N.}~\bibnamefont {{Kugland}}}, \bibinfo {author} {\bibfnamefont
  {S.}~\bibnamefont {{Le Pape}}}, \bibinfo {author} {\bibfnamefont
  {B.}~\bibnamefont {{Li}}}, \bibinfo {author} {\bibfnamefont {M.}~\bibnamefont
  {{Makita}}}, \bibinfo {author} {\bibfnamefont {J.}~\bibnamefont {{Mithen}}},
  \bibinfo {author} {\bibfnamefont {C.}~\bibnamefont {{Niemann}}}, \bibinfo
  {author} {\bibfnamefont {F.}~\bibnamefont {{N{\"u}rnberg}}}, \bibinfo
  {author} {\bibfnamefont {D.}~\bibnamefont {{Offermann}}}, \bibinfo {author}
  {\bibfnamefont {A.}~\bibnamefont {{Otten}}}, \bibinfo {author} {\bibfnamefont
  {A.}~\bibnamefont {{Pelka}}}, \bibinfo {author} {\bibfnamefont
  {D.}~\bibnamefont {{Riley}}}, \bibinfo {author} {\bibfnamefont
  {G.}~\bibnamefont {{Schaumann}}}, \bibinfo {author} {\bibfnamefont
  {M.}~\bibnamefont {{Schollmeier}}}, \bibinfo {author} {\bibfnamefont
  {J.}~\bibnamefont {{Sch{\"u}trumpf}}}, \bibinfo {author} {\bibfnamefont
  {M.}~\bibnamefont {{Tampo}}}, \bibinfo {author} {\bibfnamefont
  {A.}~\bibnamefont {{Tauschwitz}}},\ and\ \bibinfo {author} {\bibfnamefont
  {A.}~\bibnamefont {{Tauschwitz}}},\ }\bibfield  {title} {\bibinfo {title}
  {{Proton acceleration experiments and warm dense matter research using high
  power lasers}},\ }\href {https://doi.org/10.1088/0741-3335/51/12/124039}
  {\bibfield  {journal} {\bibinfo  {journal} {Plasma Phys. Control. Fusion}\
  }\textbf {\bibinfo {volume} {51}},\ \bibinfo {eid} {124039} (\bibinfo {year}
  {2009})}\BibitemShut {NoStop}%
\bibitem [{\citenamefont {Man\ifmmode \check{c}\else
  \v{c}\fi{}i\ifmmode~\acute{c}\else \'{c}\fi{}}\ \emph
  {et~al.}(2010)\citenamefont {Man\ifmmode \check{c}\else
  \v{c}\fi{}i\ifmmode~\acute{c}\else \'{c}\fi{}}, \citenamefont {L\'evy},
  \citenamefont {Harmand}, \citenamefont {Nakatsutsumi}, \citenamefont
  {Antici}, \citenamefont {Audebert}, \citenamefont {Combis}, \citenamefont
  {Fourmaux}, \citenamefont {Mazevet}, \citenamefont {Peyrusse}, \citenamefont
  {Recoules}, \citenamefont {Renaudin}, \citenamefont {Robiche}, \citenamefont
  {Dorchies},\ and\ \citenamefont {Fuchs}}]{Mancic_2010}%
  \BibitemOpen
  \bibfield  {author} {\bibinfo {author} {\bibfnamefont {A.}~\bibnamefont
  {Man\ifmmode \check{c}\else \v{c}\fi{}i\ifmmode~\acute{c}\else \'{c}\fi{}}},
  \bibinfo {author} {\bibfnamefont {A.}~\bibnamefont {L\'evy}}, \bibinfo
  {author} {\bibfnamefont {M.}~\bibnamefont {Harmand}}, \bibinfo {author}
  {\bibfnamefont {M.}~\bibnamefont {Nakatsutsumi}}, \bibinfo {author}
  {\bibfnamefont {P.}~\bibnamefont {Antici}}, \bibinfo {author} {\bibfnamefont
  {P.}~\bibnamefont {Audebert}}, \bibinfo {author} {\bibfnamefont
  {P.}~\bibnamefont {Combis}}, \bibinfo {author} {\bibfnamefont
  {S.}~\bibnamefont {Fourmaux}}, \bibinfo {author} {\bibfnamefont
  {S.}~\bibnamefont {Mazevet}}, \bibinfo {author} {\bibfnamefont
  {O.}~\bibnamefont {Peyrusse}}, \bibinfo {author} {\bibfnamefont
  {V.}~\bibnamefont {Recoules}}, \bibinfo {author} {\bibfnamefont
  {P.}~\bibnamefont {Renaudin}}, \bibinfo {author} {\bibfnamefont
  {J.}~\bibnamefont {Robiche}}, \bibinfo {author} {\bibfnamefont
  {F.}~\bibnamefont {Dorchies}},\ and\ \bibinfo {author} {\bibfnamefont
  {J.}~\bibnamefont {Fuchs}},\ }\bibfield  {title} {\bibinfo {title}
  {Picosecond short-range disordering in isochorically heated aluminum at solid
  density},\ }\href {https://doi.org/10.1103/PhysRevLett.104.035002} {\bibfield
   {journal} {\bibinfo  {journal} {Phys. Rev. Lett.}\ }\textbf {\bibinfo
  {volume} {104}},\ \bibinfo {pages} {035002} (\bibinfo {year}
  {2010})}\BibitemShut {NoStop}%
\bibitem [{\citenamefont {Roth}\ \emph {et~al.}(2013)\citenamefont {Roth},
  \citenamefont {Jung}, \citenamefont {Falk}, \citenamefont {Guler},
  \citenamefont {Deppert}, \citenamefont {Devlin}, \citenamefont {Favalli},
  \citenamefont {Fernandez}, \citenamefont {Gautier}, \citenamefont {Geissel},
  \citenamefont {Haight}, \citenamefont {Hamilton}, \citenamefont {Hegelich},
  \citenamefont {Johnson}, \citenamefont {Merrill}, \citenamefont {Schaumann},
  \citenamefont {Schoenberg}, \citenamefont {Schollmeier}, \citenamefont
  {Shimada}, \citenamefont {Taddeucci}, \citenamefont {Tybo}, \citenamefont
  {Wagner}, \citenamefont {Wender}, \citenamefont {Wilde},\ and\ \citenamefont
  {Wurden}}]{Roth_2013}%
  \BibitemOpen
  \bibfield  {author} {\bibinfo {author} {\bibfnamefont {M.}~\bibnamefont
  {Roth}}, \bibinfo {author} {\bibfnamefont {D.}~\bibnamefont {Jung}}, \bibinfo
  {author} {\bibfnamefont {K.}~\bibnamefont {Falk}}, \bibinfo {author}
  {\bibfnamefont {N.}~\bibnamefont {Guler}}, \bibinfo {author} {\bibfnamefont
  {O.}~\bibnamefont {Deppert}}, \bibinfo {author} {\bibfnamefont
  {M.}~\bibnamefont {Devlin}}, \bibinfo {author} {\bibfnamefont
  {A.}~\bibnamefont {Favalli}}, \bibinfo {author} {\bibfnamefont
  {J.}~\bibnamefont {Fernandez}}, \bibinfo {author} {\bibfnamefont
  {D.}~\bibnamefont {Gautier}}, \bibinfo {author} {\bibfnamefont
  {M.}~\bibnamefont {Geissel}}, \bibinfo {author} {\bibfnamefont
  {R.}~\bibnamefont {Haight}}, \bibinfo {author} {\bibfnamefont {C.~E.}\
  \bibnamefont {Hamilton}}, \bibinfo {author} {\bibfnamefont {B.~M.}\
  \bibnamefont {Hegelich}}, \bibinfo {author} {\bibfnamefont {R.~P.}\
  \bibnamefont {Johnson}}, \bibinfo {author} {\bibfnamefont {F.}~\bibnamefont
  {Merrill}}, \bibinfo {author} {\bibfnamefont {G.}~\bibnamefont {Schaumann}},
  \bibinfo {author} {\bibfnamefont {K.}~\bibnamefont {Schoenberg}}, \bibinfo
  {author} {\bibfnamefont {M.}~\bibnamefont {Schollmeier}}, \bibinfo {author}
  {\bibfnamefont {T.}~\bibnamefont {Shimada}}, \bibinfo {author} {\bibfnamefont
  {T.}~\bibnamefont {Taddeucci}}, \bibinfo {author} {\bibfnamefont {J.~L.}\
  \bibnamefont {Tybo}}, \bibinfo {author} {\bibfnamefont {F.}~\bibnamefont
  {Wagner}}, \bibinfo {author} {\bibfnamefont {S.~A.}\ \bibnamefont {Wender}},
  \bibinfo {author} {\bibfnamefont {C.~H.}\ \bibnamefont {Wilde}},\ and\
  \bibinfo {author} {\bibfnamefont {G.~A.}\ \bibnamefont {Wurden}},\ }\bibfield
   {title} {\bibinfo {title} {Bright laser-driven neutron source based on the
  relativistic transparency of solids},\ }\href
  {https://doi.org/10.1103/PhysRevLett.110.044802} {\bibfield  {journal}
  {\bibinfo  {journal} {Phys. Rev. Lett.}\ }\textbf {\bibinfo {volume} {110}},\
  \bibinfo {pages} {044802} (\bibinfo {year} {2013})}\BibitemShut {NoStop}%
\bibitem [{\citenamefont {{Kleinschmidt}}\ \emph {et~al.}(2018)\citenamefont
  {{Kleinschmidt}}, \citenamefont {{Bagnoud}}, \citenamefont {{Deppert}},
  \citenamefont {{Favalli}}, \citenamefont {{Frydrych}}, \citenamefont
  {{Hornung}}, \citenamefont {{Jahn}}, \citenamefont {{Schaumann}},
  \citenamefont {{Tebartz}}, \citenamefont {{Wagner}}, \citenamefont
  {{Wurden}}, \citenamefont {{Zielbauer}},\ and\ \citenamefont
  {{Roth}}}]{Kleinschmidt_2018}%
  \BibitemOpen
  \bibfield  {author} {\bibinfo {author} {\bibfnamefont {A.}~\bibnamefont
  {{Kleinschmidt}}}, \bibinfo {author} {\bibfnamefont {V.}~\bibnamefont
  {{Bagnoud}}}, \bibinfo {author} {\bibfnamefont {O.}~\bibnamefont
  {{Deppert}}}, \bibinfo {author} {\bibfnamefont {A.}~\bibnamefont
  {{Favalli}}}, \bibinfo {author} {\bibfnamefont {S.}~\bibnamefont
  {{Frydrych}}}, \bibinfo {author} {\bibfnamefont {J.}~\bibnamefont
  {{Hornung}}}, \bibinfo {author} {\bibfnamefont {D.}~\bibnamefont {{Jahn}}},
  \bibinfo {author} {\bibfnamefont {G.}~\bibnamefont {{Schaumann}}}, \bibinfo
  {author} {\bibfnamefont {A.}~\bibnamefont {{Tebartz}}}, \bibinfo {author}
  {\bibfnamefont {F.}~\bibnamefont {{Wagner}}}, \bibinfo {author}
  {\bibfnamefont {G.}~\bibnamefont {{Wurden}}}, \bibinfo {author}
  {\bibfnamefont {B.}~\bibnamefont {{Zielbauer}}},\ and\ \bibinfo {author}
  {\bibfnamefont {M.}~\bibnamefont {{Roth}}},\ }\bibfield  {title} {\bibinfo
  {title} {{Intense, directed neutron beams from a laser-driven neutron source
  at PHELIX}},\ }\href {https://doi.org/10.1063/1.5006613} {\bibfield
  {journal} {\bibinfo  {journal} {Phys. Plasmas}\ }\textbf {\bibinfo {volume}
  {25}},\ \bibinfo {eid} {053101} (\bibinfo {year} {2018})}\BibitemShut
  {NoStop}%
\bibitem [{\citenamefont {{Horn{\'y}}}\ \emph {et~al.}(2022)\citenamefont
  {{Horn{\'y}}}, \citenamefont {{Chen}}, \citenamefont {{Davoine}},
  \citenamefont {{Lelasseux}}, \citenamefont {{Gremillet}},\ and\ \citenamefont
  {{Fuchs}}}]{Horny_2022}%
  \BibitemOpen
  \bibfield  {author} {\bibinfo {author} {\bibfnamefont {V.}~\bibnamefont
  {{Horn{\'y}}}}, \bibinfo {author} {\bibfnamefont {S.~N.}\ \bibnamefont
  {{Chen}}}, \bibinfo {author} {\bibfnamefont {X.}~\bibnamefont {{Davoine}}},
  \bibinfo {author} {\bibfnamefont {V.}~\bibnamefont {{Lelasseux}}}, \bibinfo
  {author} {\bibfnamefont {L.}~\bibnamefont {{Gremillet}}},\ and\ \bibinfo
  {author} {\bibfnamefont {J.}~\bibnamefont {{Fuchs}}},\ }\bibfield  {title}
  {\bibinfo {title} {{High-flux neutron generation by laser-accelerated ions
  from single- and double-layer targets}},\ }\href
  {https://doi.org/10.1038/s41598-022-24155-z} {\bibfield  {journal} {\bibinfo
  {journal} {Sci. Rep.}\ }\textbf {\bibinfo {volume} {12}},\ \bibinfo {eid}
  {19767} (\bibinfo {year} {2022})}\BibitemShut {NoStop}%
\bibitem [{\citenamefont {{Spencer}}\ \emph {et~al.}(2001)\citenamefont
  {{Spencer}}, \citenamefont {{Ledingham}}, \citenamefont {{Singhal}},
  \citenamefont {{McCanny}}, \citenamefont {{McKenna}}, \citenamefont
  {{Clark}}, \citenamefont {{Krushelnick}}, \citenamefont {{Zepf}},
  \citenamefont {{Beg}}, \citenamefont {{Tatarakis}}, \citenamefont {{Dangor}},
  \citenamefont {{Norreys}}, \citenamefont {{Clarke}}, \citenamefont
  {{Allott}},\ and\ \citenamefont {{Ross}}}]{Spencer_2001}%
  \BibitemOpen
  \bibfield  {author} {\bibinfo {author} {\bibfnamefont {I.}~\bibnamefont
  {{Spencer}}}, \bibinfo {author} {\bibfnamefont {K.~W.~D.}\ \bibnamefont
  {{Ledingham}}}, \bibinfo {author} {\bibfnamefont {R.~P.}\ \bibnamefont
  {{Singhal}}}, \bibinfo {author} {\bibfnamefont {T.}~\bibnamefont
  {{McCanny}}}, \bibinfo {author} {\bibfnamefont {P.}~\bibnamefont
  {{McKenna}}}, \bibinfo {author} {\bibfnamefont {E.~L.}\ \bibnamefont
  {{Clark}}}, \bibinfo {author} {\bibfnamefont {K.}~\bibnamefont
  {{Krushelnick}}}, \bibinfo {author} {\bibfnamefont {M.}~\bibnamefont
  {{Zepf}}}, \bibinfo {author} {\bibfnamefont {F.~N.}\ \bibnamefont {{Beg}}},
  \bibinfo {author} {\bibfnamefont {M.}~\bibnamefont {{Tatarakis}}}, \bibinfo
  {author} {\bibfnamefont {A.~E.}\ \bibnamefont {{Dangor}}}, \bibinfo {author}
  {\bibfnamefont {P.~A.}\ \bibnamefont {{Norreys}}}, \bibinfo {author}
  {\bibfnamefont {R.~J.}\ \bibnamefont {{Clarke}}}, \bibinfo {author}
  {\bibfnamefont {R.~M.}\ \bibnamefont {{Allott}}},\ and\ \bibinfo {author}
  {\bibfnamefont {I.~N.}\ \bibnamefont {{Ross}}},\ }\bibfield  {title}
  {\bibinfo {title} {{Laser generation of proton beams for the production of
  short-lived positron emitting radioisotopes}},\ }\href
  {https://doi.org/10.1016/S0168-583X(01)00771-6} {\bibfield  {journal}
  {\bibinfo  {journal} {Nucl. Instrum. Methods Phys. Res. B}\ }\textbf
  {\bibinfo {volume} {183}},\ \bibinfo {pages} {449} (\bibinfo {year}
  {2001})}\BibitemShut {NoStop}%
\bibitem [{\citenamefont {{Fritzler}}\ \emph {et~al.}(2003)\citenamefont
  {{Fritzler}}, \citenamefont {{Malka}}, \citenamefont {{Grillon}},
  \citenamefont {{Rousseau}}, \citenamefont {{Burgy}}, \citenamefont
  {{Lefebvre}}, \citenamefont {{d'Humi{\`e}res}}, \citenamefont {{McKenna}},\
  and\ \citenamefont {{Ledingham}}}]{Fritzler_2003}%
  \BibitemOpen
  \bibfield  {author} {\bibinfo {author} {\bibfnamefont {S.}~\bibnamefont
  {{Fritzler}}}, \bibinfo {author} {\bibfnamefont {V.}~\bibnamefont {{Malka}}},
  \bibinfo {author} {\bibfnamefont {G.}~\bibnamefont {{Grillon}}}, \bibinfo
  {author} {\bibfnamefont {J.~P.}\ \bibnamefont {{Rousseau}}}, \bibinfo
  {author} {\bibfnamefont {F.}~\bibnamefont {{Burgy}}}, \bibinfo {author}
  {\bibfnamefont {E.}~\bibnamefont {{Lefebvre}}}, \bibinfo {author}
  {\bibfnamefont {E.}~\bibnamefont {{d'Humi{\`e}res}}}, \bibinfo {author}
  {\bibfnamefont {P.}~\bibnamefont {{McKenna}}},\ and\ \bibinfo {author}
  {\bibfnamefont {K.~W.~D.}\ \bibnamefont {{Ledingham}}},\ }\bibfield  {title}
  {\bibinfo {title} {{Proton beams generated with high-intensity lasers:
  Applications to medical isotope production}},\ }\href
  {https://doi.org/10.1063/1.1616661} {\bibfield  {journal} {\bibinfo
  {journal} {Appl. Phys. Lett.}\ }\textbf {\bibinfo {volume} {83}},\ \bibinfo
  {eid} {3039} (\bibinfo {year} {2003})}\BibitemShut {NoStop}%
\bibitem [{\citenamefont {Ledingham}\ \emph {et~al.}(2014)\citenamefont
  {Ledingham}, \citenamefont {Bolton}, \citenamefont {Shikazono},\ and\
  \citenamefont {Ma}}]{Ledingham_2014}%
  \BibitemOpen
  \bibfield  {author} {\bibinfo {author} {\bibfnamefont {K.~W.~D.}\
  \bibnamefont {Ledingham}}, \bibinfo {author} {\bibfnamefont {P.~R.}\
  \bibnamefont {Bolton}}, \bibinfo {author} {\bibfnamefont {N.}~\bibnamefont
  {Shikazono}},\ and\ \bibinfo {author} {\bibfnamefont {C.-M.~C.}\ \bibnamefont
  {Ma}},\ }\bibfield  {title} {\bibinfo {title} {Towards laser driven hadron
  cancer radiotherapy: A review of progress},\ }\href
  {https://doi.org/10.3390/app4030402} {\bibfield  {journal} {\bibinfo
  {journal} {Appl. Sci.}\ }\textbf {\bibinfo {volume} {4}},\ \bibinfo {pages}
  {402} (\bibinfo {year} {2014})}\BibitemShut {NoStop}%
\bibitem [{\citenamefont {Cowan}\ \emph {et~al.}(2004)\citenamefont {Cowan},
  \citenamefont {Fuchs}, \citenamefont {Ruhl}, \citenamefont {Kemp},
  \citenamefont {Audebert}, \citenamefont {Roth}, \citenamefont {Stephens},
  \citenamefont {Barton}, \citenamefont {Blazevic}, \citenamefont {Brambrink},
  \citenamefont {Cobble}, \citenamefont {Fern\'andez}, \citenamefont
  {Gauthier}, \citenamefont {Geissel}, \citenamefont {Hegelich}, \citenamefont
  {Kaae}, \citenamefont {Karsch}, \citenamefont {Le~Sage}, \citenamefont
  {Letzring}, \citenamefont {Manclossi}, \citenamefont {Meyroneinc},
  \citenamefont {Newkirk}, \citenamefont {P\'epin},\ and\ \citenamefont
  {Renard-LeGalloudec}}]{Cowan_2004}%
  \BibitemOpen
  \bibfield  {author} {\bibinfo {author} {\bibfnamefont {T.~E.}\ \bibnamefont
  {Cowan}}, \bibinfo {author} {\bibfnamefont {J.}~\bibnamefont {Fuchs}},
  \bibinfo {author} {\bibfnamefont {H.}~\bibnamefont {Ruhl}}, \bibinfo {author}
  {\bibfnamefont {A.}~\bibnamefont {Kemp}}, \bibinfo {author} {\bibfnamefont
  {P.}~\bibnamefont {Audebert}}, \bibinfo {author} {\bibfnamefont
  {M.}~\bibnamefont {Roth}}, \bibinfo {author} {\bibfnamefont {R.}~\bibnamefont
  {Stephens}}, \bibinfo {author} {\bibfnamefont {I.}~\bibnamefont {Barton}},
  \bibinfo {author} {\bibfnamefont {A.}~\bibnamefont {Blazevic}}, \bibinfo
  {author} {\bibfnamefont {E.}~\bibnamefont {Brambrink}}, \bibinfo {author}
  {\bibfnamefont {J.}~\bibnamefont {Cobble}}, \bibinfo {author} {\bibfnamefont
  {J.}~\bibnamefont {Fern\'andez}}, \bibinfo {author} {\bibfnamefont {J.-C.}\
  \bibnamefont {Gauthier}}, \bibinfo {author} {\bibfnamefont {M.}~\bibnamefont
  {Geissel}}, \bibinfo {author} {\bibfnamefont {M.}~\bibnamefont {Hegelich}},
  \bibinfo {author} {\bibfnamefont {J.}~\bibnamefont {Kaae}}, \bibinfo {author}
  {\bibfnamefont {S.}~\bibnamefont {Karsch}}, \bibinfo {author} {\bibfnamefont
  {G.~P.}\ \bibnamefont {Le~Sage}}, \bibinfo {author} {\bibfnamefont
  {S.}~\bibnamefont {Letzring}}, \bibinfo {author} {\bibfnamefont
  {M.}~\bibnamefont {Manclossi}}, \bibinfo {author} {\bibfnamefont
  {S.}~\bibnamefont {Meyroneinc}}, \bibinfo {author} {\bibfnamefont
  {A.}~\bibnamefont {Newkirk}}, \bibinfo {author} {\bibfnamefont
  {H.}~\bibnamefont {P\'epin}},\ and\ \bibinfo {author} {\bibfnamefont
  {N.}~\bibnamefont {Renard-LeGalloudec}},\ }\bibfield  {title} {\bibinfo
  {title} {Ultralow emittance, multi-mev proton beams from a laser
  virtual-cathode plasma accelerator},\ }\href
  {https://doi.org/10.1103/PhysRevLett.92.204801} {\bibfield  {journal}
  {\bibinfo  {journal} {Phys. Rev. Lett.}\ }\textbf {\bibinfo {volume} {92}},\
  \bibinfo {pages} {204801} (\bibinfo {year} {2004})}\BibitemShut {NoStop}%
\bibitem [{\citenamefont {Chen}\ \emph {et~al.}(2019)\citenamefont {Chen},
  \citenamefont {Negoita}, \citenamefont {Spohr}, \citenamefont
  {d’Humi{\`e}res}, \citenamefont {Pomerantz},\ and\ \citenamefont
  {Fuchs}}]{Chen_2019}%
  \BibitemOpen
  \bibfield  {author} {\bibinfo {author} {\bibfnamefont {S.}~\bibnamefont
  {Chen}}, \bibinfo {author} {\bibfnamefont {F.}~\bibnamefont {Negoita}},
  \bibinfo {author} {\bibfnamefont {K.}~\bibnamefont {Spohr}}, \bibinfo
  {author} {\bibfnamefont {E.}~\bibnamefont {d’Humi{\`e}res}}, \bibinfo
  {author} {\bibfnamefont {I.}~\bibnamefont {Pomerantz}},\ and\ \bibinfo
  {author} {\bibfnamefont {J.}~\bibnamefont {Fuchs}},\ }\bibfield  {title}
  {\bibinfo {title} {Extreme brightness laser-based neutron pulses as a pathway
  for investigating nucleosynthesis in the laboratory},\ }\href
  {https://doi.org/10.1063/1.5081666} {\bibfield  {journal} {\bibinfo
  {journal} {Matter Radiat. Extremes}\ }\textbf {\bibinfo {volume} {4}},\
  \bibinfo {pages} {054402} (\bibinfo {year} {2019})}\BibitemShut {NoStop}%
\bibitem [{\citenamefont {{Horn{\'y}}}\ \emph {et~al.}(2023)\citenamefont
  {{Horn{\'y}}}, \citenamefont {{Chen}}, \citenamefont {{Davoine}},
  \citenamefont {{Gremillet}},\ and\ \citenamefont {{Fuchs}}}]{Horny_2023}%
  \BibitemOpen
  \bibfield  {author} {\bibinfo {author} {\bibfnamefont {V.}~\bibnamefont
  {{Horn{\'y}}}}, \bibinfo {author} {\bibfnamefont {S.~N.}\ \bibnamefont
  {{Chen}}}, \bibinfo {author} {\bibfnamefont {X.}~\bibnamefont {{Davoine}}},
  \bibinfo {author} {\bibfnamefont {L.}~\bibnamefont {{Gremillet}}},\ and\
  \bibinfo {author} {\bibfnamefont {J.}~\bibnamefont {{Fuchs}}},\ }\bibfield
  {title} {\bibinfo {title} {{Quantitative feasibility study of astrophysical
  rapid neutron capture demonstration using intense lasers}},\ }\href
  {https://doi.org/10.48550/arXiv.2304.05981} {\bibfield  {journal} {\bibinfo
  {journal} {arXiv e-prints}\ ,\ \bibinfo {eid} {arXiv:2304.05981}} (\bibinfo
  {year} {2023})}\BibitemShut {NoStop}%
\bibitem [{\citenamefont {{G{\"u}nther}}\ \emph {et~al.}(2022)\citenamefont
  {{G{\"u}nther}}, \citenamefont {{Rosmej}}, \citenamefont {{Tavana}},
  \citenamefont {{Gyrdymov}}, \citenamefont {{Skobliakov}}, \citenamefont
  {{Kantsyrev}}, \citenamefont {{Z{\"a}hter}}, \citenamefont {{Borisenko}},
  \citenamefont {{Pukhov}},\ and\ \citenamefont {{Andreev}}}]{Gunther_2022}%
  \BibitemOpen
  \bibfield  {author} {\bibinfo {author} {\bibfnamefont {M.~M.}\ \bibnamefont
  {{G{\"u}nther}}}, \bibinfo {author} {\bibfnamefont {O.~N.}\ \bibnamefont
  {{Rosmej}}}, \bibinfo {author} {\bibfnamefont {P.}~\bibnamefont {{Tavana}}},
  \bibinfo {author} {\bibfnamefont {M.}~\bibnamefont {{Gyrdymov}}}, \bibinfo
  {author} {\bibfnamefont {A.}~\bibnamefont {{Skobliakov}}}, \bibinfo {author}
  {\bibfnamefont {A.}~\bibnamefont {{Kantsyrev}}}, \bibinfo {author}
  {\bibfnamefont {S.}~\bibnamefont {{Z{\"a}hter}}}, \bibinfo {author}
  {\bibfnamefont {N.~G.}\ \bibnamefont {{Borisenko}}}, \bibinfo {author}
  {\bibfnamefont {A.}~\bibnamefont {{Pukhov}}},\ and\ \bibinfo {author}
  {\bibfnamefont {N.~E.}\ \bibnamefont {{Andreev}}},\ }\bibfield  {title}
  {\bibinfo {title} {{Forward-looking insights in laser-generated ultra-intense
  {\ensuremath{\gamma}}-ray and neutron sources for nuclear application and
  science}},\ }\href {https://doi.org/10.1038/s41467-021-27694-7} {\bibfield
  {journal} {\bibinfo  {journal} {Nat. Commun.}\ }\textbf {\bibinfo {volume}
  {13}},\ \bibinfo {eid} {170} (\bibinfo {year} {2022})}\BibitemShut {NoStop}%
\bibitem [{\citenamefont {{Iwamoto}}\ and\ \citenamefont
  {{Kodama}}(2020)}]{Iwamoto_2020}%
  \BibitemOpen
  \bibfield  {author} {\bibinfo {author} {\bibfnamefont {A.}~\bibnamefont
  {{Iwamoto}}}\ and\ \bibinfo {author} {\bibfnamefont {R.}~\bibnamefont
  {{Kodama}}},\ }\bibfield  {title} {\bibinfo {title} {{Conceptual design of a
  subcritical research reactor for inertial fusion energy with the J-EPoCH
  facility}},\ }\href {https://doi.org/10.1016/j.hedp.2020.100842} {\bibfield
  {journal} {\bibinfo  {journal} {High Energy Density Phys.}\ }\textbf
  {\bibinfo {volume} {36}},\ \bibinfo {eid} {100842} (\bibinfo {year}
  {2020})}\BibitemShut {NoStop}%
\bibitem [{\citenamefont {Burdonov}\ \emph {et~al.}(2021)\citenamefont
  {Burdonov}, \citenamefont {Fazzini}, \citenamefont {Lelasseux}, \citenamefont
  {Albrecht}, \citenamefont {Antici}, \citenamefont {Ayoul}, \citenamefont
  {Beluze}, \citenamefont {Cavanna}, \citenamefont {Ceccotti}, \citenamefont
  {Chabanis} \emph {et~al.}}]{Burdonov_2021}%
  \BibitemOpen
  \bibfield  {author} {\bibinfo {author} {\bibfnamefont {K.}~\bibnamefont
  {Burdonov}}, \bibinfo {author} {\bibfnamefont {A.}~\bibnamefont {Fazzini}},
  \bibinfo {author} {\bibfnamefont {V.}~\bibnamefont {Lelasseux}}, \bibinfo
  {author} {\bibfnamefont {J.}~\bibnamefont {Albrecht}}, \bibinfo {author}
  {\bibfnamefont {P.}~\bibnamefont {Antici}}, \bibinfo {author} {\bibfnamefont
  {Y.}~\bibnamefont {Ayoul}}, \bibinfo {author} {\bibfnamefont
  {A.}~\bibnamefont {Beluze}}, \bibinfo {author} {\bibfnamefont
  {D.}~\bibnamefont {Cavanna}}, \bibinfo {author} {\bibfnamefont
  {T.}~\bibnamefont {Ceccotti}}, \bibinfo {author} {\bibfnamefont
  {M.}~\bibnamefont {Chabanis}}, \emph {et~al.},\ }\bibfield  {title} {\bibinfo
  {title} {{Characterization and performance of the Apollon Short-Focal-Area
  facility following its commissioning at 1 PW level}},\ }\href
  {https://doi.org/10.1063/5.0065138} {\bibfield  {journal} {\bibinfo
  {journal} {Matter Radiat. Extremes}\ }\textbf {\bibinfo {volume} {6}},\
  \bibinfo {pages} {064402} (\bibinfo {year} {2021})}\BibitemShut {NoStop}%
\bibitem [{\citenamefont {Salvadori}\ \emph {et~al.}(2022)\citenamefont
  {Salvadori}, \citenamefont {Di~Giorgio}, \citenamefont {Cipriani},
  \citenamefont {Scisciò}, \citenamefont {Verona}, \citenamefont {Andreoli},
  \citenamefont {Cristofari}, \citenamefont {De~Angelis}, \citenamefont
  {Pillon}, \citenamefont {Andreev},\ and\ \citenamefont
  {et~al.}}]{Salvadori_2022}%
  \BibitemOpen
  \bibfield  {author} {\bibinfo {author} {\bibfnamefont {M.}~\bibnamefont
  {Salvadori}}, \bibinfo {author} {\bibfnamefont {G.}~\bibnamefont
  {Di~Giorgio}}, \bibinfo {author} {\bibfnamefont {M.}~\bibnamefont
  {Cipriani}}, \bibinfo {author} {\bibfnamefont {M.}~\bibnamefont {Scisciò}},
  \bibinfo {author} {\bibfnamefont {C.}~\bibnamefont {Verona}}, \bibinfo
  {author} {\bibfnamefont {P.~L.}\ \bibnamefont {Andreoli}}, \bibinfo {author}
  {\bibfnamefont {G.}~\bibnamefont {Cristofari}}, \bibinfo {author}
  {\bibfnamefont {R.}~\bibnamefont {De~Angelis}}, \bibinfo {author}
  {\bibfnamefont {M.}~\bibnamefont {Pillon}}, \bibinfo {author} {\bibfnamefont
  {N.~E.}\ \bibnamefont {Andreev}},\ and\ \bibinfo {author} {\bibnamefont
  {et~al.}},\ }\bibfield  {title} {\bibinfo {title} {{Time-of-flight
  methodologies with large-area diamond detectors for ion characterization in
  laser-driven experiments}},\ }\href {https://doi.org/10.1017/hpl.2021.59}
  {\bibfield  {journal} {\bibinfo  {journal} {High Power Laser Sci. Eng.}\
  }\textbf {\bibinfo {volume} {10}},\ \bibinfo {pages} {e6} (\bibinfo {year}
  {2022})}\BibitemShut {NoStop}%
\bibitem [{\citenamefont {Sung}\ \emph {et~al.}(2017)\citenamefont {Sung},
  \citenamefont {Lee}, \citenamefont {Yoo}, \citenamefont {Yoon}, \citenamefont
  {Lee}, \citenamefont {Yang}, \citenamefont {Son}, \citenamefont {Jang},
  \citenamefont {Lee},\ and\ \citenamefont {Nam}}]{Sung_2017}%
  \BibitemOpen
  \bibfield  {author} {\bibinfo {author} {\bibfnamefont {J.~H.}\ \bibnamefont
  {Sung}}, \bibinfo {author} {\bibfnamefont {H.~W.}\ \bibnamefont {Lee}},
  \bibinfo {author} {\bibfnamefont {J.~Y.}\ \bibnamefont {Yoo}}, \bibinfo
  {author} {\bibfnamefont {J.~W.}\ \bibnamefont {Yoon}}, \bibinfo {author}
  {\bibfnamefont {C.~W.}\ \bibnamefont {Lee}}, \bibinfo {author} {\bibfnamefont
  {J.~M.}\ \bibnamefont {Yang}}, \bibinfo {author} {\bibfnamefont {Y.~J.}\
  \bibnamefont {Son}}, \bibinfo {author} {\bibfnamefont {Y.~H.}\ \bibnamefont
  {Jang}}, \bibinfo {author} {\bibfnamefont {S.~K.}\ \bibnamefont {Lee}},\ and\
  \bibinfo {author} {\bibfnamefont {C.~H.}\ \bibnamefont {Nam}},\ }\bibfield
  {title} {\bibinfo {title} {{4.2~PW, 20~fs Ti:sapphire laser at 0.1 Hz}},\
  }\href {https://doi.org/10.1364/OL.42.002058} {\bibfield  {journal} {\bibinfo
   {journal} {Opt. Lett.}\ }\textbf {\bibinfo {volume} {42}},\ \bibinfo {pages}
  {2058} (\bibinfo {year} {2017})}\BibitemShut {NoStop}%
\bibitem [{\citenamefont {{Rus}}\ \emph {et~al.}(2017)\citenamefont {{Rus}},
  \citenamefont {{Bakule}}, \citenamefont {{Kramer}}, \citenamefont {{Naylon}},
  \citenamefont {{Thoma}}, \citenamefont {{Fibrich}}, \citenamefont {{Green}},
  \citenamefont {{Lagron}}, \citenamefont {{Antipenkov}}, \citenamefont
  {{Barton{\'\i}{\v{c}}ek}}, \citenamefont {{Batysta}}, \citenamefont
  {{Ba{\v{s}}e}}, \citenamefont {{Boge}}, \citenamefont {{Buck}}, \citenamefont
  {{Cupal}}, \citenamefont {{Drouin}}, \citenamefont {{Dur{\'a}k}},
  \citenamefont {{Himmel}}, \citenamefont {{Havl{\'\i}{\v{c}}ek}},
  \citenamefont {{Homer}}, \citenamefont {{Honsa}}, \citenamefont
  {{Hor{\'a}{\v{c}}ek}}, \citenamefont {{Hr{\'\i}bek}}, \citenamefont
  {{Hub{\'a}{\v{c}}ek}}, \citenamefont {{Hubka}}, \citenamefont
  {{Kalinchenko}}, \citenamefont {{Kasl}}, \citenamefont {{Indra}},
  \citenamefont {{Korous}}, \citenamefont {{Ko{\v{s}}elja}}, \citenamefont
  {{Koub{\'\i}kov{\'a}}}, \citenamefont {{Laub}}, \citenamefont {{Mazanec}},
  \citenamefont {{Meadows}}, \citenamefont {{Nov{\'a}k}}, \citenamefont
  {{Peceli}}, \citenamefont {{Polan}}, \citenamefont {{Snopek}}, \citenamefont
  {{{\v{S}}obr}}, \citenamefont {{Trojek}}, \citenamefont {{Tykalewicz}},
  \citenamefont {{Velpula}}, \citenamefont {{Verhagen}}, \citenamefont
  {{Vyhl{\'\i}dka}}, \citenamefont {{Weiss}}, \citenamefont {{Haefner}},
  \citenamefont {{Bayramian}}, \citenamefont {{Betts}}, \citenamefont
  {{Erlandson}}, \citenamefont {{Jarboe}}, \citenamefont {{Johnson}},
  \citenamefont {{Horner}}, \citenamefont {{Kim}}, \citenamefont {{Koh}},
  \citenamefont {{Marshall}}, \citenamefont {{Mason}}, \citenamefont
  {{Sistrunk}}, \citenamefont {{Smith}}, \citenamefont {{Spinka}},
  \citenamefont {{Stanley}}, \citenamefont {{Stolz}}, \citenamefont
  {{Suratwala}}, \citenamefont {{Telford}}, \citenamefont {{Ditmire}},
  \citenamefont {{Gaul}}, \citenamefont {{Donovan}}, \citenamefont
  {{Frederickson}}, \citenamefont {{Friedman}}, \citenamefont {{Hammond}},
  \citenamefont {{Hidinger}}, \citenamefont {{Ch{\'e}riaux}}, \citenamefont
  {{Jochmann}}, \citenamefont {{Kepler}}, \citenamefont {{Malato}},
  \citenamefont {{Martinez}}, \citenamefont {{Metzger}}, \citenamefont
  {{Schultze}}, \citenamefont {{Mason}}, \citenamefont {{Ertel}}, \citenamefont
  {{Lintern}}, \citenamefont {{Edwards}}, \citenamefont {{Hernandez-Gomez}},\
  and\ \citenamefont {{Collier}}}]{Rus_2017}%
  \BibitemOpen
  \bibfield  {author} {\bibinfo {author} {\bibfnamefont {B.}~\bibnamefont
  {{Rus}}}, \bibinfo {author} {\bibfnamefont {P.}~\bibnamefont {{Bakule}}},
  \bibinfo {author} {\bibfnamefont {D.}~\bibnamefont {{Kramer}}}, \bibinfo
  {author} {\bibfnamefont {J.}~\bibnamefont {{Naylon}}}, \bibinfo {author}
  {\bibfnamefont {J.}~\bibnamefont {{Thoma}}}, \bibinfo {author} {\bibfnamefont
  {M.}~\bibnamefont {{Fibrich}}}, \bibinfo {author} {\bibfnamefont {J.~T.}\
  \bibnamefont {{Green}}}, \bibinfo {author} {\bibfnamefont {J.~C.}\
  \bibnamefont {{Lagron}}}, \bibinfo {author} {\bibfnamefont {R.}~\bibnamefont
  {{Antipenkov}}}, \bibinfo {author} {\bibfnamefont {J.}~\bibnamefont
  {{Barton{\'\i}{\v{c}}ek}}}, \bibinfo {author} {\bibfnamefont
  {F.}~\bibnamefont {{Batysta}}}, \bibinfo {author} {\bibfnamefont
  {R.}~\bibnamefont {{Ba{\v{s}}e}}}, \bibinfo {author} {\bibfnamefont
  {R.}~\bibnamefont {{Boge}}}, \bibinfo {author} {\bibfnamefont
  {S.}~\bibnamefont {{Buck}}}, \bibinfo {author} {\bibfnamefont
  {J.}~\bibnamefont {{Cupal}}}, \bibinfo {author} {\bibfnamefont {M.~A.}\
  \bibnamefont {{Drouin}}}, \bibinfo {author} {\bibfnamefont {M.}~\bibnamefont
  {{Dur{\'a}k}}}, \bibinfo {author} {\bibfnamefont {B.}~\bibnamefont
  {{Himmel}}}, \bibinfo {author} {\bibfnamefont {T.}~\bibnamefont
  {{Havl{\'\i}{\v{c}}ek}}}, \bibinfo {author} {\bibfnamefont {P.}~\bibnamefont
  {{Homer}}}, \bibinfo {author} {\bibfnamefont {A.}~\bibnamefont {{Honsa}}},
  \bibinfo {author} {\bibfnamefont {M.}~\bibnamefont {{Hor{\'a}{\v{c}}ek}}},
  \bibinfo {author} {\bibfnamefont {P.}~\bibnamefont {{Hr{\'\i}bek}}}, \bibinfo
  {author} {\bibfnamefont {J.}~\bibnamefont {{Hub{\'a}{\v{c}}ek}}}, \bibinfo
  {author} {\bibfnamefont {Z.}~\bibnamefont {{Hubka}}}, \bibinfo {author}
  {\bibfnamefont {G.}~\bibnamefont {{Kalinchenko}}}, \bibinfo {author}
  {\bibfnamefont {K.}~\bibnamefont {{Kasl}}}, \bibinfo {author} {\bibfnamefont
  {L.}~\bibnamefont {{Indra}}}, \bibinfo {author} {\bibfnamefont
  {P.}~\bibnamefont {{Korous}}}, \bibinfo {author} {\bibfnamefont
  {M.}~\bibnamefont {{Ko{\v{s}}elja}}}, \bibinfo {author} {\bibfnamefont
  {L.}~\bibnamefont {{Koub{\'\i}kov{\'a}}}}, \bibinfo {author} {\bibfnamefont
  {M.}~\bibnamefont {{Laub}}}, \bibinfo {author} {\bibfnamefont
  {T.}~\bibnamefont {{Mazanec}}}, \bibinfo {author} {\bibfnamefont
  {A.}~\bibnamefont {{Meadows}}}, \bibinfo {author} {\bibfnamefont
  {J.}~\bibnamefont {{Nov{\'a}k}}}, \bibinfo {author} {\bibfnamefont
  {D.}~\bibnamefont {{Peceli}}}, \bibinfo {author} {\bibfnamefont
  {J.}~\bibnamefont {{Polan}}}, \bibinfo {author} {\bibfnamefont
  {D.}~\bibnamefont {{Snopek}}}, \bibinfo {author} {\bibfnamefont
  {V.}~\bibnamefont {{{\v{S}}obr}}}, \bibinfo {author} {\bibfnamefont
  {P.}~\bibnamefont {{Trojek}}}, \bibinfo {author} {\bibfnamefont
  {B.}~\bibnamefont {{Tykalewicz}}}, \bibinfo {author} {\bibfnamefont
  {P.}~\bibnamefont {{Velpula}}}, \bibinfo {author} {\bibfnamefont
  {E.}~\bibnamefont {{Verhagen}}}, \bibinfo {author} {\bibfnamefont
  {{\r{A}}.}~\bibnamefont {{Vyhl{\'\i}dka}}}, \bibinfo {author} {\bibfnamefont
  {J.}~\bibnamefont {{Weiss}}}, \bibinfo {author} {\bibfnamefont
  {C.}~\bibnamefont {{Haefner}}}, \bibinfo {author} {\bibfnamefont
  {A.}~\bibnamefont {{Bayramian}}}, \bibinfo {author} {\bibfnamefont
  {S.}~\bibnamefont {{Betts}}}, \bibinfo {author} {\bibfnamefont
  {A.}~\bibnamefont {{Erlandson}}}, \bibinfo {author} {\bibfnamefont
  {J.}~\bibnamefont {{Jarboe}}}, \bibinfo {author} {\bibfnamefont
  {G.}~\bibnamefont {{Johnson}}}, \bibinfo {author} {\bibfnamefont
  {J.}~\bibnamefont {{Horner}}}, \bibinfo {author} {\bibfnamefont
  {D.}~\bibnamefont {{Kim}}}, \bibinfo {author} {\bibfnamefont
  {E.}~\bibnamefont {{Koh}}}, \bibinfo {author} {\bibfnamefont
  {C.}~\bibnamefont {{Marshall}}}, \bibinfo {author} {\bibfnamefont
  {D.}~\bibnamefont {{Mason}}}, \bibinfo {author} {\bibfnamefont
  {E.}~\bibnamefont {{Sistrunk}}}, \bibinfo {author} {\bibfnamefont
  {D.}~\bibnamefont {{Smith}}}, \bibinfo {author} {\bibfnamefont
  {T.}~\bibnamefont {{Spinka}}}, \bibinfo {author} {\bibfnamefont
  {J.}~\bibnamefont {{Stanley}}}, \bibinfo {author} {\bibfnamefont
  {C.}~\bibnamefont {{Stolz}}}, \bibinfo {author} {\bibfnamefont
  {T.}~\bibnamefont {{Suratwala}}}, \bibinfo {author} {\bibfnamefont
  {S.}~\bibnamefont {{Telford}}}, \bibinfo {author} {\bibfnamefont
  {T.}~\bibnamefont {{Ditmire}}}, \bibinfo {author} {\bibfnamefont
  {E.}~\bibnamefont {{Gaul}}}, \bibinfo {author} {\bibfnamefont
  {M.}~\bibnamefont {{Donovan}}}, \bibinfo {author} {\bibfnamefont
  {C.}~\bibnamefont {{Frederickson}}}, \bibinfo {author} {\bibfnamefont
  {G.}~\bibnamefont {{Friedman}}}, \bibinfo {author} {\bibfnamefont
  {D.}~\bibnamefont {{Hammond}}}, \bibinfo {author} {\bibfnamefont
  {D.}~\bibnamefont {{Hidinger}}}, \bibinfo {author} {\bibfnamefont
  {G.}~\bibnamefont {{Ch{\'e}riaux}}}, \bibinfo {author} {\bibfnamefont
  {A.}~\bibnamefont {{Jochmann}}}, \bibinfo {author} {\bibfnamefont
  {M.}~\bibnamefont {{Kepler}}}, \bibinfo {author} {\bibfnamefont
  {C.}~\bibnamefont {{Malato}}}, \bibinfo {author} {\bibfnamefont
  {M.}~\bibnamefont {{Martinez}}}, \bibinfo {author} {\bibfnamefont
  {T.}~\bibnamefont {{Metzger}}}, \bibinfo {author} {\bibfnamefont
  {M.}~\bibnamefont {{Schultze}}}, \bibinfo {author} {\bibfnamefont
  {P.}~\bibnamefont {{Mason}}}, \bibinfo {author} {\bibfnamefont
  {K.}~\bibnamefont {{Ertel}}}, \bibinfo {author} {\bibfnamefont
  {A.}~\bibnamefont {{Lintern}}}, \bibinfo {author} {\bibfnamefont
  {C.}~\bibnamefont {{Edwards}}}, \bibinfo {author} {\bibfnamefont
  {C.}~\bibnamefont {{Hernandez-Gomez}}},\ and\ \bibinfo {author}
  {\bibfnamefont {J.}~\bibnamefont {{Collier}}},\ }\bibfield  {title} {\bibinfo
  {title} {{ELI-Beamlines: progress in development of next generation
  short-pulse laser systems}},\ }in\ \href {https://doi.org/10.1117/12.2269818}
  {\emph {\bibinfo {booktitle} {Society of Photo-Optical Instrumentation
  Engineers (SPIE) Conference Series}}},\ \bibinfo {series} {Society of
  Photo-Optical Instrumentation Engineers (SPIE) Conference Series}, Vol.\
  \bibinfo {volume} {10241},\ \bibinfo {editor} {edited by\ \bibinfo {editor}
  {\bibfnamefont {G.}~\bibnamefont {{Korn}}}\ and\ \bibinfo {editor}
  {\bibfnamefont {L.~O.}\ \bibnamefont {{Silva}}}}\ (\bibinfo {year} {2017})\
  p.\ \bibinfo {pages} {102410J}\BibitemShut {NoStop}%
\bibitem [{\citenamefont {{Gales}}\ \emph {et~al.}(2018)\citenamefont
  {{Gales}}, \citenamefont {{Tanaka}}, \citenamefont {{Balabanski}},
  \citenamefont {{Negoita}}, \citenamefont {{Stutman}}, \citenamefont
  {{Tesileanu}}, \citenamefont {{Ur}}, \citenamefont {{Ursescu}}, \citenamefont
  {{Andrei}}, \citenamefont {{Ataman}}, \citenamefont {{Cernaianu}},
  \citenamefont {{D'Alessi}}, \citenamefont {{Dancus}}, \citenamefont
  {{Diaconescu}}, \citenamefont {{Djourelov}}, \citenamefont {{Filipescu}},
  \citenamefont {{Ghenuche}}, \citenamefont {{Ghita}}, \citenamefont {{Matei}},
  \citenamefont {{Seto}}, \citenamefont {{Zeng}},\ and\ \citenamefont
  {{Zamfir}}}]{Gales_2018}%
  \BibitemOpen
  \bibfield  {author} {\bibinfo {author} {\bibfnamefont {S.}~\bibnamefont
  {{Gales}}}, \bibinfo {author} {\bibfnamefont {K.~A.}\ \bibnamefont
  {{Tanaka}}}, \bibinfo {author} {\bibfnamefont {D.~L.}\ \bibnamefont
  {{Balabanski}}}, \bibinfo {author} {\bibfnamefont {F.}~\bibnamefont
  {{Negoita}}}, \bibinfo {author} {\bibfnamefont {D.}~\bibnamefont
  {{Stutman}}}, \bibinfo {author} {\bibfnamefont {O.}~\bibnamefont
  {{Tesileanu}}}, \bibinfo {author} {\bibfnamefont {C.~A.}\ \bibnamefont
  {{Ur}}}, \bibinfo {author} {\bibfnamefont {D.}~\bibnamefont {{Ursescu}}},
  \bibinfo {author} {\bibfnamefont {I.}~\bibnamefont {{Andrei}}}, \bibinfo
  {author} {\bibfnamefont {S.}~\bibnamefont {{Ataman}}}, \bibinfo {author}
  {\bibfnamefont {M.~O.}\ \bibnamefont {{Cernaianu}}}, \bibinfo {author}
  {\bibfnamefont {L.}~\bibnamefont {{D'Alessi}}}, \bibinfo {author}
  {\bibfnamefont {I.}~\bibnamefont {{Dancus}}}, \bibinfo {author}
  {\bibfnamefont {B.}~\bibnamefont {{Diaconescu}}}, \bibinfo {author}
  {\bibfnamefont {N.}~\bibnamefont {{Djourelov}}}, \bibinfo {author}
  {\bibfnamefont {D.}~\bibnamefont {{Filipescu}}}, \bibinfo {author}
  {\bibfnamefont {P.}~\bibnamefont {{Ghenuche}}}, \bibinfo {author}
  {\bibfnamefont {D.~G.}\ \bibnamefont {{Ghita}}}, \bibinfo {author}
  {\bibfnamefont {C.}~\bibnamefont {{Matei}}}, \bibinfo {author} {\bibfnamefont
  {K.}~\bibnamefont {{Seto}}}, \bibinfo {author} {\bibfnamefont
  {M.}~\bibnamefont {{Zeng}}},\ and\ \bibinfo {author} {\bibfnamefont {N.~V.}\
  \bibnamefont {{Zamfir}}},\ }\bibfield  {title} {\bibinfo {title} {{The
  extreme light infrastructure{\textemdash}nuclear physics (ELI-NP) facility:
  new horizons in physics with 10 PW ultra-intense lasers and 20 MeV brilliant
  gamma beams}},\ }\href {https://doi.org/10.1088/1361-6633/aacfe8} {\bibfield
  {journal} {\bibinfo  {journal} {Rep. Prog. Phys.}\ }\textbf {\bibinfo
  {volume} {81}},\ \bibinfo {eid} {094301} (\bibinfo {year}
  {2018})}\BibitemShut {NoStop}%
\bibitem [{\citenamefont {Hakimi}\ \emph {et~al.}(2022)\citenamefont {Hakimi},
  \citenamefont {Obst-Huebl}, \citenamefont {Huebl}, \citenamefont {Nakamura},
  \citenamefont {Bulanov}, \citenamefont {Steinke}, \citenamefont {Leemans},
  \citenamefont {Kober}, \citenamefont {Ostermayr}, \citenamefont {Schenkel}
  \emph {et~al.}}]{Hakimi_2022}%
  \BibitemOpen
  \bibfield  {author} {\bibinfo {author} {\bibfnamefont {S.}~\bibnamefont
  {Hakimi}}, \bibinfo {author} {\bibfnamefont {L.}~\bibnamefont {Obst-Huebl}},
  \bibinfo {author} {\bibfnamefont {A.}~\bibnamefont {Huebl}}, \bibinfo
  {author} {\bibfnamefont {K.}~\bibnamefont {Nakamura}}, \bibinfo {author}
  {\bibfnamefont {S.~S.}\ \bibnamefont {Bulanov}}, \bibinfo {author}
  {\bibfnamefont {S.}~\bibnamefont {Steinke}}, \bibinfo {author} {\bibfnamefont
  {W.~P.}\ \bibnamefont {Leemans}}, \bibinfo {author} {\bibfnamefont
  {Z.}~\bibnamefont {Kober}}, \bibinfo {author} {\bibfnamefont {T.~M.}\
  \bibnamefont {Ostermayr}}, \bibinfo {author} {\bibfnamefont {T.}~\bibnamefont
  {Schenkel}}, \emph {et~al.},\ }\bibfield  {title} {\bibinfo {title}
  {Laser--solid interaction studies enabled by the new capabilities of the ip2
  bella pw beamline},\ }\href {https://doi.org/10.1063/5.0089331} {\bibfield
  {journal} {\bibinfo  {journal} {Phys. Plasmas}\ }\textbf {\bibinfo {volume}
  {29}},\ \bibinfo {pages} {083102} (\bibinfo {year} {2022})}\BibitemShut
  {NoStop}%
\bibitem [{\citenamefont {{Maksimchuk}}\ \emph {et~al.}(2021)\citenamefont
  {{Maksimchuk}}, \citenamefont {{Nees}}, \citenamefont {{Kalinchenko}},
  \citenamefont {{Hou}}, \citenamefont {{Ma}}, \citenamefont {{McKelvey}},
  \citenamefont {{Shi}}, \citenamefont {{Campbell}}, \citenamefont {{Antoine}},
  \citenamefont {{Balcazar}}, \citenamefont {{Cardarelli}}, \citenamefont
  {{Ernst}}, \citenamefont {{Fitzgarrald}}, \citenamefont {{Graham}},
  \citenamefont {{Qian}}, \citenamefont {{Jovanovic}}, \citenamefont
  {{Kuranz}}, \citenamefont {{Thomas}}, \citenamefont {{Willingale}},\ and\
  \citenamefont {{Krushelnick}}}]{Maksimchuk_2021}%
  \BibitemOpen
  \bibfield  {author} {\bibinfo {author} {\bibfnamefont {A.}~\bibnamefont
  {{Maksimchuk}}}, \bibinfo {author} {\bibfnamefont {J.}~\bibnamefont
  {{Nees}}}, \bibinfo {author} {\bibfnamefont {G.}~\bibnamefont
  {{Kalinchenko}}}, \bibinfo {author} {\bibfnamefont {B.}~\bibnamefont
  {{Hou}}}, \bibinfo {author} {\bibfnamefont {Y.}~\bibnamefont {{Ma}}},
  \bibinfo {author} {\bibfnamefont {A.}~\bibnamefont {{McKelvey}}}, \bibinfo
  {author} {\bibfnamefont {T.}~\bibnamefont {{Shi}}}, \bibinfo {author}
  {\bibfnamefont {P.}~\bibnamefont {{Campbell}}}, \bibinfo {author}
  {\bibfnamefont {A.}~\bibnamefont {{Antoine}}}, \bibinfo {author}
  {\bibfnamefont {M.}~\bibnamefont {{Balcazar}}}, \bibinfo {author}
  {\bibfnamefont {J.}~\bibnamefont {{Cardarelli}}}, \bibinfo {author}
  {\bibfnamefont {N.}~\bibnamefont {{Ernst}}}, \bibinfo {author} {\bibfnamefont
  {R.}~\bibnamefont {{Fitzgarrald}}}, \bibinfo {author} {\bibfnamefont
  {C.}~\bibnamefont {{Graham}}}, \bibinfo {author} {\bibfnamefont
  {Q.}~\bibnamefont {{Qian}}}, \bibinfo {author} {\bibfnamefont
  {I.}~\bibnamefont {{Jovanovic}}}, \bibinfo {author} {\bibfnamefont
  {C.}~\bibnamefont {{Kuranz}}}, \bibinfo {author} {\bibfnamefont
  {A.}~\bibnamefont {{Thomas}}}, \bibinfo {author} {\bibfnamefont
  {L.}~\bibnamefont {{Willingale}}},\ and\ \bibinfo {author} {\bibfnamefont
  {K.}~\bibnamefont {{Krushelnick}}},\ }\bibfield  {title} {\bibinfo {title}
  {{Status report on the construction of Zettawatt-Equivalent Ultrashort pulse
  laser System (ZEUS) at the University of Michigan}},\ }in\ \href@noop {}
  {\emph {\bibinfo {booktitle} {APS Division of Plasma Physics Meeting
  Abstracts}}},\ \bibinfo {series} {APS Meeting Abstracts}, Vol.\ \bibinfo
  {volume} {2021}\ (\bibinfo {year} {2021})\ p.\ \bibinfo {pages}
  {BP11.065}\BibitemShut {NoStop}%
\bibitem [{\citenamefont {Prencipe}\ \emph {et~al.}(2017)\citenamefont
  {Prencipe}, \citenamefont {Fuchs}, \citenamefont {Pascarelli}, \citenamefont
  {Schumacher}, \citenamefont {Stephens}, \citenamefont {Alexander},
  \citenamefont {Briggs}, \citenamefont {Büscher}, \citenamefont {Cernaianu},
  \citenamefont {Choukourov},\ and\ \citenamefont {et~al.}}]{Prencipe_2017}%
  \BibitemOpen
  \bibfield  {author} {\bibinfo {author} {\bibfnamefont {I.}~\bibnamefont
  {Prencipe}}, \bibinfo {author} {\bibfnamefont {J.}~\bibnamefont {Fuchs}},
  \bibinfo {author} {\bibfnamefont {S.}~\bibnamefont {Pascarelli}}, \bibinfo
  {author} {\bibfnamefont {D.~W.}\ \bibnamefont {Schumacher}}, \bibinfo
  {author} {\bibfnamefont {R.~B.}\ \bibnamefont {Stephens}}, \bibinfo {author}
  {\bibfnamefont {N.~B.}\ \bibnamefont {Alexander}}, \bibinfo {author}
  {\bibfnamefont {R.}~\bibnamefont {Briggs}}, \bibinfo {author} {\bibfnamefont
  {M.}~\bibnamefont {Büscher}}, \bibinfo {author} {\bibfnamefont {M.~O.}\
  \bibnamefont {Cernaianu}}, \bibinfo {author} {\bibfnamefont {A.}~\bibnamefont
  {Choukourov}},\ and\ \bibinfo {author} {\bibnamefont {et~al.}},\ }\bibfield
  {title} {\bibinfo {title} {Targets for high repetition rate laser facilities:
  needs, challenges and perspectives},\ }\href
  {https://doi.org/10.1017/hpl.2017.18} {\bibfield  {journal} {\bibinfo
  {journal} {High Power Laser Sci. Eng.}\ }\textbf {\bibinfo {volume} {5}},\
  \bibinfo {pages} {e17} (\bibinfo {year} {2017})}\BibitemShut {NoStop}%
\bibitem [{\citenamefont {{Ehret}}\ \emph {et~al.}(2020)\citenamefont
  {{Ehret}}, \citenamefont {{Salgado-Lopez}}, \citenamefont
  {{Ospina-Bohorquez}}, \citenamefont {{Perez-Hernandez}}, \citenamefont
  {{Huault}}, \citenamefont {{de Marco}}, \citenamefont {{Apinaniz}},
  \citenamefont {{Hannachi}}, \citenamefont {{De Luis}}, \citenamefont
  {{Hernandez Toro}}, \citenamefont {{Arana}}, \citenamefont {{Mendez}},
  \citenamefont {{Varela}}, \citenamefont {{Debayle}}, \citenamefont
  {{Gremillet}}, \citenamefont {{Nguyen-Bui}}, \citenamefont {{Olivier}},
  \citenamefont {{Revet}}, \citenamefont {{Bukharskii}}, \citenamefont
  {{Larreur}}, \citenamefont {{Caron}}, \citenamefont {{Vlachos}},
  \citenamefont {{Ceccotti}}, \citenamefont {{Raffestin}}, \citenamefont
  {{Nicolai}}, \citenamefont {{Feugeas}}, \citenamefont {{Roth}}, \citenamefont
  {{Vaisseau}}, \citenamefont {{Gatti}}, \citenamefont {{Volpe}},\ and\
  \citenamefont {{Santos}}}]{Ehret_2020}%
  \BibitemOpen
  \bibfield  {author} {\bibinfo {author} {\bibfnamefont {M.}~\bibnamefont
  {{Ehret}}}, \bibinfo {author} {\bibfnamefont {C.}~\bibnamefont
  {{Salgado-Lopez}}}, \bibinfo {author} {\bibfnamefont {V.}~\bibnamefont
  {{Ospina-Bohorquez}}}, \bibinfo {author} {\bibfnamefont {J.~A.}\ \bibnamefont
  {{Perez-Hernandez}}}, \bibinfo {author} {\bibfnamefont {M.}~\bibnamefont
  {{Huault}}}, \bibinfo {author} {\bibfnamefont {M.}~\bibnamefont {{de
  Marco}}}, \bibinfo {author} {\bibfnamefont {J.~I.}\ \bibnamefont
  {{Apinaniz}}}, \bibinfo {author} {\bibfnamefont {F.}~\bibnamefont
  {{Hannachi}}}, \bibinfo {author} {\bibfnamefont {D.}~\bibnamefont {{De
  Luis}}}, \bibinfo {author} {\bibfnamefont {J.}~\bibnamefont {{Hernandez
  Toro}}}, \bibinfo {author} {\bibfnamefont {D.}~\bibnamefont {{Arana}}},
  \bibinfo {author} {\bibfnamefont {C.}~\bibnamefont {{Mendez}}}, \bibinfo
  {author} {\bibfnamefont {O.}~\bibnamefont {{Varela}}}, \bibinfo {author}
  {\bibfnamefont {A.}~\bibnamefont {{Debayle}}}, \bibinfo {author}
  {\bibfnamefont {L.}~\bibnamefont {{Gremillet}}}, \bibinfo {author}
  {\bibfnamefont {T.~H.}\ \bibnamefont {{Nguyen-Bui}}}, \bibinfo {author}
  {\bibfnamefont {E.}~\bibnamefont {{Olivier}}}, \bibinfo {author}
  {\bibfnamefont {G.}~\bibnamefont {{Revet}}}, \bibinfo {author} {\bibfnamefont
  {N.~D.}\ \bibnamefont {{Bukharskii}}}, \bibinfo {author} {\bibfnamefont
  {H.}~\bibnamefont {{Larreur}}}, \bibinfo {author} {\bibfnamefont
  {J.}~\bibnamefont {{Caron}}}, \bibinfo {author} {\bibfnamefont
  {C.}~\bibnamefont {{Vlachos}}}, \bibinfo {author} {\bibfnamefont
  {T.}~\bibnamefont {{Ceccotti}}}, \bibinfo {author} {\bibfnamefont
  {D.}~\bibnamefont {{Raffestin}}}, \bibinfo {author} {\bibfnamefont
  {P.}~\bibnamefont {{Nicolai}}}, \bibinfo {author} {\bibfnamefont {J.~L.}\
  \bibnamefont {{Feugeas}}}, \bibinfo {author} {\bibfnamefont {M.}~\bibnamefont
  {{Roth}}}, \bibinfo {author} {\bibfnamefont {X.}~\bibnamefont {{Vaisseau}}},
  \bibinfo {author} {\bibfnamefont {G.}~\bibnamefont {{Gatti}}}, \bibinfo
  {author} {\bibfnamefont {L.}~\bibnamefont {{Volpe}}},\ and\ \bibinfo {author}
  {\bibfnamefont {J.~J.}\ \bibnamefont {{Santos}}},\ }\bibfield  {title}
  {\bibinfo {title} {{Ion acceleration by an ultrashort laser pulse interacting
  with a near-critical-density gas jet}},\ }\href@noop {} {\bibfield  {journal}
  {\bibinfo  {journal} {arXiv e-prints}\ ,\ \bibinfo {eid} {arXiv:2012.09455}}
  (\bibinfo {year} {2020})},\ \Eprint {https://arxiv.org/abs/2012.09455}
  {arXiv:2012.09455} \BibitemShut {NoStop}%
\bibitem [{\citenamefont {{Bradford \textit{et al.,}}}()}]{Bradford_2023}%
  \BibitemOpen
  \bibfield  {author} {\bibinfo {author} {\bibfnamefont {P.~W.}\ \bibnamefont
  {{Bradford \textit{et al.,}}}},\ }\bibfield  {title} {\bibinfo {title}
  {{Laser interactions with gas jets: EMP emission and nozzle damage}},\
  }\href@noop {} {\bibinfo  {journal} {In preparation}\ }\BibitemShut {NoStop}%
\bibitem [{\citenamefont {{Krushelnick}}\ \emph {et~al.}(1999)\citenamefont
  {{Krushelnick}}, \citenamefont {{Clark}}, \citenamefont {{Najmudin}},
  \citenamefont {{Salvati}}, \citenamefont {{Santala}}, \citenamefont
  {{Tatarakis}}, \citenamefont {{Dangor}}, \citenamefont {{Malka}},
  \citenamefont {{Neely}}, \citenamefont {{Allott}},\ and\ \citenamefont
  {{Danson}}}]{Krushelnick_1999}%
  \BibitemOpen
\bibfield  {journal} {  }\bibfield  {author} {\bibinfo {author} {\bibfnamefont
  {K.}~\bibnamefont {{Krushelnick}}}, \bibinfo {author} {\bibfnamefont {E.~L.}\
  \bibnamefont {{Clark}}}, \bibinfo {author} {\bibfnamefont {Z.}~\bibnamefont
  {{Najmudin}}}, \bibinfo {author} {\bibfnamefont {M.}~\bibnamefont
  {{Salvati}}}, \bibinfo {author} {\bibfnamefont {M.~I.~K.}\ \bibnamefont
  {{Santala}}}, \bibinfo {author} {\bibfnamefont {M.}~\bibnamefont
  {{Tatarakis}}}, \bibinfo {author} {\bibfnamefont {A.~E.}\ \bibnamefont
  {{Dangor}}}, \bibinfo {author} {\bibfnamefont {V.}~\bibnamefont {{Malka}}},
  \bibinfo {author} {\bibfnamefont {D.}~\bibnamefont {{Neely}}}, \bibinfo
  {author} {\bibfnamefont {R.}~\bibnamefont {{Allott}}},\ and\ \bibinfo
  {author} {\bibfnamefont {C.}~\bibnamefont {{Danson}}},\ }\bibfield  {title}
  {\bibinfo {title} {{Multi-MeV Ion Production from High-Intensity Laser
  Interactions with Underdense Plasmas}},\ }\href
  {https://doi.org/10.1103/PhysRevLett.83.737} {\bibfield  {journal} {\bibinfo
  {journal} {\prl}\ }\textbf {\bibinfo {volume} {83}},\ \bibinfo {pages} {737}
  (\bibinfo {year} {1999})}\BibitemShut {NoStop}%
\bibitem [{\citenamefont {Sentoku}\ \emph {et~al.}(2000)\citenamefont
  {Sentoku}, \citenamefont {Liseikina}, \citenamefont {Esirkepov},
  \citenamefont {Califano}, \citenamefont {Naumova}, \citenamefont {Ueshima},
  \citenamefont {Vshivkov}, \citenamefont {Kato}, \citenamefont {Mima},
  \citenamefont {Nishihara}, \citenamefont {Pegoraro},\ and\ \citenamefont
  {Bulanov}}]{Sentoku_2000}%
  \BibitemOpen
  \bibfield  {author} {\bibinfo {author} {\bibfnamefont {Y.}~\bibnamefont
  {Sentoku}}, \bibinfo {author} {\bibfnamefont {T.~V.}\ \bibnamefont
  {Liseikina}}, \bibinfo {author} {\bibfnamefont {T.~Z.}\ \bibnamefont
  {Esirkepov}}, \bibinfo {author} {\bibfnamefont {F.}~\bibnamefont {Califano}},
  \bibinfo {author} {\bibfnamefont {N.~M.}\ \bibnamefont {Naumova}}, \bibinfo
  {author} {\bibfnamefont {Y.}~\bibnamefont {Ueshima}}, \bibinfo {author}
  {\bibfnamefont {V.~A.}\ \bibnamefont {Vshivkov}}, \bibinfo {author}
  {\bibfnamefont {Y.}~\bibnamefont {Kato}}, \bibinfo {author} {\bibfnamefont
  {K.}~\bibnamefont {Mima}}, \bibinfo {author} {\bibfnamefont {K.}~\bibnamefont
  {Nishihara}}, \bibinfo {author} {\bibfnamefont {F.}~\bibnamefont
  {Pegoraro}},\ and\ \bibinfo {author} {\bibfnamefont {S.~V.}\ \bibnamefont
  {Bulanov}},\ }\bibfield  {title} {\bibinfo {title} {High density collimated
  beams of relativistic ions produced by petawatt laser pulses in plasmas},\
  }\href {https://doi.org/10.1103/PhysRevE.62.7271} {\bibfield  {journal}
  {\bibinfo  {journal} {Phys. Rev. E}\ }\textbf {\bibinfo {volume} {62}},\
  \bibinfo {pages} {7271} (\bibinfo {year} {2000})}\BibitemShut {NoStop}%
\bibitem [{\citenamefont {{Chen}}\ \emph {et~al.}(2017)\citenamefont {{Chen}},
  \citenamefont {{Vranic}}, \citenamefont {{Gangolf}}, \citenamefont
  {{Boella}}, \citenamefont {{Antici}}, \citenamefont {{Bailly-Grandvaux}},
  \citenamefont {{Loiseau}}, \citenamefont {{P{\'e}pin}}, \citenamefont
  {{Revet}}, \citenamefont {{Santos}}, \citenamefont {{Schroer}}, \citenamefont
  {{Starodubtsev}}, \citenamefont {{Willi}}, \citenamefont {{Silva}},
  \citenamefont {{d'Humi{\`e}res}},\ and\ \citenamefont {{Fuchs}}}]{Chen_2017}%
  \BibitemOpen
  \bibfield  {author} {\bibinfo {author} {\bibfnamefont {S.~N.}\ \bibnamefont
  {{Chen}}}, \bibinfo {author} {\bibfnamefont {M.}~\bibnamefont {{Vranic}}},
  \bibinfo {author} {\bibfnamefont {T.}~\bibnamefont {{Gangolf}}}, \bibinfo
  {author} {\bibfnamefont {E.}~\bibnamefont {{Boella}}}, \bibinfo {author}
  {\bibfnamefont {P.}~\bibnamefont {{Antici}}}, \bibinfo {author}
  {\bibfnamefont {M.}~\bibnamefont {{Bailly-Grandvaux}}}, \bibinfo {author}
  {\bibfnamefont {P.}~\bibnamefont {{Loiseau}}}, \bibinfo {author}
  {\bibfnamefont {H.}~\bibnamefont {{P{\'e}pin}}}, \bibinfo {author}
  {\bibfnamefont {G.}~\bibnamefont {{Revet}}}, \bibinfo {author} {\bibfnamefont
  {J.~J.}\ \bibnamefont {{Santos}}}, \bibinfo {author} {\bibfnamefont {A.~M.}\
  \bibnamefont {{Schroer}}}, \bibinfo {author} {\bibfnamefont {M.}~\bibnamefont
  {{Starodubtsev}}}, \bibinfo {author} {\bibfnamefont {O.}~\bibnamefont
  {{Willi}}}, \bibinfo {author} {\bibfnamefont {L.~O.}\ \bibnamefont
  {{Silva}}}, \bibinfo {author} {\bibfnamefont {E.}~\bibnamefont
  {{d'Humi{\`e}res}}},\ and\ \bibinfo {author} {\bibfnamefont {J.}~\bibnamefont
  {{Fuchs}}},\ }\bibfield  {title} {\bibinfo {title} {{Collimated protons
  accelerated from an overdense gas jet irradiated by a 1 um wavelength
  high-intensity short-pulse laser}},\ }\href
  {https://doi.org/10.1038/s41598-017-12910-6} {\bibfield  {journal} {\bibinfo
  {journal} {Sci. Rep.}\ }\textbf {\bibinfo {volume} {7}},\ \bibinfo {eid}
  {13505} (\bibinfo {year} {2017})}\BibitemShut {NoStop}%
\bibitem [{\citenamefont {{Fiuza}}\ \emph {et~al.}(2012)\citenamefont
  {{Fiuza}}, \citenamefont {{Stockem}}, \citenamefont {{Boella}}, \citenamefont
  {{Fonseca}}, \citenamefont {{Silva}}, \citenamefont {{Haberberger}},
  \citenamefont {{Tochitsky}}, \citenamefont {{Gong}}, \citenamefont {{Mori}},\
  and\ \citenamefont {{Joshi}}}]{Fiuza_2012}%
  \BibitemOpen
  \bibfield  {author} {\bibinfo {author} {\bibfnamefont {F.}~\bibnamefont
  {{Fiuza}}}, \bibinfo {author} {\bibfnamefont {A.}~\bibnamefont {{Stockem}}},
  \bibinfo {author} {\bibfnamefont {E.}~\bibnamefont {{Boella}}}, \bibinfo
  {author} {\bibfnamefont {R.~A.}\ \bibnamefont {{Fonseca}}}, \bibinfo {author}
  {\bibfnamefont {L.~O.}\ \bibnamefont {{Silva}}}, \bibinfo {author}
  {\bibfnamefont {D.}~\bibnamefont {{Haberberger}}}, \bibinfo {author}
  {\bibfnamefont {S.}~\bibnamefont {{Tochitsky}}}, \bibinfo {author}
  {\bibfnamefont {C.}~\bibnamefont {{Gong}}}, \bibinfo {author} {\bibfnamefont
  {W.~B.}\ \bibnamefont {{Mori}}},\ and\ \bibinfo {author} {\bibfnamefont
  {C.}~\bibnamefont {{Joshi}}},\ }\bibfield  {title} {\bibinfo {title}
  {{Laser-Driven Shock Acceleration of Monoenergetic Ion Beams}},\ }\href
  {https://doi.org/10.1103/PhysRevLett.109.215001} {\bibfield  {journal}
  {\bibinfo  {journal} {Phys. Rev. Lett.}\ }\textbf {\bibinfo {volume} {109}},\
  \bibinfo {eid} {215001} (\bibinfo {year} {2012})}\BibitemShut {NoStop}%
\bibitem [{\citenamefont {{Fiuza}}\ \emph {et~al.}(2013)\citenamefont
  {{Fiuza}}, \citenamefont {{Stockem}}, \citenamefont {{Boella}}, \citenamefont
  {{Fonseca}}, \citenamefont {{Silva}}, \citenamefont {{Haberberger}},
  \citenamefont {{Tochitsky}}, \citenamefont {{Mori}},\ and\ \citenamefont
  {{Joshi}}}]{Fiuza_2013}%
  \BibitemOpen
  \bibfield  {author} {\bibinfo {author} {\bibfnamefont {F.}~\bibnamefont
  {{Fiuza}}}, \bibinfo {author} {\bibfnamefont {A.}~\bibnamefont {{Stockem}}},
  \bibinfo {author} {\bibfnamefont {E.}~\bibnamefont {{Boella}}}, \bibinfo
  {author} {\bibfnamefont {R.~A.}\ \bibnamefont {{Fonseca}}}, \bibinfo {author}
  {\bibfnamefont {L.~O.}\ \bibnamefont {{Silva}}}, \bibinfo {author}
  {\bibfnamefont {D.}~\bibnamefont {{Haberberger}}}, \bibinfo {author}
  {\bibfnamefont {S.}~\bibnamefont {{Tochitsky}}}, \bibinfo {author}
  {\bibfnamefont {W.~B.}\ \bibnamefont {{Mori}}},\ and\ \bibinfo {author}
  {\bibfnamefont {C.}~\bibnamefont {{Joshi}}},\ }\bibfield  {title} {\bibinfo
  {title} {{Ion acceleration from laser-driven electrostatic shocks}},\ }\href
  {https://doi.org/10.1063/1.4801526} {\bibfield  {journal} {\bibinfo
  {journal} {Phys. Plasmas}\ }\textbf {\bibinfo {volume} {20}},\ \bibinfo {eid}
  {056304} (\bibinfo {year} {2013})}\BibitemShut {NoStop}%
\bibitem [{\citenamefont {{Dieckmann}}\ \emph
  {et~al.}(2013{\natexlab{a}})\citenamefont {{Dieckmann}}, \citenamefont
  {{Ahmed}}, \citenamefont {{Sarri}}, \citenamefont {{Doria}}, \citenamefont
  {{Kourakis}}, \citenamefont {{Romagnani}}, \citenamefont {{Pohl}},\ and\
  \citenamefont {{Borghesi}}}]{Dieckmann_2013a}%
  \BibitemOpen
  \bibfield  {author} {\bibinfo {author} {\bibfnamefont {M.~E.}\ \bibnamefont
  {{Dieckmann}}}, \bibinfo {author} {\bibfnamefont {H.}~\bibnamefont
  {{Ahmed}}}, \bibinfo {author} {\bibfnamefont {G.}~\bibnamefont {{Sarri}}},
  \bibinfo {author} {\bibfnamefont {D.}~\bibnamefont {{Doria}}}, \bibinfo
  {author} {\bibfnamefont {I.}~\bibnamefont {{Kourakis}}}, \bibinfo {author}
  {\bibfnamefont {L.}~\bibnamefont {{Romagnani}}}, \bibinfo {author}
  {\bibfnamefont {M.}~\bibnamefont {{Pohl}}},\ and\ \bibinfo {author}
  {\bibfnamefont {M.}~\bibnamefont {{Borghesi}}},\ }\bibfield  {title}
  {\bibinfo {title} {{Parametric study of non-relativistic electrostatic shocks
  and the structure of their transition layer}},\ }\href
  {https://doi.org/10.1063/1.4801447} {\bibfield  {journal} {\bibinfo
  {journal} {Phys. Plasmas}\ }\textbf {\bibinfo {volume} {20}},\ \bibinfo {eid}
  {042111} (\bibinfo {year} {2013}{\natexlab{a}})}\BibitemShut {NoStop}%
\bibitem [{\citenamefont {{Dieckmann}}\ \emph
  {et~al.}(2013{\natexlab{b}})\citenamefont {{Dieckmann}}, \citenamefont
  {{Sarri}}, \citenamefont {{Doria}}, \citenamefont {{Pohl}},\ and\
  \citenamefont {{Borghesi}}}]{Dieckmann_2013b}%
  \BibitemOpen
  \bibfield  {author} {\bibinfo {author} {\bibfnamefont {M.~E.}\ \bibnamefont
  {{Dieckmann}}}, \bibinfo {author} {\bibfnamefont {G.}~\bibnamefont
  {{Sarri}}}, \bibinfo {author} {\bibfnamefont {D.}~\bibnamefont {{Doria}}},
  \bibinfo {author} {\bibfnamefont {M.}~\bibnamefont {{Pohl}}},\ and\ \bibinfo
  {author} {\bibfnamefont {M.}~\bibnamefont {{Borghesi}}},\ }\bibfield  {title}
  {\bibinfo {title} {{Modification of the formation of high-Mach number
  electrostatic shock-like structures by the ion acoustic instability}},\
  }\href {https://doi.org/10.1063/1.4825339} {\bibfield  {journal} {\bibinfo
  {journal} {Phys. Plasmas}\ }\textbf {\bibinfo {volume} {20}},\ \bibinfo {eid}
  {102112} (\bibinfo {year} {2013}{\natexlab{b}})}\BibitemShut {NoStop}%
\bibitem [{\citenamefont {{L{\'e}cz}}\ and\ \citenamefont
  {{Andreev}}(2015)}]{Lecz_2015}%
  \BibitemOpen
  \bibfield  {author} {\bibinfo {author} {\bibfnamefont {Z.}~\bibnamefont
  {{L{\'e}cz}}}\ and\ \bibinfo {author} {\bibfnamefont {A.}~\bibnamefont
  {{Andreev}}},\ }\bibfield  {title} {\bibinfo {title} {{Shock wave
  acceleration of protons in inhomogeneous plasma interacting with ultrashort
  intense laser pulses}},\ }\href {https://doi.org/10.1063/1.4913438}
  {\bibfield  {journal} {\bibinfo  {journal} {Phys. Plasmas}\ }\textbf
  {\bibinfo {volume} {22}},\ \bibinfo {eid} {043103} (\bibinfo {year}
  {2015})}\BibitemShut {NoStop}%
\bibitem [{\citenamefont {{Liu}}\ \emph {et~al.}(2016)\citenamefont {{Liu}},
  \citenamefont {{Weng}}, \citenamefont {{Li}}, \citenamefont {{Yuan}},
  \citenamefont {{Chen}}, \citenamefont {{Mulser}}, \citenamefont {{Sheng}},
  \citenamefont {{Murakami}}, \citenamefont {{Yu}}, \citenamefont {{Zheng}},\
  and\ \citenamefont {{Zhang}}}]{Liu_2016}%
  \BibitemOpen
  \bibfield  {author} {\bibinfo {author} {\bibfnamefont {M.}~\bibnamefont
  {{Liu}}}, \bibinfo {author} {\bibfnamefont {S.~M.}\ \bibnamefont {{Weng}}},
  \bibinfo {author} {\bibfnamefont {Y.~T.}\ \bibnamefont {{Li}}}, \bibinfo
  {author} {\bibfnamefont {D.~W.}\ \bibnamefont {{Yuan}}}, \bibinfo {author}
  {\bibfnamefont {M.}~\bibnamefont {{Chen}}}, \bibinfo {author} {\bibfnamefont
  {P.}~\bibnamefont {{Mulser}}}, \bibinfo {author} {\bibfnamefont {Z.~M.}\
  \bibnamefont {{Sheng}}}, \bibinfo {author} {\bibfnamefont {M.}~\bibnamefont
  {{Murakami}}}, \bibinfo {author} {\bibfnamefont {L.~L.}\ \bibnamefont
  {{Yu}}}, \bibinfo {author} {\bibfnamefont {X.~L.}\ \bibnamefont {{Zheng}}},\
  and\ \bibinfo {author} {\bibfnamefont {J.}~\bibnamefont {{Zhang}}},\
  }\bibfield  {title} {\bibinfo {title} {{Collisionless electrostatic shock
  formation and ion acceleration in intense laser interactions with near
  critical density plasmas}},\ }\href {https://doi.org/10.1063/1.4967946}
  {\bibfield  {journal} {\bibinfo  {journal} {Phys. Plasmas}\ }\textbf
  {\bibinfo {volume} {23}},\ \bibinfo {eid} {113103} (\bibinfo {year}
  {2016})}\BibitemShut {NoStop}%
\bibitem [{\citenamefont {{Haberberger}}\ \emph {et~al.}(2012)\citenamefont
  {{Haberberger}}, \citenamefont {{Tochitsky}}, \citenamefont {{Fiuza}},
  \citenamefont {{Gong}}, \citenamefont {{Fonseca}}, \citenamefont {{Silva}},
  \citenamefont {{Mori}},\ and\ \citenamefont {{Joshi}}}]{Haberberger_2012}%
  \BibitemOpen
  \bibfield  {author} {\bibinfo {author} {\bibfnamefont {D.}~\bibnamefont
  {{Haberberger}}}, \bibinfo {author} {\bibfnamefont {S.}~\bibnamefont
  {{Tochitsky}}}, \bibinfo {author} {\bibfnamefont {F.}~\bibnamefont
  {{Fiuza}}}, \bibinfo {author} {\bibfnamefont {C.}~\bibnamefont {{Gong}}},
  \bibinfo {author} {\bibfnamefont {R.~A.}\ \bibnamefont {{Fonseca}}}, \bibinfo
  {author} {\bibfnamefont {L.~O.}\ \bibnamefont {{Silva}}}, \bibinfo {author}
  {\bibfnamefont {W.~B.}\ \bibnamefont {{Mori}}},\ and\ \bibinfo {author}
  {\bibfnamefont {C.}~\bibnamefont {{Joshi}}},\ }\bibfield  {title} {\bibinfo
  {title} {{Collisionless shocks in laser-produced plasma generate
  monoenergetic high-energy proton beams}},\ }\href
  {https://doi.org/10.1038/nphys2130} {\bibfield  {journal} {\bibinfo
  {journal} {Nat. Phys.}\ }\textbf {\bibinfo {volume} {8}},\ \bibinfo {pages}
  {95} (\bibinfo {year} {2012})}\BibitemShut {NoStop}%
\bibitem [{\citenamefont {{Puyuelo-Valdes}}\ \emph
  {et~al.}(2019{\natexlab{a}})\citenamefont {{Puyuelo-Valdes}}, \citenamefont
  {{Henares}}, \citenamefont {{Hannachi}}, \citenamefont {{Ceccotti}},
  \citenamefont {{Domange}}, \citenamefont {{Ehret}}, \citenamefont
  {{d'Humieres}}, \citenamefont {{Lancia}}, \citenamefont {{Marqu{\`e}s}},
  \citenamefont {{Ribeyre}}, \citenamefont {{Santos}}, \citenamefont
  {{Tikhonchuk}},\ and\ \citenamefont {{Tarisien}}}]{Puyuelo-Valdes_2019}%
  \BibitemOpen
  \bibfield  {author} {\bibinfo {author} {\bibfnamefont {P.}~\bibnamefont
  {{Puyuelo-Valdes}}}, \bibinfo {author} {\bibfnamefont {J.~L.}\ \bibnamefont
  {{Henares}}}, \bibinfo {author} {\bibfnamefont {F.}~\bibnamefont
  {{Hannachi}}}, \bibinfo {author} {\bibfnamefont {T.}~\bibnamefont
  {{Ceccotti}}}, \bibinfo {author} {\bibfnamefont {J.}~\bibnamefont
  {{Domange}}}, \bibinfo {author} {\bibfnamefont {M.}~\bibnamefont {{Ehret}}},
  \bibinfo {author} {\bibfnamefont {E.}~\bibnamefont {{d'Humieres}}}, \bibinfo
  {author} {\bibfnamefont {L.}~\bibnamefont {{Lancia}}}, \bibinfo {author}
  {\bibfnamefont {J.~R.}\ \bibnamefont {{Marqu{\`e}s}}}, \bibinfo {author}
  {\bibfnamefont {X.}~\bibnamefont {{Ribeyre}}}, \bibinfo {author}
  {\bibfnamefont {J.~J.}\ \bibnamefont {{Santos}}}, \bibinfo {author}
  {\bibfnamefont {V.}~\bibnamefont {{Tikhonchuk}}},\ and\ \bibinfo {author}
  {\bibfnamefont {M.}~\bibnamefont {{Tarisien}}},\ }\bibfield  {title}
  {\bibinfo {title} {{Proton acceleration by collisionless shocks using a
  supersonic H$_{2}$ gas-jet target and high-power infrared laser pulses}},\
  }\href {https://doi.org/10.1063/1.5116337} {\bibfield  {journal} {\bibinfo
  {journal} {Phys. Plasmas}\ }\textbf {\bibinfo {volume} {26}},\ \bibinfo {eid}
  {123109} (\bibinfo {year} {2019}{\natexlab{a}})}\BibitemShut {NoStop}%
\bibitem [{\citenamefont {{Wilks}}\ \emph {et~al.}(1992)\citenamefont
  {{Wilks}}, \citenamefont {{Kruer}}, \citenamefont {{Tabak}},\ and\
  \citenamefont {{Langdon}}}]{Wilks_1992}%
  \BibitemOpen
  \bibfield  {author} {\bibinfo {author} {\bibfnamefont {S.~C.}\ \bibnamefont
  {{Wilks}}}, \bibinfo {author} {\bibfnamefont {W.~L.}\ \bibnamefont
  {{Kruer}}}, \bibinfo {author} {\bibfnamefont {M.}~\bibnamefont {{Tabak}}},\
  and\ \bibinfo {author} {\bibfnamefont {A.~B.}\ \bibnamefont {{Langdon}}},\
  }\bibfield  {title} {\bibinfo {title} {{Absorption of ultra-intense laser
  pulses}},\ }\href {https://doi.org/10.1103/PhysRevLett.69.1383} {\bibfield
  {journal} {\bibinfo  {journal} {Phys. Rev. Lett.}\ }\textbf {\bibinfo
  {volume} {69}},\ \bibinfo {pages} {1383} (\bibinfo {year}
  {1992})}\BibitemShut {NoStop}%
\bibitem [{\citenamefont {{Puyuelo-Valdes}}\ \emph
  {et~al.}(2019{\natexlab{b}})\citenamefont {{Puyuelo-Valdes}}, \citenamefont
  {{Henares}}, \citenamefont {{Hannachi}}, \citenamefont {{Ceccotti}},
  \citenamefont {{Domange}}, \citenamefont {{Ehret}}, \citenamefont
  {{D'Humieres}}, \citenamefont {{Lancia}}, \citenamefont {{Marqu{\`e}s}},
  \citenamefont {{Santos}},\ and\ \citenamefont
  {{Tarisien}}}]{Puyuelo-Valdes_2019b}%
  \BibitemOpen
  \bibfield  {author} {\bibinfo {author} {\bibfnamefont {P.}~\bibnamefont
  {{Puyuelo-Valdes}}}, \bibinfo {author} {\bibfnamefont {J.~L.}\ \bibnamefont
  {{Henares}}}, \bibinfo {author} {\bibfnamefont {F.}~\bibnamefont
  {{Hannachi}}}, \bibinfo {author} {\bibfnamefont {T.}~\bibnamefont
  {{Ceccotti}}}, \bibinfo {author} {\bibfnamefont {J.}~\bibnamefont
  {{Domange}}}, \bibinfo {author} {\bibfnamefont {M.}~\bibnamefont {{Ehret}}},
  \bibinfo {author} {\bibfnamefont {E.}~\bibnamefont {{D'Humieres}}}, \bibinfo
  {author} {\bibfnamefont {L.}~\bibnamefont {{Lancia}}}, \bibinfo {author}
  {\bibfnamefont {J.~R.}\ \bibnamefont {{Marqu{\`e}s}}}, \bibinfo {author}
  {\bibfnamefont {J.}~\bibnamefont {{Santos}}},\ and\ \bibinfo {author}
  {\bibfnamefont {M.}~\bibnamefont {{Tarisien}}},\ }\bibfield  {title}
  {\bibinfo {title} {{Laser driven ion acceleration in high-density gas
  jets}},\ }in\ \href {https://doi.org/10.1117/12.2520799} {\emph {\bibinfo
  {booktitle} {Laser Acceleration of Electrons, Protons, and Ions V}}},\
  \bibinfo {series} {Society of Photo-Optical Instrumentation Engineers (SPIE)
  Conference Series}, Vol.\ \bibinfo {volume} {11037},\ \bibinfo {editor}
  {edited by\ \bibinfo {editor} {\bibfnamefont {E.}~\bibnamefont {{Esarey}}},
  \bibinfo {editor} {\bibfnamefont {C.~B.}\ \bibnamefont {{Schroeder}}},\ and\
  \bibinfo {editor} {\bibfnamefont {J.}~\bibnamefont {{Schreiber}}}}\ (\bibinfo
  {year} {2019})\ p.\ \bibinfo {pages} {110370B}\BibitemShut {NoStop}%
\bibitem [{\citenamefont {{Sylla}}\ \emph {et~al.}(2013)\citenamefont
  {{Sylla}}, \citenamefont {{Flacco}}, \citenamefont {{Kahaly}}, \citenamefont
  {{Veltcheva}}, \citenamefont {{Lifschitz}}, \citenamefont {{Malka}},
  \citenamefont {{d'Humi{\`e}res}}, \citenamefont {{Andriyash}},\ and\
  \citenamefont {{Tikhonchuk}}}]{Sylla_2013}%
  \BibitemOpen
  \bibfield  {author} {\bibinfo {author} {\bibfnamefont {F.}~\bibnamefont
  {{Sylla}}}, \bibinfo {author} {\bibfnamefont {A.}~\bibnamefont {{Flacco}}},
  \bibinfo {author} {\bibfnamefont {S.}~\bibnamefont {{Kahaly}}}, \bibinfo
  {author} {\bibfnamefont {M.}~\bibnamefont {{Veltcheva}}}, \bibinfo {author}
  {\bibfnamefont {A.}~\bibnamefont {{Lifschitz}}}, \bibinfo {author}
  {\bibfnamefont {V.}~\bibnamefont {{Malka}}}, \bibinfo {author} {\bibfnamefont
  {E.}~\bibnamefont {{d'Humi{\`e}res}}}, \bibinfo {author} {\bibfnamefont
  {I.}~\bibnamefont {{Andriyash}}},\ and\ \bibinfo {author} {\bibfnamefont
  {V.}~\bibnamefont {{Tikhonchuk}}},\ }\bibfield  {title} {\bibinfo {title}
  {{Short Intense Laser Pulse Collapse in Near-Critical Plasma}},\ }\href
  {https://doi.org/10.1103/PhysRevLett.110.085001} {\bibfield  {journal}
  {\bibinfo  {journal} {\prl}\ }\textbf {\bibinfo {volume} {110}},\ \bibinfo
  {eid} {085001} (\bibinfo {year} {2013})}\BibitemShut {NoStop}%
\bibitem [{\citenamefont {Singh}\ \emph {et~al.}(2020)\citenamefont {Singh},
  \citenamefont {Pathak}, \citenamefont {Shin}, \citenamefont {Choi},
  \citenamefont {Nakajima}, \citenamefont {Lee}, \citenamefont {Sung},
  \citenamefont {Lee}, \citenamefont {Rhee}, \citenamefont {Aniculaesei},
  \citenamefont {Kim}, \citenamefont {Pae}, \citenamefont {Cho}, \citenamefont
  {Hojbota}, \citenamefont {Lee}, \citenamefont {Mollica}, \citenamefont
  {Malka}, \citenamefont {Ryu}, \citenamefont {Kim},\ and\ \citenamefont
  {Nam}}]{Singh_2020}%
  \BibitemOpen
  \bibfield  {author} {\bibinfo {author} {\bibfnamefont {P.~K.}\ \bibnamefont
  {Singh}}, \bibinfo {author} {\bibfnamefont {V.~B.}\ \bibnamefont {Pathak}},
  \bibinfo {author} {\bibfnamefont {J.~H.}\ \bibnamefont {Shin}}, \bibinfo
  {author} {\bibfnamefont {I.~W.}\ \bibnamefont {Choi}}, \bibinfo {author}
  {\bibfnamefont {K.}~\bibnamefont {Nakajima}}, \bibinfo {author}
  {\bibfnamefont {S.~K.}\ \bibnamefont {Lee}}, \bibinfo {author} {\bibfnamefont
  {J.~H.}\ \bibnamefont {Sung}}, \bibinfo {author} {\bibfnamefont {H.~W.}\
  \bibnamefont {Lee}}, \bibinfo {author} {\bibfnamefont {Y.~J.}\ \bibnamefont
  {Rhee}}, \bibinfo {author} {\bibfnamefont {C.}~\bibnamefont {Aniculaesei}},
  \bibinfo {author} {\bibfnamefont {C.~M.}\ \bibnamefont {Kim}}, \bibinfo
  {author} {\bibfnamefont {K.~H.}\ \bibnamefont {Pae}}, \bibinfo {author}
  {\bibfnamefont {M.~C.}\ \bibnamefont {Cho}}, \bibinfo {author} {\bibfnamefont
  {C.}~\bibnamefont {Hojbota}}, \bibinfo {author} {\bibfnamefont {S.~G.}\
  \bibnamefont {Lee}}, \bibinfo {author} {\bibfnamefont {F.}~\bibnamefont
  {Mollica}}, \bibinfo {author} {\bibfnamefont {V.}~\bibnamefont {Malka}},
  \bibinfo {author} {\bibfnamefont {C.-M.}\ \bibnamefont {Ryu}}, \bibinfo
  {author} {\bibfnamefont {H.~T.}\ \bibnamefont {Kim}},\ and\ \bibinfo {author}
  {\bibfnamefont {C.~H.}\ \bibnamefont {Nam}},\ }\bibfield  {title} {\bibinfo
  {title} {{Electrostatic shock acceleration of ions in near-critical-density
  plasma driven by a femtosecond petawatt laser}},\ }\href
  {https://doi.org/10.1038/s41598-020-75455-1} {\bibfield  {journal} {\bibinfo
  {journal} {Sci. Rep.}\ }\textbf {\bibinfo {volume} {10}},\ \bibinfo {pages}
  {18452} (\bibinfo {year} {2020})}\BibitemShut {NoStop}%
\bibitem [{\citenamefont {{Arefiev}}\ \emph {et~al.}(2016)\citenamefont
  {{Arefiev}}, \citenamefont {{Khudik}}, \citenamefont {{Robinson}},
  \citenamefont {{Shvets}}, \citenamefont {{Willingale}},\ and\ \citenamefont
  {{Schollmeier}}}]{Arefiev_2016}%
  \BibitemOpen
  \bibfield  {author} {\bibinfo {author} {\bibfnamefont {A.~V.}\ \bibnamefont
  {{Arefiev}}}, \bibinfo {author} {\bibfnamefont {V.~N.}\ \bibnamefont
  {{Khudik}}}, \bibinfo {author} {\bibfnamefont {A.~P.~L.}\ \bibnamefont
  {{Robinson}}}, \bibinfo {author} {\bibfnamefont {G.}~\bibnamefont
  {{Shvets}}}, \bibinfo {author} {\bibfnamefont {L.}~\bibnamefont
  {{Willingale}}},\ and\ \bibinfo {author} {\bibfnamefont {M.}~\bibnamefont
  {{Schollmeier}}},\ }\bibfield  {title} {\bibinfo {title} {{Beyond the
  ponderomotive limit: Direct laser acceleration of relativistic electrons in
  sub-critical plasmas}},\ }\href {https://doi.org/10.1063/1.4946024}
  {\bibfield  {journal} {\bibinfo  {journal} {Phys. Plasmas}\ }\textbf
  {\bibinfo {volume} {23}},\ \bibinfo {eid} {056704} (\bibinfo {year}
  {2016})}\BibitemShut {NoStop}%
\bibitem [{\citenamefont {{Hussein}}\ \emph {et~al.}(2021)\citenamefont
  {{Hussein}}, \citenamefont {{Arefiev}}, \citenamefont {{Batson}},
  \citenamefont {{Chen}}, \citenamefont {{Craxton}}, \citenamefont {{Davies}},
  \citenamefont {{Froula}}, \citenamefont {{Gong}}, \citenamefont
  {{Haberberger}}, \citenamefont {{Ma}}, \citenamefont {{Nilson}},
  \citenamefont {{Theobald}}, \citenamefont {{Wang}}, \citenamefont
  {{Weichman}}, \citenamefont {{Williams}},\ and\ \citenamefont
  {{Willingale}}}]{Hussein_2021}%
  \BibitemOpen
  \bibfield  {author} {\bibinfo {author} {\bibfnamefont {A.~E.}\ \bibnamefont
  {{Hussein}}}, \bibinfo {author} {\bibfnamefont {A.~V.}\ \bibnamefont
  {{Arefiev}}}, \bibinfo {author} {\bibfnamefont {T.}~\bibnamefont {{Batson}}},
  \bibinfo {author} {\bibfnamefont {H.}~\bibnamefont {{Chen}}}, \bibinfo
  {author} {\bibfnamefont {R.~S.}\ \bibnamefont {{Craxton}}}, \bibinfo {author}
  {\bibfnamefont {A.~S.}\ \bibnamefont {{Davies}}}, \bibinfo {author}
  {\bibfnamefont {D.~H.}\ \bibnamefont {{Froula}}}, \bibinfo {author}
  {\bibfnamefont {Z.}~\bibnamefont {{Gong}}}, \bibinfo {author} {\bibfnamefont
  {D.}~\bibnamefont {{Haberberger}}}, \bibinfo {author} {\bibfnamefont
  {Y.}~\bibnamefont {{Ma}}}, \bibinfo {author} {\bibfnamefont {P.~M.}\
  \bibnamefont {{Nilson}}}, \bibinfo {author} {\bibfnamefont {W.}~\bibnamefont
  {{Theobald}}}, \bibinfo {author} {\bibfnamefont {T.}~\bibnamefont {{Wang}}},
  \bibinfo {author} {\bibfnamefont {K.}~\bibnamefont {{Weichman}}}, \bibinfo
  {author} {\bibfnamefont {G.~J.}\ \bibnamefont {{Williams}}},\ and\ \bibinfo
  {author} {\bibfnamefont {L.}~\bibnamefont {{Willingale}}},\ }\bibfield
  {title} {\bibinfo {title} {{Towards the optimisation of direct laser
  acceleration}},\ }\href {https://doi.org/10.1088/1367-2630/abdf9a} {\bibfield
   {journal} {\bibinfo  {journal} {New J. Phys.}\ }\textbf {\bibinfo {volume}
  {23}},\ \bibinfo {eid} {023031} (\bibinfo {year} {2021})}\BibitemShut
  {NoStop}%
\bibitem [{\citenamefont {{Shaw}}\ \emph {et~al.}(2017)\citenamefont {{Shaw}},
  \citenamefont {{Lemos}}, \citenamefont {{Amorim}}, \citenamefont
  {{Vafaei-Najafabadi}}, \citenamefont {{Marsh}}, \citenamefont {{Tsung}},
  \citenamefont {{Mori}},\ and\ \citenamefont {{Joshi}}}]{Shaw_2017}%
  \BibitemOpen
  \bibfield  {author} {\bibinfo {author} {\bibfnamefont {J.~L.}\ \bibnamefont
  {{Shaw}}}, \bibinfo {author} {\bibfnamefont {N.}~\bibnamefont {{Lemos}}},
  \bibinfo {author} {\bibfnamefont {L.~D.}\ \bibnamefont {{Amorim}}}, \bibinfo
  {author} {\bibfnamefont {N.}~\bibnamefont {{Vafaei-Najafabadi}}}, \bibinfo
  {author} {\bibfnamefont {K.~A.}\ \bibnamefont {{Marsh}}}, \bibinfo {author}
  {\bibfnamefont {F.~S.}\ \bibnamefont {{Tsung}}}, \bibinfo {author}
  {\bibfnamefont {W.~B.}\ \bibnamefont {{Mori}}},\ and\ \bibinfo {author}
  {\bibfnamefont {C.}~\bibnamefont {{Joshi}}},\ }\bibfield  {title} {\bibinfo
  {title} {{Role of Direct Laser Acceleration of Electrons in a Laser Wakefield
  Accelerator with Ionization Injection}},\ }\href
  {https://doi.org/10.1103/PhysRevLett.118.064801} {\bibfield  {journal}
  {\bibinfo  {journal} {Phys. Rev. Lett.}\ }\textbf {\bibinfo {volume} {118}},\
  \bibinfo {eid} {064801} (\bibinfo {year} {2017})}\BibitemShut {NoStop}%
\bibitem [{\citenamefont {{Shaw}}\ \emph {et~al.}(2018)\citenamefont {{Shaw}},
  \citenamefont {{Lemos}}, \citenamefont {{Marsh}}, \citenamefont {{Froula}},\
  and\ \citenamefont {{Joshi}}}]{Shaw_2018}%
  \BibitemOpen
  \bibfield  {author} {\bibinfo {author} {\bibfnamefont {J.~L.}\ \bibnamefont
  {{Shaw}}}, \bibinfo {author} {\bibfnamefont {N.}~\bibnamefont {{Lemos}}},
  \bibinfo {author} {\bibfnamefont {K.~A.}\ \bibnamefont {{Marsh}}}, \bibinfo
  {author} {\bibfnamefont {D.~H.}\ \bibnamefont {{Froula}}},\ and\ \bibinfo
  {author} {\bibfnamefont {C.}~\bibnamefont {{Joshi}}},\ }\bibfield  {title}
  {\bibinfo {title} {{Experimental signatures of
  direct-laser-acceleration-assisted laser wakefield acceleration}},\ }\href
  {https://doi.org/10.1088/1361-6587/aaade1} {\bibfield  {journal} {\bibinfo
  {journal} {Plasma Phys. Control. Fusion}\ }\textbf {\bibinfo {volume} {60}},\
  \bibinfo {eid} {044012} (\bibinfo {year} {2018})}\BibitemShut {NoStop}%
\bibitem [{\citenamefont {{Rovige}}\ \emph {et~al.}(2021)\citenamefont
  {{Rovige}}, \citenamefont {{Huijts}}, \citenamefont {{Vernier}},
  \citenamefont {{Andriyash}}, \citenamefont {{Sylla}}, \citenamefont
  {{Tomkus}}, \citenamefont {{Girdauskas}}, \citenamefont {{Raciukaitis}},
  \citenamefont {{Dudutis}}, \citenamefont {{Stankevic}}, \citenamefont
  {{Gecys}},\ and\ \citenamefont {{Faure}}}]{Rovige_2021}%
  \BibitemOpen
  \bibfield  {author} {\bibinfo {author} {\bibfnamefont {L.}~\bibnamefont
  {{Rovige}}}, \bibinfo {author} {\bibfnamefont {J.}~\bibnamefont {{Huijts}}},
  \bibinfo {author} {\bibfnamefont {A.}~\bibnamefont {{Vernier}}}, \bibinfo
  {author} {\bibfnamefont {I.}~\bibnamefont {{Andriyash}}}, \bibinfo {author}
  {\bibfnamefont {F.}~\bibnamefont {{Sylla}}}, \bibinfo {author} {\bibfnamefont
  {V.}~\bibnamefont {{Tomkus}}}, \bibinfo {author} {\bibfnamefont
  {V.}~\bibnamefont {{Girdauskas}}}, \bibinfo {author} {\bibfnamefont
  {G.}~\bibnamefont {{Raciukaitis}}}, \bibinfo {author} {\bibfnamefont
  {J.}~\bibnamefont {{Dudutis}}}, \bibinfo {author} {\bibfnamefont
  {V.}~\bibnamefont {{Stankevic}}}, \bibinfo {author} {\bibfnamefont
  {P.}~\bibnamefont {{Gecys}}},\ and\ \bibinfo {author} {\bibfnamefont
  {J.}~\bibnamefont {{Faure}}},\ }\bibfield  {title} {\bibinfo {title}
  {{Symmetric and asymmetric shocked gas jets for laser-plasma experiments}},\
  }\href {https://doi.org/10.1063/5.0051173} {\bibfield  {journal} {\bibinfo
  {journal} {Rev. Sci. Instrum.}\ }\textbf {\bibinfo {volume} {92}},\ \bibinfo
  {eid} {083302} (\bibinfo {year} {2021})}\BibitemShut {NoStop}%
\bibitem [{Sou()}]{SourceLab}%
  \BibitemOpen
  \href@noop {} {\bibinfo {title} {{SourceLAB}}},\ \bibinfo {howpublished}
  {\url{https://www.sourcelab-plasma.com}}\BibitemShut {NoStop}%
\bibitem [{\citenamefont {Mart\'{i}n-L\'{o}pez}(2021)}]{MartinLopez_2021}%
  \BibitemOpen
  \bibfield  {author} {\bibinfo {author} {\bibfnamefont {A.}~\bibnamefont
  {Mart\'{i}n-L\'{o}pez}},\ }\emph {\bibinfo {title} {Safety system automation
  for gaseous targets experiments and real-time leak detection in the vacuum
  system}},\ \href@noop {} {Master's thesis},\ \bibinfo  {school} {University
  of Salamanca} (\bibinfo {year} {2021})\BibitemShut {NoStop}%
\bibitem [{\citenamefont {{Puyuelo-Valdes}}(2020)}]{Puyuelo-Valdes_2020}%
  \BibitemOpen
  \bibfield  {author} {\bibinfo {author} {\bibfnamefont {P.}~\bibnamefont
  {{Puyuelo-Valdes}}},\ }\emph {\bibinfo {title} {{Laser-driven ion
  acceleration with high-density gas-jet targets and application to elemental
  analysis}}},\ \href {https://tel.archives-ouvertes.fr/tel-03053720} {Ph.D.
  thesis},\ \bibinfo  {school} {{Universit{\'e} de Bordeaux (France); Institut
  national de la recherche scientifique (Canada)}} (\bibinfo {year}
  {2020})\BibitemShut {NoStop}%
\bibitem [{\citenamefont {{Henares}}\ \emph {et~al.}(2019)\citenamefont
  {{Henares}}, \citenamefont {{Puyuelo-Valdes}}, \citenamefont {{Hannachi}},
  \citenamefont {{Ceccotti}}, \citenamefont {{Ehret}}, \citenamefont {{Gobet}},
  \citenamefont {{Lancia}}, \citenamefont {{Marqu{\`e}s}}, \citenamefont
  {{Santos}}, \citenamefont {{Versteegen}},\ and\ \citenamefont
  {{Tarisien}}}]{Henares_2019}%
  \BibitemOpen
  \bibfield  {author} {\bibinfo {author} {\bibfnamefont {J.~L.}\ \bibnamefont
  {{Henares}}}, \bibinfo {author} {\bibfnamefont {P.}~\bibnamefont
  {{Puyuelo-Valdes}}}, \bibinfo {author} {\bibfnamefont {F.}~\bibnamefont
  {{Hannachi}}}, \bibinfo {author} {\bibfnamefont {T.}~\bibnamefont
  {{Ceccotti}}}, \bibinfo {author} {\bibfnamefont {M.}~\bibnamefont {{Ehret}}},
  \bibinfo {author} {\bibfnamefont {F.}~\bibnamefont {{Gobet}}}, \bibinfo
  {author} {\bibfnamefont {L.}~\bibnamefont {{Lancia}}}, \bibinfo {author}
  {\bibfnamefont {J.~R.}\ \bibnamefont {{Marqu{\`e}s}}}, \bibinfo {author}
  {\bibfnamefont {J.~J.}\ \bibnamefont {{Santos}}}, \bibinfo {author}
  {\bibfnamefont {M.}~\bibnamefont {{Versteegen}}},\ and\ \bibinfo {author}
  {\bibfnamefont {M.}~\bibnamefont {{Tarisien}}},\ }\bibfield  {title}
  {\bibinfo {title} {{Development of gas jet targets for laser-plasma
  experiments at near-critical density}},\ }\href
  {https://doi.org/10.1063/1.5093613} {\bibfield  {journal} {\bibinfo
  {journal} {Rev. Sci. Instrum.}\ }\textbf {\bibinfo {volume} {90}},\ \bibinfo
  {eid} {063302} (\bibinfo {year} {2019})}\BibitemShut {NoStop}%
\bibitem [{\citenamefont {{Pisarczyk}}\ \emph {et~al.}(2019)\citenamefont
  {{Pisarczyk}}, \citenamefont {{Santos}}, \citenamefont {{Dudzak}},
  \citenamefont {{Zaras-Szyd{\l}owska}}, \citenamefont {{Ehret}}, \citenamefont
  {{Rusiniak}}, \citenamefont {{Dostal}}, \citenamefont {{Chodukowski}},
  \citenamefont {{Renner}}, \citenamefont {{Gus'kov}}, \citenamefont
  {{Korneev}}, \citenamefont {{Burian}}, \citenamefont {{Vlachos}},
  \citenamefont {{Kochetkov}}, \citenamefont {{Makaruk}}, \citenamefont
  {{Rosinski}}, \citenamefont {{Kalal}}, \citenamefont {{Krupka}},
  \citenamefont {{Pfeifer}}, \citenamefont {{Klir}}, \citenamefont
  {{Cikhardt}}, \citenamefont {{Krasa}}, \citenamefont {{Singh}}, \citenamefont
  {{Borodziuk}}, \citenamefont {{Krus}}, \citenamefont {{Juha}}, \citenamefont
  {{Hrebicek}}, \citenamefont {{Golasowski}},\ and\ \citenamefont
  {{Skala}}}]{Pisarczyk_2019}%
  \BibitemOpen
  \bibfield  {author} {\bibinfo {author} {\bibfnamefont {T.}~\bibnamefont
  {{Pisarczyk}}}, \bibinfo {author} {\bibfnamefont {J.~J.}\ \bibnamefont
  {{Santos}}}, \bibinfo {author} {\bibfnamefont {R.}~\bibnamefont {{Dudzak}}},
  \bibinfo {author} {\bibfnamefont {A.}~\bibnamefont {{Zaras-Szyd{\l}owska}}},
  \bibinfo {author} {\bibfnamefont {M.}~\bibnamefont {{Ehret}}}, \bibinfo
  {author} {\bibfnamefont {Z.}~\bibnamefont {{Rusiniak}}}, \bibinfo {author}
  {\bibfnamefont {J.}~\bibnamefont {{Dostal}}}, \bibinfo {author}
  {\bibfnamefont {T.}~\bibnamefont {{Chodukowski}}}, \bibinfo {author}
  {\bibfnamefont {O.}~\bibnamefont {{Renner}}}, \bibinfo {author}
  {\bibfnamefont {S.~Y.}\ \bibnamefont {{Gus'kov}}}, \bibinfo {author}
  {\bibfnamefont {P.}~\bibnamefont {{Korneev}}}, \bibinfo {author}
  {\bibfnamefont {T.}~\bibnamefont {{Burian}}}, \bibinfo {author}
  {\bibfnamefont {C.}~\bibnamefont {{Vlachos}}}, \bibinfo {author}
  {\bibfnamefont {I.}~\bibnamefont {{Kochetkov}}}, \bibinfo {author}
  {\bibfnamefont {D.}~\bibnamefont {{Makaruk}}}, \bibinfo {author}
  {\bibfnamefont {M.}~\bibnamefont {{Rosinski}}}, \bibinfo {author}
  {\bibfnamefont {M.}~\bibnamefont {{Kalal}}}, \bibinfo {author} {\bibfnamefont
  {M.}~\bibnamefont {{Krupka}}}, \bibinfo {author} {\bibfnamefont
  {M.}~\bibnamefont {{Pfeifer}}}, \bibinfo {author} {\bibfnamefont
  {D.}~\bibnamefont {{Klir}}}, \bibinfo {author} {\bibfnamefont
  {J.}~\bibnamefont {{Cikhardt}}}, \bibinfo {author} {\bibfnamefont
  {J.}~\bibnamefont {{Krasa}}}, \bibinfo {author} {\bibfnamefont
  {S.}~\bibnamefont {{Singh}}}, \bibinfo {author} {\bibfnamefont
  {S.}~\bibnamefont {{Borodziuk}}}, \bibinfo {author} {\bibfnamefont
  {M.}~\bibnamefont {{Krus}}}, \bibinfo {author} {\bibfnamefont
  {L.}~\bibnamefont {{Juha}}}, \bibinfo {author} {\bibfnamefont
  {J.}~\bibnamefont {{Hrebicek}}}, \bibinfo {author} {\bibfnamefont
  {J.}~\bibnamefont {{Golasowski}}},\ and\ \bibinfo {author} {\bibfnamefont
  {J.}~\bibnamefont {{Skala}}},\ }\bibfield  {title} {\bibinfo {title}
  {{Elaboration of 3-frame complex interferometry for optimization studies of
  capacitor-coil optical magnetic field generators}},\ }\href
  {https://doi.org/10.1088/17480221/14/11/C11024} {\bibfield  {journal}
  {\bibinfo  {journal} {J. Instrum.}\ }\textbf {\bibinfo {volume} {14}}\bibinfo
   {number} { (11)},\ \bibinfo {pages} {C11024}}\BibitemShut {NoStop}%
\bibitem [{\citenamefont {{Zara{\'s}-Szyd{\l}owska}}\ \emph
  {et~al.}(2020)\citenamefont {{Zara{\'s}-Szyd{\l}owska}}, \citenamefont
  {{Pisarczyk}}, \citenamefont {{Chodukowski}}, \citenamefont {{Rusiniak}},
  \citenamefont {{Dudzak}}, \citenamefont {{Dostal}}, \citenamefont {{Kalal}},
  \citenamefont {{Kochetkov}}, \citenamefont {{Krupka}}, \citenamefont
  {{Borodziuk}},\ and\ \citenamefont {{Pisarczyk}}}]{Zaras_2020}%
  \BibitemOpen
\bibfield  {number} {  }\bibfield  {author} {\bibinfo {author} {\bibfnamefont
  {A.}~\bibnamefont {{Zara{\'s}-Szyd{\l}owska}}}, \bibinfo {author}
  {\bibfnamefont {T.}~\bibnamefont {{Pisarczyk}}}, \bibinfo {author}
  {\bibfnamefont {T.}~\bibnamefont {{Chodukowski}}}, \bibinfo {author}
  {\bibfnamefont {Z.}~\bibnamefont {{Rusiniak}}}, \bibinfo {author}
  {\bibfnamefont {R.}~\bibnamefont {{Dudzak}}}, \bibinfo {author}
  {\bibfnamefont {J.}~\bibnamefont {{Dostal}}}, \bibinfo {author}
  {\bibfnamefont {M.}~\bibnamefont {{Kalal}}}, \bibinfo {author} {\bibfnamefont
  {I.}~\bibnamefont {{Kochetkov}}}, \bibinfo {author} {\bibfnamefont
  {M.}~\bibnamefont {{Krupka}}}, \bibinfo {author} {\bibfnamefont
  {S.}~\bibnamefont {{Borodziuk}}},\ and\ \bibinfo {author} {\bibfnamefont
  {P.}~\bibnamefont {{Pisarczyk}}},\ }\bibfield  {title} {\bibinfo {title}
  {{Implementation of amplitude-phase analysis of complex interferograms for
  measurement of spontaneous magnetic fields in laser generated plasma}},\
  }\href {https://doi.org/10.1063/5.0020511} {\bibfield  {journal} {\bibinfo
  {journal} {AIP Advances}\ }\textbf {\bibinfo {volume} {10}},\ \bibinfo {eid}
  {115201} (\bibinfo {year} {2020})}\BibitemShut {NoStop}%
\bibitem [{\citenamefont {{Verona}}\ \emph {et~al.}(2020)\citenamefont
  {{Verona}}, \citenamefont {{Marinelli}}, \citenamefont {{Palomba}},
  \citenamefont {{Verona-Rinati}}, \citenamefont {{Salvadori}}, \citenamefont
  {{Consoli}}, \citenamefont {{Cipriani}}, \citenamefont {{Antici}},
  \citenamefont {{Migliorati}}, \citenamefont {{Bisesto}},\ and\ \citenamefont
  {{Pompili}}}]{Verona_2020}%
  \BibitemOpen
  \bibfield  {author} {\bibinfo {author} {\bibfnamefont {C.}~\bibnamefont
  {{Verona}}}, \bibinfo {author} {\bibfnamefont {M.}~\bibnamefont
  {{Marinelli}}}, \bibinfo {author} {\bibfnamefont {S.}~\bibnamefont
  {{Palomba}}}, \bibinfo {author} {\bibfnamefont {G.}~\bibnamefont
  {{Verona-Rinati}}}, \bibinfo {author} {\bibfnamefont {M.}~\bibnamefont
  {{Salvadori}}}, \bibinfo {author} {\bibfnamefont {F.}~\bibnamefont
  {{Consoli}}}, \bibinfo {author} {\bibfnamefont {M.}~\bibnamefont
  {{Cipriani}}}, \bibinfo {author} {\bibfnamefont {P.}~\bibnamefont
  {{Antici}}}, \bibinfo {author} {\bibfnamefont {M.}~\bibnamefont
  {{Migliorati}}}, \bibinfo {author} {\bibfnamefont {F.}~\bibnamefont
  {{Bisesto}}},\ and\ \bibinfo {author} {\bibfnamefont {R.}~\bibnamefont
  {{Pompili}}},\ }\bibfield  {title} {\bibinfo {title} {{Comparison of single
  crystal diamond TOF detectors in planar and transverse configuration}},\
  }\href {https://doi.org/10.1088/1748-0221/15/09/C09066} {\bibfield  {journal}
  {\bibinfo  {journal} {J. Instrum.}\ }\textbf {\bibinfo {volume} {15}}\bibinfo
   {number} { (9)},\ \bibinfo {pages} {C09066}}\BibitemShut {NoStop}%
\bibitem [{\citenamefont {Salvadori}\ \emph {et~al.}(2021)\citenamefont
  {Salvadori}, \citenamefont {Consoli}, \citenamefont {Verona}, \citenamefont
  {Cipriani}, \citenamefont {Anania}, \citenamefont {Andreoli}, \citenamefont
  {Antici}, \citenamefont {Bisesto}, \citenamefont {Costa}, \citenamefont
  {Cristofari} \emph {et~al.}}]{Salvadori_2021}%
  \BibitemOpen
\bibfield  {number} {  }\bibfield  {author} {\bibinfo {author} {\bibfnamefont
  {M.}~\bibnamefont {Salvadori}}, \bibinfo {author} {\bibfnamefont
  {F.}~\bibnamefont {Consoli}}, \bibinfo {author} {\bibfnamefont
  {C.}~\bibnamefont {Verona}}, \bibinfo {author} {\bibfnamefont
  {M.}~\bibnamefont {Cipriani}}, \bibinfo {author} {\bibfnamefont
  {M.}~\bibnamefont {Anania}}, \bibinfo {author} {\bibfnamefont
  {P.}~\bibnamefont {Andreoli}}, \bibinfo {author} {\bibfnamefont
  {P.}~\bibnamefont {Antici}}, \bibinfo {author} {\bibfnamefont
  {F.}~\bibnamefont {Bisesto}}, \bibinfo {author} {\bibfnamefont
  {G.}~\bibnamefont {Costa}}, \bibinfo {author} {\bibfnamefont
  {G.}~\bibnamefont {Cristofari}}, \emph {et~al.},\ }\bibfield  {title}
  {\bibinfo {title} {Accurate spectra for high energy ions by advanced
  time-of-flight diamond-detector schemes in experiments with high energy and
  intensity lasers},\ }\href {https://doi.org/10.1038/s41598-021-82655-w}
  {\bibfield  {journal} {\bibinfo  {journal} {Sci. Rep.}\ }\textbf {\bibinfo
  {volume} {11}},\ \bibinfo {pages} {3071} (\bibinfo {year}
  {2021})}\BibitemShut {NoStop}%
\bibitem [{\citenamefont {Krupka}\ \emph {et~al.}(2021)\citenamefont {Krupka},
  \citenamefont {Singh}, \citenamefont {Pisarczyk}, \citenamefont {Dostal},
  \citenamefont {Kalal}, \citenamefont {Krasa}, \citenamefont {Dudzak},
  \citenamefont {Burian}, \citenamefont {Jelinek}, \citenamefont {Chodukowski}
  \emph {et~al.}}]{Krupka_2021}%
  \BibitemOpen
  \bibfield  {author} {\bibinfo {author} {\bibfnamefont {M.}~\bibnamefont
  {Krupka}}, \bibinfo {author} {\bibfnamefont {S.}~\bibnamefont {Singh}},
  \bibinfo {author} {\bibfnamefont {T.}~\bibnamefont {Pisarczyk}}, \bibinfo
  {author} {\bibfnamefont {J.}~\bibnamefont {Dostal}}, \bibinfo {author}
  {\bibfnamefont {M.}~\bibnamefont {Kalal}}, \bibinfo {author} {\bibfnamefont
  {J.}~\bibnamefont {Krasa}}, \bibinfo {author} {\bibfnamefont
  {R.}~\bibnamefont {Dudzak}}, \bibinfo {author} {\bibfnamefont
  {T.}~\bibnamefont {Burian}}, \bibinfo {author} {\bibfnamefont
  {S.}~\bibnamefont {Jelinek}}, \bibinfo {author} {\bibfnamefont
  {T.}~\bibnamefont {Chodukowski}}, \emph {et~al.},\ }\bibfield  {title}
  {\bibinfo {title} {Design of modular multi-channel electron spectrometers for
  application in laser matter interaction experiments at prague asterix laser
  system},\ }\href@noop {} {\bibfield  {journal} {\bibinfo  {journal} {Review
  of Scientific Instruments}\ }\textbf {\bibinfo {volume} {92}},\ \bibinfo
  {pages} {023514} (\bibinfo {year} {2021})}\BibitemShut {NoStop}%
\bibitem [{\citenamefont {{Salgado-L{\'o}pez}}\ \emph
  {et~al.}(2022)\citenamefont {{Salgado-L{\'o}pez}}, \citenamefont
  {{Api{\~n}aniz}}, \citenamefont {{Henares}}, \citenamefont
  {{P{\'e}rez-Hern{\'a}ndez}}, \citenamefont {{de Luis}}, \citenamefont
  {{Volpe}},\ and\ \citenamefont {{Gatti}}}]{Salgado_2022}%
  \BibitemOpen
  \bibfield  {author} {\bibinfo {author} {\bibfnamefont {C.}~\bibnamefont
  {{Salgado-L{\'o}pez}}}, \bibinfo {author} {\bibfnamefont {J.~I.}\
  \bibnamefont {{Api{\~n}aniz}}}, \bibinfo {author} {\bibfnamefont {J.~L.}\
  \bibnamefont {{Henares}}}, \bibinfo {author} {\bibfnamefont {J.~A.}\
  \bibnamefont {{P{\'e}rez-Hern{\'a}ndez}}}, \bibinfo {author} {\bibfnamefont
  {D.}~\bibnamefont {{de Luis}}}, \bibinfo {author} {\bibfnamefont
  {L.}~\bibnamefont {{Volpe}}},\ and\ \bibinfo {author} {\bibfnamefont
  {G.}~\bibnamefont {{Gatti}}},\ }\bibfield  {title} {\bibinfo {title}
  {{Angular-Resolved Thomson Parabola Spectrometer for Laser-Driven Ion
  Accelerators}},\ }\href {https://doi.org/10.3390/s22093239} {\bibfield
  {journal} {\bibinfo  {journal} {Sensors}\ }\textbf {\bibinfo {volume} {22}},\
  \bibinfo {pages} {3239} (\bibinfo {year} {2022})}\BibitemShut {NoStop}%
\bibitem [{\citenamefont {Ospina-Boh{\'o}rquez}(2022)}]{Ospina_2022}%
  \BibitemOpen
  \bibfield  {author} {\bibinfo {author} {\bibfnamefont {V.}~\bibnamefont
  {Ospina-Boh{\'o}rquez}},\ }\emph {\bibinfo {title} {Ion acceleration from
  high-intensity laser interactions with quasi-critical-gas targets:
  experiments and numerical simulations}},\ \href@noop {} {Ph.D. thesis},\
  \bibinfo  {school} {Universit{\'e} de Bordeaux (France); Universidad de
  Salamanca (Spain)} (\bibinfo {year} {2022}),\ \bibinfo {note} {available at
  \url{https://theses.hal.science/tel-04011154}}\BibitemShut {NoStop}%
\bibitem [{\citenamefont {{Sprangle}}\ \emph {et~al.}(1987)\citenamefont
  {{Sprangle}}, \citenamefont {{Tang}},\ and\ \citenamefont
  {{Esarey}}}]{Sprangle_1987}%
  \BibitemOpen
  \bibfield  {author} {\bibinfo {author} {\bibfnamefont {P.}~\bibnamefont
  {{Sprangle}}}, \bibinfo {author} {\bibfnamefont {C.-M.}\ \bibnamefont
  {{Tang}}},\ and\ \bibinfo {author} {\bibfnamefont {E.}~\bibnamefont
  {{Esarey}}},\ }\bibfield  {title} {\bibinfo {title} {{Relativistic
  self-focusing of short-pulse radiation beams in plasmas}},\ }\href
  {https://doi.org/10.1109/TPS.1987.4316677} {\bibfield  {journal} {\bibinfo
  {journal} {IEEE Transactions on Plasma Science}\ }\textbf {\bibinfo {volume}
  {15}},\ \bibinfo {pages} {145} (\bibinfo {year} {1987})}\BibitemShut
  {NoStop}%
\bibitem [{\citenamefont {{Lefebvre}}\ \emph {et~al.}(2003)\citenamefont
  {{Lefebvre}}, \citenamefont {{Cochet}}, \citenamefont {{Fritzler}},
  \citenamefont {{Malka}}, \citenamefont {{Al{\'e}onard}}, \citenamefont
  {{Chemin}}, \citenamefont {{Darbon}}, \citenamefont {{Disdier}},
  \citenamefont {{Faure}}, \citenamefont {{Fedotoff}}, \citenamefont
  {{Landoas}}, \citenamefont {{Malka}}, \citenamefont {{M{\'e}ot}},
  \citenamefont {{Morel}}, \citenamefont {{Rabec LeGloahec}}, \citenamefont
  {{Rouyer}}, \citenamefont {{Rubbelynck}}, \citenamefont {{Tikhonchuk}},
  \citenamefont {{Wrobel}}, \citenamefont {{Audebert}},\ and\ \citenamefont
  {{Rousseaux}}}]{Lefebvre_2003}%
  \BibitemOpen
  \bibfield  {author} {\bibinfo {author} {\bibfnamefont {E.}~\bibnamefont
  {{Lefebvre}}}, \bibinfo {author} {\bibfnamefont {N.}~\bibnamefont
  {{Cochet}}}, \bibinfo {author} {\bibfnamefont {S.}~\bibnamefont
  {{Fritzler}}}, \bibinfo {author} {\bibfnamefont {V.}~\bibnamefont {{Malka}}},
  \bibinfo {author} {\bibfnamefont {M.~M.}\ \bibnamefont {{Al{\'e}onard}}},
  \bibinfo {author} {\bibfnamefont {J.~F.}\ \bibnamefont {{Chemin}}}, \bibinfo
  {author} {\bibfnamefont {S.}~\bibnamefont {{Darbon}}}, \bibinfo {author}
  {\bibfnamefont {L.}~\bibnamefont {{Disdier}}}, \bibinfo {author}
  {\bibfnamefont {J.}~\bibnamefont {{Faure}}}, \bibinfo {author} {\bibfnamefont
  {A.}~\bibnamefont {{Fedotoff}}}, \bibinfo {author} {\bibfnamefont
  {O.}~\bibnamefont {{Landoas}}}, \bibinfo {author} {\bibfnamefont
  {G.}~\bibnamefont {{Malka}}}, \bibinfo {author} {\bibfnamefont
  {V.}~\bibnamefont {{M{\'e}ot}}}, \bibinfo {author} {\bibfnamefont
  {P.}~\bibnamefont {{Morel}}}, \bibinfo {author} {\bibfnamefont
  {M.}~\bibnamefont {{Rabec LeGloahec}}}, \bibinfo {author} {\bibfnamefont
  {A.}~\bibnamefont {{Rouyer}}}, \bibinfo {author} {\bibfnamefont
  {C.}~\bibnamefont {{Rubbelynck}}}, \bibinfo {author} {\bibfnamefont
  {V.}~\bibnamefont {{Tikhonchuk}}}, \bibinfo {author} {\bibfnamefont
  {R.}~\bibnamefont {{Wrobel}}}, \bibinfo {author} {\bibfnamefont
  {P.}~\bibnamefont {{Audebert}}},\ and\ \bibinfo {author} {\bibfnamefont
  {C.}~\bibnamefont {{Rousseaux}}},\ }\bibfield  {title} {\bibinfo {title}
  {{Electron and photon production from relativistic laser plasma
  interactions}},\ }\href {https://doi.org/10.1088/0029-5515/43/7/317}
  {\bibfield  {journal} {\bibinfo  {journal} {Nucl. Fusion}\ }\textbf {\bibinfo
  {volume} {43}},\ \bibinfo {pages} {629} (\bibinfo {year} {2003})}\BibitemShut
  {NoStop}%
\bibitem [{\citenamefont {Debayle}\ \emph {et~al.}(2017)\citenamefont
  {Debayle}, \citenamefont {Mollica}, \citenamefont {Vauzour}, \citenamefont
  {Wan}, \citenamefont {Flacco}, \citenamefont {Malka}, \citenamefont
  {Davoine},\ and\ \citenamefont {Gremillet}}]{Debayle_2017}%
  \BibitemOpen
  \bibfield  {author} {\bibinfo {author} {\bibfnamefont {A.}~\bibnamefont
  {Debayle}}, \bibinfo {author} {\bibfnamefont {F.}~\bibnamefont {Mollica}},
  \bibinfo {author} {\bibfnamefont {B.}~\bibnamefont {Vauzour}}, \bibinfo
  {author} {\bibfnamefont {Y.}~\bibnamefont {Wan}}, \bibinfo {author}
  {\bibfnamefont {A.}~\bibnamefont {Flacco}}, \bibinfo {author} {\bibfnamefont
  {V.}~\bibnamefont {Malka}}, \bibinfo {author} {\bibfnamefont
  {X.}~\bibnamefont {Davoine}},\ and\ \bibinfo {author} {\bibfnamefont
  {L.}~\bibnamefont {Gremillet}},\ }\bibfield  {title} {\bibinfo {title}
  {Electron heating by intense short-pulse lasers propagating through
  near-critical plasmas},\ }\href {https://doi.org/10.1088/1367-2630/aa953f}
  {\bibfield  {journal} {\bibinfo  {journal} {New J. Phys.}\ }\textbf {\bibinfo
  {volume} {19}},\ \bibinfo {pages} {123013} (\bibinfo {year}
  {2017})}\BibitemShut {NoStop}%
\bibitem [{\citenamefont {{P{\'e}rez}}\ \emph {et~al.}(2012)\citenamefont
  {{P{\'e}rez}}, \citenamefont {{Gremillet}}, \citenamefont {{Decoster}},
  \citenamefont {{Drouin}},\ and\ \citenamefont {{Lefebvre}}}]{Perez_2012}%
  \BibitemOpen
  \bibfield  {author} {\bibinfo {author} {\bibfnamefont {F.}~\bibnamefont
  {{P{\'e}rez}}}, \bibinfo {author} {\bibfnamefont {L.}~\bibnamefont
  {{Gremillet}}}, \bibinfo {author} {\bibfnamefont {A.}~\bibnamefont
  {{Decoster}}}, \bibinfo {author} {\bibfnamefont {M.}~\bibnamefont
  {{Drouin}}},\ and\ \bibinfo {author} {\bibfnamefont {E.}~\bibnamefont
  {{Lefebvre}}},\ }\bibfield  {title} {\bibinfo {title} {{Improved modeling of
  relativistic collisions and collisional ionization in particle-in-cell
  codes}},\ }\href {https://doi.org/10.1063/1.4742167} {\bibfield  {journal}
  {\bibinfo  {journal} {Phys. Plasmas}\ }\textbf {\bibinfo {volume} {19}},\
  \bibinfo {eid} {083104} (\bibinfo {year} {2012})}\BibitemShut {NoStop}%
\bibitem [{\citenamefont {{Nuter}}\ \emph {et~al.}(2011)\citenamefont
  {{Nuter}}, \citenamefont {{Gremillet}}, \citenamefont {{Lefebvre}},
  \citenamefont {{L{\'e}vy}}, \citenamefont {{Ceccotti}},\ and\ \citenamefont
  {{Martin}}}]{Nuter_2011}%
  \BibitemOpen
  \bibfield  {author} {\bibinfo {author} {\bibfnamefont {R.}~\bibnamefont
  {{Nuter}}}, \bibinfo {author} {\bibfnamefont {L.}~\bibnamefont
  {{Gremillet}}}, \bibinfo {author} {\bibfnamefont {E.}~\bibnamefont
  {{Lefebvre}}}, \bibinfo {author} {\bibfnamefont {A.}~\bibnamefont
  {{L{\'e}vy}}}, \bibinfo {author} {\bibfnamefont {T.}~\bibnamefont
  {{Ceccotti}}},\ and\ \bibinfo {author} {\bibfnamefont {P.}~\bibnamefont
  {{Martin}}},\ }\bibfield  {title} {\bibinfo {title} {{Field ionization model
  implemented in Particle In Cell code and applied to laser-accelerated carbon
  ions}},\ }\href {https://doi.org/10.1063/1.3559494} {\bibfield  {journal}
  {\bibinfo  {journal} {Phys. Plasmas}\ }\textbf {\bibinfo {volume} {18}},\
  \bibinfo {eid} {033107} (\bibinfo {year} {2011})}\BibitemShut {NoStop}%
\bibitem [{\citenamefont {{Giulietti}}\ \emph {et~al.}(2013)\citenamefont
  {{Giulietti}}, \citenamefont {{Andr{\'e}}}, \citenamefont {{Dobosz
  Dufr{\'e}noy}}, \citenamefont {{Giulietti}}, \citenamefont {{Hosokai}},
  \citenamefont {{Koester}}, \citenamefont {{Kotaki}}, \citenamefont
  {{Labate}}, \citenamefont {{Levato}}, \citenamefont {{Nuter}}, \citenamefont
  {{Pathak}}, \citenamefont {{Monot}},\ and\ \citenamefont
  {{Gizzi}}}]{Giuletti_2013}%
  \BibitemOpen
  \bibfield  {author} {\bibinfo {author} {\bibfnamefont {A.}~\bibnamefont
  {{Giulietti}}}, \bibinfo {author} {\bibfnamefont {A.}~\bibnamefont
  {{Andr{\'e}}}}, \bibinfo {author} {\bibfnamefont {S.}~\bibnamefont {{Dobosz
  Dufr{\'e}noy}}}, \bibinfo {author} {\bibfnamefont {D.}~\bibnamefont
  {{Giulietti}}}, \bibinfo {author} {\bibfnamefont {T.}~\bibnamefont
  {{Hosokai}}}, \bibinfo {author} {\bibfnamefont {P.}~\bibnamefont
  {{Koester}}}, \bibinfo {author} {\bibfnamefont {H.}~\bibnamefont {{Kotaki}}},
  \bibinfo {author} {\bibfnamefont {L.}~\bibnamefont {{Labate}}}, \bibinfo
  {author} {\bibfnamefont {T.}~\bibnamefont {{Levato}}}, \bibinfo {author}
  {\bibfnamefont {R.}~\bibnamefont {{Nuter}}}, \bibinfo {author} {\bibfnamefont
  {N.~C.}\ \bibnamefont {{Pathak}}}, \bibinfo {author} {\bibfnamefont
  {P.}~\bibnamefont {{Monot}}},\ and\ \bibinfo {author} {\bibfnamefont {L.~A.}\
  \bibnamefont {{Gizzi}}},\ }\bibfield  {title} {\bibinfo {title} {{Space- and
  time-resolved observation of extreme laser frequency upshifting during
  ultrafast-ionization}},\ }\href {https://doi.org/10.1063/1.4818602}
  {\bibfield  {journal} {\bibinfo  {journal} {Phys. Plasmas}\ }\textbf
  {\bibinfo {volume} {20}},\ \bibinfo {eid} {082307} (\bibinfo {year}
  {2013})}\BibitemShut {NoStop}%
\bibitem [{\citenamefont {{Nicholls}}\ \emph {et~al.}(1974)\citenamefont
  {{Nicholls}}, \citenamefont {{Sichel}}, \citenamefont {{Fry}},\ and\
  \citenamefont {{Glass}}}]{Nicholls_1974}%
  \BibitemOpen
  \bibfield  {author} {\bibinfo {author} {\bibfnamefont {J.~A.}\ \bibnamefont
  {{Nicholls}}}, \bibinfo {author} {\bibfnamefont {M.}~\bibnamefont
  {{Sichel}}}, \bibinfo {author} {\bibfnamefont {R.}~\bibnamefont {{Fry}}},\
  and\ \bibinfo {author} {\bibfnamefont {D.~R.}\ \bibnamefont {{Glass}}},\
  }\bibfield  {title} {\bibinfo {title} {{Theoretical and experimental study of
  cylindrical shock and heterogeneous detonation waves}},\ }\href
  {https://doi.org/10.1016/0094-5765(74)90105-2} {\bibfield  {journal}
  {\bibinfo  {journal} {Acta Astronautica}\ }\textbf {\bibinfo {volume} {1}},\
  \bibinfo {pages} {385} (\bibinfo {year} {1974})}\BibitemShut {NoStop}%
\bibitem [{\citenamefont {{Borghesi}}\ \emph {et~al.}(1998)\citenamefont
  {{Borghesi}}, \citenamefont {{MacKinnon}}, \citenamefont {{Gaillard}},
  \citenamefont {{Willi}}, \citenamefont {{Pukhov}},\ and\ \citenamefont
  {{Meyer-Ter-Vehn}}}]{Borghesi_1998}%
  \BibitemOpen
  \bibfield  {author} {\bibinfo {author} {\bibfnamefont {M.}~\bibnamefont
  {{Borghesi}}}, \bibinfo {author} {\bibfnamefont {A.~J.}\ \bibnamefont
  {{MacKinnon}}}, \bibinfo {author} {\bibfnamefont {R.}~\bibnamefont
  {{Gaillard}}}, \bibinfo {author} {\bibfnamefont {O.}~\bibnamefont {{Willi}}},
  \bibinfo {author} {\bibfnamefont {A.}~\bibnamefont {{Pukhov}}},\ and\
  \bibinfo {author} {\bibfnamefont {J.}~\bibnamefont {{Meyer-Ter-Vehn}}},\
  }\bibfield  {title} {\bibinfo {title} {{Large Quasistatic Magnetic Fields
  Generated by a Relativistically Intense Laser Pulse Propagating in a
  Preionized Plasma}},\ }\href {https://doi.org/10.1103/PhysRevLett.80.5137}
  {\bibfield  {journal} {\bibinfo  {journal} {Phys. Rev. Lett.}\ }\textbf
  {\bibinfo {volume} {80}},\ \bibinfo {pages} {5137} (\bibinfo {year}
  {1998})}\BibitemShut {NoStop}%
\bibitem [{\citenamefont {{Sarri}}\ \emph {et~al.}(2012)\citenamefont
  {{Sarri}}, \citenamefont {{Macchi}}, \citenamefont {{Cecchetti}},
  \citenamefont {{Kar}}, \citenamefont {{Liseykina}}, \citenamefont {{Yang}},
  \citenamefont {{Dieckmann}}, \citenamefont {{Fuchs}}, \citenamefont
  {{Galimberti}}, \citenamefont {{Gizzi}}, \citenamefont {{Jung}},
  \citenamefont {{Kourakis}}, \citenamefont {{Osterholz}}, \citenamefont
  {{Pegoraro}}, \citenamefont {{Robinson}}, \citenamefont {{Romagnani}},
  \citenamefont {{Willi}},\ and\ \citenamefont {{Borghesi}}}]{Sarri_2012}%
  \BibitemOpen
  \bibfield  {author} {\bibinfo {author} {\bibfnamefont {G.}~\bibnamefont
  {{Sarri}}}, \bibinfo {author} {\bibfnamefont {A.}~\bibnamefont {{Macchi}}},
  \bibinfo {author} {\bibfnamefont {C.~A.}\ \bibnamefont {{Cecchetti}}},
  \bibinfo {author} {\bibfnamefont {S.}~\bibnamefont {{Kar}}}, \bibinfo
  {author} {\bibfnamefont {T.~V.}\ \bibnamefont {{Liseykina}}}, \bibinfo
  {author} {\bibfnamefont {X.~H.}\ \bibnamefont {{Yang}}}, \bibinfo {author}
  {\bibfnamefont {M.~E.}\ \bibnamefont {{Dieckmann}}}, \bibinfo {author}
  {\bibfnamefont {J.}~\bibnamefont {{Fuchs}}}, \bibinfo {author} {\bibfnamefont
  {M.}~\bibnamefont {{Galimberti}}}, \bibinfo {author} {\bibfnamefont {L.~A.}\
  \bibnamefont {{Gizzi}}}, \bibinfo {author} {\bibfnamefont {R.}~\bibnamefont
  {{Jung}}}, \bibinfo {author} {\bibfnamefont {I.}~\bibnamefont {{Kourakis}}},
  \bibinfo {author} {\bibfnamefont {J.}~\bibnamefont {{Osterholz}}}, \bibinfo
  {author} {\bibfnamefont {F.}~\bibnamefont {{Pegoraro}}}, \bibinfo {author}
  {\bibfnamefont {A.~P.~L.}\ \bibnamefont {{Robinson}}}, \bibinfo {author}
  {\bibfnamefont {L.}~\bibnamefont {{Romagnani}}}, \bibinfo {author}
  {\bibfnamefont {O.}~\bibnamefont {{Willi}}},\ and\ \bibinfo {author}
  {\bibfnamefont {M.}~\bibnamefont {{Borghesi}}},\ }\bibfield  {title}
  {\bibinfo {title} {{Dynamics of Self-Generated, Large Amplitude Magnetic
  Fields Following High-Intensity Laser Matter Interaction}},\ }\href
  {https://doi.org/10.1103/PhysRevLett.109.205002} {\bibfield  {journal}
  {\bibinfo  {journal} {Phys. Rev. Lett.}\ }\textbf {\bibinfo {volume} {109}},\
  \bibinfo {eid} {205002} (\bibinfo {year} {2012})}\BibitemShut {NoStop}%
\bibitem [{\citenamefont {{Tatarakis}}\ \emph {et~al.}(2003)\citenamefont
  {{Tatarakis}}, \citenamefont {{Beg}}, \citenamefont {{Clark}}, \citenamefont
  {{Dangor}}, \citenamefont {{Edwards}}, \citenamefont {{Evans}}, \citenamefont
  {{Goldsack}}, \citenamefont {{Ledingham}}, \citenamefont {{Norreys}},
  \citenamefont {{Sinclair}}, \citenamefont {{Wei}}, \citenamefont {{Zepf}},\
  and\ \citenamefont {{Krushelnick}}}]{Tatarakis_2003}%
  \BibitemOpen
  \bibfield  {author} {\bibinfo {author} {\bibfnamefont {M.}~\bibnamefont
  {{Tatarakis}}}, \bibinfo {author} {\bibfnamefont {F.~N.}\ \bibnamefont
  {{Beg}}}, \bibinfo {author} {\bibfnamefont {E.~L.}\ \bibnamefont {{Clark}}},
  \bibinfo {author} {\bibfnamefont {A.~E.}\ \bibnamefont {{Dangor}}}, \bibinfo
  {author} {\bibfnamefont {R.~D.}\ \bibnamefont {{Edwards}}}, \bibinfo {author}
  {\bibfnamefont {R.~G.}\ \bibnamefont {{Evans}}}, \bibinfo {author}
  {\bibfnamefont {T.~J.}\ \bibnamefont {{Goldsack}}}, \bibinfo {author}
  {\bibfnamefont {K.~W.}\ \bibnamefont {{Ledingham}}}, \bibinfo {author}
  {\bibfnamefont {P.~A.}\ \bibnamefont {{Norreys}}}, \bibinfo {author}
  {\bibfnamefont {M.~A.}\ \bibnamefont {{Sinclair}}}, \bibinfo {author}
  {\bibfnamefont {M.~S.}\ \bibnamefont {{Wei}}}, \bibinfo {author}
  {\bibfnamefont {M.}~\bibnamefont {{Zepf}}},\ and\ \bibinfo {author}
  {\bibfnamefont {K.}~\bibnamefont {{Krushelnick}}},\ }\bibfield  {title}
  {\bibinfo {title} {{Propagation Instabilities of High-Intensity
  Laser-Produced Electron Beams}},\ }\href
  {https://doi.org/10.1103/PhysRevLett.90.175001} {\bibfield  {journal}
  {\bibinfo  {journal} {Phys. Rev. Lett.}\ }\textbf {\bibinfo {volume} {90}},\
  \bibinfo {eid} {175001} (\bibinfo {year} {2003})}\BibitemShut {NoStop}%
\bibitem [{\citenamefont {{G{\"o}de}}\ \emph {et~al.}(2017)\citenamefont
  {{G{\"o}de}}, \citenamefont {{R{\"o}del}}, \citenamefont {{Zeil}},
  \citenamefont {{Mishra}}, \citenamefont {{Gauthier}}, \citenamefont
  {{Brack}}, \citenamefont {{Kluge}}, \citenamefont {{MacDonald}},
  \citenamefont {{Metzkes}}, \citenamefont {{Obst}}, \citenamefont {{Rehwald}},
  \citenamefont {{Ruyer}}, \citenamefont {{Schlenvoigt}}, \citenamefont
  {{Schumaker}}, \citenamefont {{Sommer}}, \citenamefont {{Cowan}},
  \citenamefont {{Schramm}}, \citenamefont {{Glenzer}},\ and\ \citenamefont
  {{Fiuza}}}]{Gode_2017}%
  \BibitemOpen
  \bibfield  {author} {\bibinfo {author} {\bibfnamefont {S.}~\bibnamefont
  {{G{\"o}de}}}, \bibinfo {author} {\bibfnamefont {C.}~\bibnamefont
  {{R{\"o}del}}}, \bibinfo {author} {\bibfnamefont {K.}~\bibnamefont {{Zeil}}},
  \bibinfo {author} {\bibfnamefont {R.}~\bibnamefont {{Mishra}}}, \bibinfo
  {author} {\bibfnamefont {M.}~\bibnamefont {{Gauthier}}}, \bibinfo {author}
  {\bibfnamefont {F.~E.}\ \bibnamefont {{Brack}}}, \bibinfo {author}
  {\bibfnamefont {T.}~\bibnamefont {{Kluge}}}, \bibinfo {author} {\bibfnamefont
  {M.~J.}\ \bibnamefont {{MacDonald}}}, \bibinfo {author} {\bibfnamefont
  {J.}~\bibnamefont {{Metzkes}}}, \bibinfo {author} {\bibfnamefont
  {L.}~\bibnamefont {{Obst}}}, \bibinfo {author} {\bibfnamefont
  {M.}~\bibnamefont {{Rehwald}}}, \bibinfo {author} {\bibfnamefont
  {C.}~\bibnamefont {{Ruyer}}}, \bibinfo {author} {\bibfnamefont {H.~P.}\
  \bibnamefont {{Schlenvoigt}}}, \bibinfo {author} {\bibfnamefont
  {W.}~\bibnamefont {{Schumaker}}}, \bibinfo {author} {\bibfnamefont
  {P.}~\bibnamefont {{Sommer}}}, \bibinfo {author} {\bibfnamefont {T.~E.}\
  \bibnamefont {{Cowan}}}, \bibinfo {author} {\bibfnamefont {U.}~\bibnamefont
  {{Schramm}}}, \bibinfo {author} {\bibfnamefont {S.}~\bibnamefont
  {{Glenzer}}},\ and\ \bibinfo {author} {\bibfnamefont {F.}~\bibnamefont
  {{Fiuza}}},\ }\bibfield  {title} {\bibinfo {title} {{Relativistic Electron
  Streaming Instabilities Modulate Proton Beams Accelerated in Laser-Plasma
  Interactions}},\ }\href {https://doi.org/10.1103/PhysRevLett.118.194801}
  {\bibfield  {journal} {\bibinfo  {journal} {Phys. Rev. Lett.}\ }\textbf
  {\bibinfo {volume} {118}},\ \bibinfo {eid} {194801} (\bibinfo {year}
  {2017})}\BibitemShut {NoStop}%
\bibitem [{\citenamefont {{Zhou}}\ \emph {et~al.}(2018)\citenamefont {{Zhou}},
  \citenamefont {{Bai}}, \citenamefont {{Tian}}, \citenamefont {{Sun}},
  \citenamefont {{Cao}},\ and\ \citenamefont {{Liu}}}]{Zhou_2018}%
  \BibitemOpen
  \bibfield  {author} {\bibinfo {author} {\bibfnamefont {S.}~\bibnamefont
  {{Zhou}}}, \bibinfo {author} {\bibfnamefont {Y.}~\bibnamefont {{Bai}}},
  \bibinfo {author} {\bibfnamefont {Y.}~\bibnamefont {{Tian}}}, \bibinfo
  {author} {\bibfnamefont {H.}~\bibnamefont {{Sun}}}, \bibinfo {author}
  {\bibfnamefont {L.}~\bibnamefont {{Cao}}},\ and\ \bibinfo {author}
  {\bibfnamefont {J.}~\bibnamefont {{Liu}}},\ }\bibfield  {title} {\bibinfo
  {title} {{Self-Organized Kilotesla Magnetic-Tube Array in an Expanding
  Spherical Plasma Irradiated by kHz Femtosecond Laser Pulses}},\ }\href
  {https://doi.org/10.1103/PhysRevLett.121.255002} {\bibfield  {journal}
  {\bibinfo  {journal} {Phys. Rev. Lett.}\ }\textbf {\bibinfo {volume} {121}},\
  \bibinfo {eid} {255002} (\bibinfo {year} {2018})}\BibitemShut {NoStop}%
\bibitem [{\citenamefont {{Ruyer}}\ \emph {et~al.}(2020)\citenamefont
  {{Ruyer}}, \citenamefont {{Bola{\~n}os}}, \citenamefont {{Albertazzi}},
  \citenamefont {{Chen}}, \citenamefont {{Antici}}, \citenamefont
  {{B{\"o}ker}}, \citenamefont {{Dervieux}}, \citenamefont {{Lancia}},
  \citenamefont {{Nakatsutsumi}}, \citenamefont {{Romagnani}}, \citenamefont
  {{Shepherd}}, \citenamefont {{Swantusch}}, \citenamefont {{Borghesi}},
  \citenamefont {{Willi}}, \citenamefont {{P{\'e}pin}}, \citenamefont
  {{Starodubtsev}}, \citenamefont {{Grech}}, \citenamefont {{Riconda}},
  \citenamefont {{Gremillet}},\ and\ \citenamefont {{Fuchs}}}]{Ruyer_2020}%
  \BibitemOpen
  \bibfield  {author} {\bibinfo {author} {\bibfnamefont {C.}~\bibnamefont
  {{Ruyer}}}, \bibinfo {author} {\bibfnamefont {S.}~\bibnamefont
  {{Bola{\~n}os}}}, \bibinfo {author} {\bibfnamefont {B.}~\bibnamefont
  {{Albertazzi}}}, \bibinfo {author} {\bibfnamefont {S.~N.}\ \bibnamefont
  {{Chen}}}, \bibinfo {author} {\bibfnamefont {P.}~\bibnamefont {{Antici}}},
  \bibinfo {author} {\bibfnamefont {J.}~\bibnamefont {{B{\"o}ker}}}, \bibinfo
  {author} {\bibfnamefont {V.}~\bibnamefont {{Dervieux}}}, \bibinfo {author}
  {\bibfnamefont {L.}~\bibnamefont {{Lancia}}}, \bibinfo {author}
  {\bibfnamefont {M.}~\bibnamefont {{Nakatsutsumi}}}, \bibinfo {author}
  {\bibfnamefont {L.}~\bibnamefont {{Romagnani}}}, \bibinfo {author}
  {\bibfnamefont {R.}~\bibnamefont {{Shepherd}}}, \bibinfo {author}
  {\bibfnamefont {M.}~\bibnamefont {{Swantusch}}}, \bibinfo {author}
  {\bibfnamefont {M.}~\bibnamefont {{Borghesi}}}, \bibinfo {author}
  {\bibfnamefont {O.}~\bibnamefont {{Willi}}}, \bibinfo {author} {\bibfnamefont
  {H.}~\bibnamefont {{P{\'e}pin}}}, \bibinfo {author} {\bibfnamefont
  {M.}~\bibnamefont {{Starodubtsev}}}, \bibinfo {author} {\bibfnamefont
  {M.}~\bibnamefont {{Grech}}}, \bibinfo {author} {\bibfnamefont
  {C.}~\bibnamefont {{Riconda}}}, \bibinfo {author} {\bibfnamefont
  {L.}~\bibnamefont {{Gremillet}}},\ and\ \bibinfo {author} {\bibfnamefont
  {J.}~\bibnamefont {{Fuchs}}},\ }\bibfield  {title} {\bibinfo {title} {{Growth
  of concomitant laser-driven collisionless and resistive electron
  filamentation instabilities over large spatiotemporal scales}},\ }\href
  {https://doi.org/10.1038/s41567-020-0913-x} {\bibfield  {journal} {\bibinfo
  {journal} {Nat. Phys.}\ }\textbf {\bibinfo {volume} {16}},\ \bibinfo {pages}
  {983} (\bibinfo {year} {2020})}\BibitemShut {NoStop}%
\bibitem [{\citenamefont {{Mora}}(2003)}]{Mora_2003}%
  \BibitemOpen
  \bibfield  {author} {\bibinfo {author} {\bibfnamefont {P.}~\bibnamefont
  {{Mora}}},\ }\bibfield  {title} {\bibinfo {title} {{Plasma Expansion into a
  Vacuum}},\ }\href {https://doi.org/10.1103/PhysRevLett.90.185002} {\bibfield
  {journal} {\bibinfo  {journal} {\prl}\ }\textbf {\bibinfo {volume} {90}},\
  \bibinfo {eid} {185002} (\bibinfo {year} {2003})}\BibitemShut {NoStop}%
\bibitem [{\citenamefont {{Grismayer}}\ and\ \citenamefont
  {{Mora}}(2006)}]{Grismayer_2006}%
  \BibitemOpen
  \bibfield  {author} {\bibinfo {author} {\bibfnamefont {T.}~\bibnamefont
  {{Grismayer}}}\ and\ \bibinfo {author} {\bibfnamefont {P.}~\bibnamefont
  {{Mora}}},\ }\bibfield  {title} {\bibinfo {title} {{Influence of a finite
  initial ion density gradient on plasma expansion into a vacuum}},\ }\href
  {https://doi.org/10.1063/1.2178653} {\bibfield  {journal} {\bibinfo
  {journal} {Phys. Plasmas}\ }\textbf {\bibinfo {volume} {13}},\ \bibinfo {eid}
  {032103} (\bibinfo {year} {2006})}\BibitemShut {NoStop}%
\bibitem [{\citenamefont {{Robinson}}\ \emph {et~al.}(2011)\citenamefont
  {{Robinson}}, \citenamefont {{Trines}}, \citenamefont {{Polz}},\ and\
  \citenamefont {{Kaluza}}}]{Robinson_2011}%
  \BibitemOpen
  \bibfield  {author} {\bibinfo {author} {\bibfnamefont {A.~P.~L.}\
  \bibnamefont {{Robinson}}}, \bibinfo {author} {\bibfnamefont {R.~M.~G.~M.}\
  \bibnamefont {{Trines}}}, \bibinfo {author} {\bibfnamefont {J.}~\bibnamefont
  {{Polz}}},\ and\ \bibinfo {author} {\bibfnamefont {M.}~\bibnamefont
  {{Kaluza}}},\ }\bibfield  {title} {\bibinfo {title} {{Absorption of
  circularly polarized laser pulses in near-critical plasmas}},\ }\href
  {https://doi.org/10.1088/0741-3335/53/6/065019} {\bibfield  {journal}
  {\bibinfo  {journal} {Plasma Phys. Control. Fusion}\ }\textbf {\bibinfo
  {volume} {53}},\ \bibinfo {eid} {065019} (\bibinfo {year}
  {2011})}\BibitemShut {NoStop}%
\bibitem [{\citenamefont {{Rosmej}}\ \emph {et~al.}(2019)\citenamefont
  {{Rosmej}}, \citenamefont {{Andreev}}, \citenamefont {{Zaehter}},
  \citenamefont {{Zahn}}, \citenamefont {{Christ}}, \citenamefont {{Borm}},
  \citenamefont {{Radon}}, \citenamefont {{Sokolov}}, \citenamefont
  {{Pugachev}}, \citenamefont {{Khaghani}}, \citenamefont {{Horst}},
  \citenamefont {{Borisenko}}, \citenamefont {{Sklizkov}},\ and\ \citenamefont
  {{Pimenov}}}]{Rosmej_2019}%
  \BibitemOpen
  \bibfield  {author} {\bibinfo {author} {\bibfnamefont {O.~N.}\ \bibnamefont
  {{Rosmej}}}, \bibinfo {author} {\bibfnamefont {N.~E.}\ \bibnamefont
  {{Andreev}}}, \bibinfo {author} {\bibfnamefont {S.}~\bibnamefont
  {{Zaehter}}}, \bibinfo {author} {\bibfnamefont {N.}~\bibnamefont {{Zahn}}},
  \bibinfo {author} {\bibfnamefont {P.}~\bibnamefont {{Christ}}}, \bibinfo
  {author} {\bibfnamefont {B.}~\bibnamefont {{Borm}}}, \bibinfo {author}
  {\bibfnamefont {T.}~\bibnamefont {{Radon}}}, \bibinfo {author} {\bibfnamefont
  {A.}~\bibnamefont {{Sokolov}}}, \bibinfo {author} {\bibfnamefont {L.~P.}\
  \bibnamefont {{Pugachev}}}, \bibinfo {author} {\bibfnamefont
  {D.}~\bibnamefont {{Khaghani}}}, \bibinfo {author} {\bibfnamefont
  {F.}~\bibnamefont {{Horst}}}, \bibinfo {author} {\bibfnamefont {N.~G.}\
  \bibnamefont {{Borisenko}}}, \bibinfo {author} {\bibfnamefont
  {G.}~\bibnamefont {{Sklizkov}}},\ and\ \bibinfo {author} {\bibfnamefont
  {V.~G.}\ \bibnamefont {{Pimenov}}},\ }\bibfield  {title} {\bibinfo {title}
  {{Interaction of relativistically intense laser pulses with long-scale near
  critical plasmas for optimization of laser based sources of MeV electrons and
  gamma-rays}},\ }\href {https://doi.org/10.1088/1367-2630/ab1047} {\bibfield
  {journal} {\bibinfo  {journal} {New J. Phys.}\ }\textbf {\bibinfo {volume}
  {21}},\ \bibinfo {eid} {043044} (\bibinfo {year} {2019})}\BibitemShut
  {NoStop}%
\bibitem [{\citenamefont {Ehret}(2021)}]{Ehret_2021}%
  \BibitemOpen
  \bibfield  {author} {\bibinfo {author} {\bibfnamefont {M.}~\bibnamefont
  {Ehret}},\ }\emph {\bibinfo {title} {Charged particle beam acceleration and
  strong discharge currents' fields generation by laser-a study on laser-driven
  ion sources and beam transport suited for application in high-energy-density
  physics experiments}},\ \href@noop {} {Ph.D. thesis},\ \bibinfo  {school}
  {Universit{\'e} de Bordeaux (France); Technicshe Universitat Darmstadt
  (Germany)} (\bibinfo {year} {2021}),\ \bibinfo {note} {available at
  \url{https://theses.hal.science/tel-03463918}}\BibitemShut {NoStop}%
\bibitem [{\citenamefont {{Ziegler}}\ \emph {et~al.}(2010)\citenamefont
  {{Ziegler}}, \citenamefont {{Ziegler}},\ and\ \citenamefont
  {{Biersack}}}]{Ziegler_2010}%
  \BibitemOpen
  \bibfield  {author} {\bibinfo {author} {\bibfnamefont {J.~F.}\ \bibnamefont
  {{Ziegler}}}, \bibinfo {author} {\bibfnamefont {M.~D.}\ \bibnamefont
  {{Ziegler}}},\ and\ \bibinfo {author} {\bibfnamefont {J.~P.}\ \bibnamefont
  {{Biersack}}},\ }\bibfield  {title} {\bibinfo {title} {{SRIM - The stopping
  and range of ions in matter (2010)}},\ }\href
  {https://doi.org/10.1016/j.nimb.2010.02.091} {\bibfield  {journal} {\bibinfo
  {journal} {Nuclear Instruments and Methods in Physics Research B}\ }\textbf
  {\bibinfo {volume} {268}},\ \bibinfo {pages} {1818} (\bibinfo {year}
  {2010})}\BibitemShut {NoStop}%
\bibitem [{\citenamefont {Agostinelli}\ \emph {et~al.}(2003)\citenamefont
  {Agostinelli}, \citenamefont {Allison}, \citenamefont {Amako}, \citenamefont
  {Apostolakis}, \citenamefont {Araujo}, \citenamefont {Arce}, \citenamefont
  {Asai}, \citenamefont {Axen}, \citenamefont {Banerjee}, \citenamefont
  {Barrand} \emph {et~al.}}]{Agostinelli_2003}%
  \BibitemOpen
  \bibfield  {author} {\bibinfo {author} {\bibfnamefont {S.}~\bibnamefont
  {Agostinelli}}, \bibinfo {author} {\bibfnamefont {J.}~\bibnamefont
  {Allison}}, \bibinfo {author} {\bibfnamefont {K.~a.}\ \bibnamefont {Amako}},
  \bibinfo {author} {\bibfnamefont {J.}~\bibnamefont {Apostolakis}}, \bibinfo
  {author} {\bibfnamefont {H.}~\bibnamefont {Araujo}}, \bibinfo {author}
  {\bibfnamefont {P.}~\bibnamefont {Arce}}, \bibinfo {author} {\bibfnamefont
  {M.}~\bibnamefont {Asai}}, \bibinfo {author} {\bibfnamefont {D.}~\bibnamefont
  {Axen}}, \bibinfo {author} {\bibfnamefont {S.}~\bibnamefont {Banerjee}},
  \bibinfo {author} {\bibfnamefont {G.}~\bibnamefont {Barrand}}, \emph
  {et~al.},\ }\bibfield  {title} {\bibinfo {title} {Geant4—a simulation
  toolkit},\ }\href {https://doi.org/10.1016/S0168-9002(03)01368-8} {\bibfield
  {journal} {\bibinfo  {journal} {Nucl. Instrum. Methods Phys. Res. A}\
  }\textbf {\bibinfo {volume} {506}},\ \bibinfo {pages} {250} (\bibinfo {year}
  {2003})}\BibitemShut {NoStop}%
\bibitem [{\citenamefont {{Malko}}\ \emph {et~al.}(2022)\citenamefont
  {{Malko}}, \citenamefont {{Cayzac}}, \citenamefont {{Ospina-Boh{\'o}rquez}},
  \citenamefont {{Bhutwala}}, \citenamefont {{Bailly-Grandvaux}}, \citenamefont
  {{McGuffey}}, \citenamefont {{Fedosejevs}}, \citenamefont {{Vaisseau}},
  \citenamefont {{Tauschwitz}}, \citenamefont {{Api{\~n}aniz}}, \citenamefont
  {{De Luis Blanco}}, \citenamefont {{Gatti}}, \citenamefont {{Huault}},
  \citenamefont {{Hernandez}}, \citenamefont {{Hu}}, \citenamefont {{White}},
  \citenamefont {{Collins}}, \citenamefont {{Nichols}}, \citenamefont
  {{Neumayer}}, \citenamefont {{Faussurier}}, \citenamefont {{Vorberger}},
  \citenamefont {{Prestopino}}, \citenamefont {{Verona}}, \citenamefont
  {{Santos}}, \citenamefont {{Batani}}, \citenamefont {{Beg}}, \citenamefont
  {{Roso}},\ and\ \citenamefont {{Volpe}}}]{Malko_2022}%
  \BibitemOpen
  \bibfield  {author} {\bibinfo {author} {\bibfnamefont {S.}~\bibnamefont
  {{Malko}}}, \bibinfo {author} {\bibfnamefont {W.}~\bibnamefont {{Cayzac}}},
  \bibinfo {author} {\bibfnamefont {V.}~\bibnamefont {{Ospina-Boh{\'o}rquez}}},
  \bibinfo {author} {\bibfnamefont {K.}~\bibnamefont {{Bhutwala}}}, \bibinfo
  {author} {\bibfnamefont {M.}~\bibnamefont {{Bailly-Grandvaux}}}, \bibinfo
  {author} {\bibfnamefont {C.}~\bibnamefont {{McGuffey}}}, \bibinfo {author}
  {\bibfnamefont {R.}~\bibnamefont {{Fedosejevs}}}, \bibinfo {author}
  {\bibfnamefont {X.}~\bibnamefont {{Vaisseau}}}, \bibinfo {author}
  {\bibfnamefont {A.}~\bibnamefont {{Tauschwitz}}}, \bibinfo {author}
  {\bibfnamefont {J.~I.}\ \bibnamefont {{Api{\~n}aniz}}}, \bibinfo {author}
  {\bibfnamefont {D.}~\bibnamefont {{De Luis Blanco}}}, \bibinfo {author}
  {\bibfnamefont {G.}~\bibnamefont {{Gatti}}}, \bibinfo {author} {\bibfnamefont
  {M.}~\bibnamefont {{Huault}}}, \bibinfo {author} {\bibfnamefont {J.~A.~P.}\
  \bibnamefont {{Hernandez}}}, \bibinfo {author} {\bibfnamefont {S.~X.}\
  \bibnamefont {{Hu}}}, \bibinfo {author} {\bibfnamefont {A.~J.}\ \bibnamefont
  {{White}}}, \bibinfo {author} {\bibfnamefont {L.~A.}\ \bibnamefont
  {{Collins}}}, \bibinfo {author} {\bibfnamefont {K.}~\bibnamefont
  {{Nichols}}}, \bibinfo {author} {\bibfnamefont {P.}~\bibnamefont
  {{Neumayer}}}, \bibinfo {author} {\bibfnamefont {G.}~\bibnamefont
  {{Faussurier}}}, \bibinfo {author} {\bibfnamefont {J.}~\bibnamefont
  {{Vorberger}}}, \bibinfo {author} {\bibfnamefont {G.}~\bibnamefont
  {{Prestopino}}}, \bibinfo {author} {\bibfnamefont {C.}~\bibnamefont
  {{Verona}}}, \bibinfo {author} {\bibfnamefont {J.~J.}\ \bibnamefont
  {{Santos}}}, \bibinfo {author} {\bibfnamefont {D.}~\bibnamefont {{Batani}}},
  \bibinfo {author} {\bibfnamefont {F.~N.}\ \bibnamefont {{Beg}}}, \bibinfo
  {author} {\bibfnamefont {L.}~\bibnamefont {{Roso}}},\ and\ \bibinfo {author}
  {\bibfnamefont {L.}~\bibnamefont {{Volpe}}},\ }\bibfield  {title} {\bibinfo
  {title} {{Proton stopping measurements at low velocity in warm dense
  carbon}},\ }\href {https://doi.org/10.1038/s41467-022-30472-8} {\bibfield
  {journal} {\bibinfo  {journal} {Nat. Commun.}\ }\textbf {\bibinfo {volume}
  {13}},\ \bibinfo {eid} {2893} (\bibinfo {year} {2022})}\BibitemShut {NoStop}%
\bibitem [{\citenamefont {{Zylstra}}\ and\ \citenamefont
  {{Hurricane}}(2019)}]{Zylstra_2019}%
  \BibitemOpen
  \bibfield  {author} {\bibinfo {author} {\bibfnamefont {A.~B.}\ \bibnamefont
  {{Zylstra}}}\ and\ \bibinfo {author} {\bibfnamefont {O.~A.}\ \bibnamefont
  {{Hurricane}}},\ }\bibfield  {title} {\bibinfo {title} {{On alpha-particle
  transport in inertial fusion}},\ }\href {https://doi.org/10.1063/1.5101074}
  {\bibfield  {journal} {\bibinfo  {journal} {Physics of Plasmas}\ }\textbf
  {\bibinfo {volume} {26}},\ \bibinfo {eid} {062701} (\bibinfo {year}
  {2019})}\BibitemShut {NoStop}%
\bibitem [{\citenamefont {{Zylstra}}\ \emph {et~al.}(2022)\citenamefont
  {{Zylstra}}, \citenamefont {{Hurricane}}, \citenamefont {{Callahan}},
  \citenamefont {{Kritcher}}, \citenamefont {{Ralph}}, \citenamefont {{Robey}},
  \citenamefont {{Ross}}, \citenamefont {{Young}}, \citenamefont {{Baker}},
  \citenamefont {{Casey}} \emph {et~al.}}]{Zylstra_2022}%
  \BibitemOpen
  \bibfield  {author} {\bibinfo {author} {\bibfnamefont {A.~B.}\ \bibnamefont
  {{Zylstra}}}, \bibinfo {author} {\bibfnamefont {O.~A.}\ \bibnamefont
  {{Hurricane}}}, \bibinfo {author} {\bibfnamefont {D.~A.}\ \bibnamefont
  {{Callahan}}}, \bibinfo {author} {\bibfnamefont {A.~L.}\ \bibnamefont
  {{Kritcher}}}, \bibinfo {author} {\bibfnamefont {J.~E.}\ \bibnamefont
  {{Ralph}}}, \bibinfo {author} {\bibfnamefont {H.~F.}\ \bibnamefont
  {{Robey}}}, \bibinfo {author} {\bibfnamefont {J.~S.}\ \bibnamefont {{Ross}}},
  \bibinfo {author} {\bibfnamefont {C.~V.}\ \bibnamefont {{Young}}}, \bibinfo
  {author} {\bibfnamefont {K.~L.}\ \bibnamefont {{Baker}}}, \bibinfo {author}
  {\bibfnamefont {D.~T.}\ \bibnamefont {{Casey}}}, \emph {et~al.},\ }\bibfield
  {title} {\bibinfo {title} {{Burning plasma achieved in inertial fusion}},\
  }\href {https://doi.org/10.1038/s41586-021-04281-w} {\bibfield  {journal}
  {\bibinfo  {journal} {\nat}\ }\textbf {\bibinfo {volume} {601}},\ \bibinfo
  {pages} {542} (\bibinfo {year} {2022})}\BibitemShut {NoStop}%
\end{thebibliography}%

\end{document}